\documentclass[11pt]{article}
\usepackage{titlesec}

\usepackage{graphicx}
\usepackage{tikz,amsmath,amsfonts, amssymb}
\usetikzlibrary{trees,er,snakes,shapes,mindmap}
\usetikzlibrary{decorations.pathmorphing}
\usetikzlibrary{decorations.markings}
\usepackage{bbold}
\usepackage{multicol,multirow}
\usepackage{eulervm}
\def\beqar {\begin{eqnarray}}
\def\eeqar {\end{eqnarray}}
\def\beq {\begin{equation}}
\def\eeq {\end{equation}}
\def\sloppy{\tolerance=100000\hfuzz=\maxdimen\vfuzz=\maxdimen}
\def \beq  {\begin{equation}}
\def \eeq  {\end{equation}}
\def \beqar {\begin{eqnarray}}
\def \eeqar {\end{eqnarray}}
\def\sqr#1#2{{\vcenter{\vbox{\hrule height.#2pt
\hbox{\vrule width.#2pt height#1pt \kern#1pt
\vrule width.#2pt}\hrule height.#2pt}}}}

\def\la {{\langle}}
\def\ra {{\rangle}}
\def\vx {{\vec x}}
\def\vy {{\vec y}}

\def\vf {{\varphi}}

\def\Tr {{\rm Tr}}

\def\bu {\bar{u}}

\def\bw {\bar{w}}

\def\vx {{\vec x}}

\def\vy{\vec{y}}

\def\vu {\vec{u}}

\def\vf {{\varphi}}

\def\del {\partial}
\def\bdel{\bar{\partial}}
\def\a {\alpha}

\def\D {{\cal D}}
\def\bz {{\bar{z}}}

\def\A {{\cal A}}

\def\H {{\cal H}}

\def\P {{\cal P}}
\def\S {{\cal S}}

\def\half{\textstyle{1\over 2}}

\allowdisplaybreaks
\begin{document}
\begin{titlepage}
\pagestyle{empty}
\vskip .2in
\begin{center}
\vskip.3in
~\\
~\\
{\Large \bf{Elements of Geometric Quantization}}
~\\
\vskip .05in
\scalebox{1.2}[1.4]{\large \bf  \&}\\
{\Large \bf Applications to Fields and Fluids}\\
\vskip .5in
 {\large\sc V.P. Nair}\\
\vskip .1in
{Physics Department}\\
{City College of the CUNY}\\
{New York, NY 10031}\\
\begin{tabular}{r l}
E-mail:&{\fontfamily{cmtt}\fontsize{11pt}{15pt}\selectfont  \!\!\!vpnair@ccny.cuny.edu}
\end{tabular}\\
\vspace{.5in}
{Lecture Notes for the}\\
{Second Autumn School on}
{High Energy Physics \& Quantum Field Theory}\\
{Yerevan, Armenia}\\
{October 6-10, 2014}\\
\vspace{.8in}
{Abstract}\\
\end{center} 
These lecture notes (from the Second Autumn School in High Energy Physics and
Quantum Field Theory, Yerevan 2014) cover a number of topics related to geometric
quantization. Most of the material is presented from a physicist's point of view.
The original notes are posted at
\verb+http://theorphyslab-ysu.info/VW_ASW-2014/uploads/ArmeniaLectures.pdf+.
The have been revised with some additions and changes, although referencing is
still somewhat dated.
These notes are posted here as they may be good 
background material
for some recent papers.

\vskip 1in

\end{titlepage}

\pagestyle{plain}
\pagecolor{white}
\def\thepage{\roman{page}}
\setcounter{page}{1}
\tableofcontents

\newpage
\def\thepage{\arabic{page}}
\pagestyle{plain} \setcounter{page}{1}
\section{Introduction}

A physical theory, as a logical explanation of physical phenomena, is to be constructed 
taking account of general principles and incorporating data and information
from experiments. Any effects we attribute to the quantum nature of phenomena should be
included from the outset. A classical description may then be obtained, in a suitable regime of parameters, as a useful and simpler working approximation. 
The flow of logic should thus be
\beq
\left. \begin{matrix}
{\rm General ~principles ~+}\\
{\rm experimental ~input}\\
\end{matrix} \right\} \Longrightarrow ~{\rm Quantum~ theory} ~\Longrightarrow
~{\rm Classical ~approximation}
\nonumber
\eeq
But the build-up of a theory along these lines
 is almost never done in practice.
Primarily, this is because, at the human level of direct experience,
 most phenomena are well described by classical dynamics, and hence our intuition about
physical systems is mostly classical, and so we tend to start there and try to quantize the classical theory. 
This is a process with many ambiguities, but over the course of many years, we have learned to
understand the structure of this procedure of quantization.
In these lectures, I will attempt to describe some aspects of geometric quantization and consider a few examples or applications.

Quantum theory may be defined as a unitary irreducible representation (UIR) of the algebra of observables, the latter being selected by physical criteria. The algebra itself must satisfy certain conditions so as to have the correct physical requirements.
Generally it ends up as a $C^*$-algebra with further additional conditions equivalent to symmetries (such as Lorentz invariance) and so on.
We are not going to pursue such a line of development here. Instead, we will consider the essential geometry (which has to do with the symplectic structure)
of the classical theory and work out how a quantum theory can be constructed.
This will be done in the language of Hamiltonians and Hilbert space. There is yet another approach to the quantum theory, the functional integral approach, which is formulated directly in terms of the action and can be made manifestly covariant if the theory of interest has relativistic invariance.
 We will not discuss it here, but some points of overlap will be pointed out as the occasion arises.

\section{Symplectic form and Poisson brackets}

We start with the formulation of theories in the symplectic
language \cite{{arnold}, {geom}, {nair}}.
 Later, we will briefly discuss how this is connected to the action which may be used to specify the physical theory.

\subsection{Symplectic structure}

In the analytical formulation of classical physics, the key concept is the phase space,
which is a smooth even dimensional manifold
$M$ endowed with a symplectic structure $\Omega$. 
By this we mean that there is a differential 2-form
$\Omega$ defined on $M$ which is closed and nondegenerate.
Closure means that $d\Omega =0$.
The qualification ``nondegenerate" refers to the fact that
for any vector field $\xi$
on
$M$, if $i_{\xi}\Omega=0$ then $\xi$ must be zero.
In local coordinates $q^\mu$, on $M$, we can write
\beq
\Omega~=~ {\textstyle {1\over 2}}
~\Omega_{\mu \nu} \,dq^{\mu}\wedge dq^{\nu}\label{ana1}
\eeq
The closure condition $d\Omega =0$ can be written out as
\beqar
d \Omega &\equiv& {\del \Omega_{\mu\nu} \over \del q^\alpha} \,dq^\alpha \wedge dq^\mu 
\wedge dq^\nu \nonumber\\
&=&{1\over 3} \left[ {\del
\Omega_{\mu\nu} \over \del q^\alpha} +{\del \Omega_{\alpha\mu} \over \del q^\nu}+
{\del \Omega_{\nu\alpha} \over \del q^\mu}\right] dq^\alpha \wedge dq^\mu 
\wedge dq^\nu\nonumber\\
&=&0
\label{ana1.1}
\eeqar
The contraction of $\Omega$ with a vector field
$\xi =\xi^\mu {(\del / \del q^\mu)}$ is given by
\beq
i_\xi \Omega = \xi^\mu \Omega_{\mu\nu}\, dq^\nu, \hskip .3in
\xi =\xi^\mu\, {\del ~~\over \del q^\mu}
\label{ana2}
\eeq
Thus in terms of components, the equation
$i_\xi \Omega =0$  becomes $\xi^\mu \Omega_{\mu\nu}=0$.
Nondegeneracy of
$\Omega$ is then seen to be equivalent to the invertibility of
$\Omega_{\mu\nu}$ as a matrix, so that $\xi^\mu \Omega_{\mu\nu}=0$
implies $\xi^\mu =0$; in other words, $\Omega_{\mu\nu}$, viewed as a matrix, does not have
an eigenstate of eigenvalue equal to zero.

The inverse of
$\Omega_{\mu\nu}$, which will be needed for some equations,
will be denoted by $\Omega^{\mu\nu}$, i.e.,
\beq
\Omega_{\mu\nu}~\Omega^{\nu\alpha} =\delta_\mu^{~\alpha}
\label{ana3}
\eeq
For now, we will take $\Omega$ to be nondegenerate.
There are cases where the action will lead to a degenerate
$\Omega_{\mu\nu}$; this occurs when the theory has a gauge symmetry.
Elimination of certain components of the gauge field via gauge-fixing is needed to define
a nondegenerate $\Omega$; we
will consider such cases briefly later.
With the structure $\Omega$ defined on it,
$M$ is a symplectic manifold.

Since $\Omega$ is closed, at least locally we can write
\beq
\Omega = d \A \label{ana4}
\eeq
The one-form $\A$ defined by this equation is called the canonical one-form or symplectic
potential. There is an ambiguity in the definition of
$\A$ since $\A$ and $\A +d \Lambda$ will give the same
$\Omega$ for any function $\Lambda$ on $M$. As we shall see shortly, this
corresponds to the freedom of canonical transformations.

There are two types of features associated with the topology of the
phase space which are apparent at this stage.
The first question is: Is every 2-form $\Omega$ which is closed
(i.e., $d\, \Omega =0$) the exterior derivative of a 1-form $\A$?
The set of linearly independent 2-forms which cannot be expressed as 
$d \A$ for some 1-form $\A$ is the second cohomology group of
the manifold $M$; this is denoted by $\H^2 (M)$.
Thus, if the phase space
$M$ has nontrivial second cohomology, i.e., if $\H^2(M)\neq 0$, then there
are possible choices for
$\Omega$ for which there is no globally defined potential $\A$.
There are examples of physical interest where this happens.
They correspond to the Wess-Zumino terms in the action and are
related to anomalies and also to central (and other)
extensions of the algebra of observables. 

Even when $\H^2(M)=0$, there can be topological issues in defining $\A$.
If the first cohomology $\H^{1}(M)\neq 0$, this means, by definition, that 
there are 1-forms $A$ whose derivative is zero, but which are not of the form
$d$ of a function on $M$. Thus $\A$ and $\A + A$ will give the same
$\Omega$, but the difference is not just $d \Lambda$ for some function
$\Lambda$, since 
$A$ does not have to be of the form $d \Lambda$, globally.
In other words, there are inequivalent
$\A$'s for the same $\Omega$. In these cases,
one can consider the integral of $\A$ around
closed noncontractible curves on $M$.
The values of these
integrals or holonomies will be important in the quantum theory as vacuum
angles. The standard $\theta$-vacuum of nonabelian gauge theories
is an example.
We take up these topological issues in more
detail later.\footnote{To be very specific, the
 cohomologies we are talking about are over $\mathbb{R}$, this refined statement will not be needed in much of what we discuss.}

Given the symplectic structure, transformations which
preserve
$\Omega$ are evidently special; these are called {\it canonical
transformations}.
In other words, a canonical transformation is a
diffeomorphism (or coordinate transformation) of $M$
which preserves
$\Omega$. Infinitesimally, the coordinate transformation may be taken to be
$q^\mu \rightarrow q^\mu + \xi^\mu (q)$. The change in $\Omega$ due to this is given by
\beqar
\delta_\xi \,\Omega &=& 
{1\over 2} \Omega_{\mu\nu} (q + \xi) \, d(q^\mu + \xi^\mu ) \wedge 
d (q^\nu + \xi^\nu ) - {1\over 2} \Omega_{\mu\nu} \, dq^\mu \wedge dq ^\nu\nonumber\\
&=& {1\over 2} \left[\xi^\alpha {\del \Omega_{\mu\nu}\over \del q^\alpha} 
+ \Omega_{\alpha \nu} {\del \xi^\alpha \over \del q^\mu} 
+ \Omega_{\mu \alpha } {\del \xi^\alpha \over \del q^\nu} \right] \, dq^\mu \wedge dq^\nu
\nonumber\\
&= & {1\over 2} \xi^\alpha \left[ {\del \Omega_{\mu\nu}\over \del q^\alpha} 
+ {\del \Omega_{\alpha\mu}\over \del q^\nu} +{\del \Omega_{\nu\alpha}\over \del q^\mu} 
\right] \, dq^\mu \wedge dq^\nu \nonumber\\
&&\hskip .2in + {1\over 2} \left[ \del_\mu ( \xi^\alpha \Omega_{\alpha \nu})
- \del_\nu ( \xi^\alpha \Omega_{\alpha \mu}) \right] \, dq^\mu \wedge dq^\nu
\nonumber\\
&=& i_\xi (d \Omega ) + d ( i_\xi \Omega )
\label{ana4a}
\eeqar
For a canonical transformation, this change must be zero.\footnote{The right hand side of (\ref{ana4a}) is the Lie derivative of $\Omega$ with respect to
the vector field $\xi^\alpha (\del /\del q^\alpha )$. We will not use this terminology to avoid introducing
too many definitions.} Since $d \Omega = 0$, this means that
canonical transformations are generated by
vector fields
$\xi$ such that 
\beq
d~(i_\xi \Omega ) = 0
\label{ana5}
\eeq
Thus for canonical transformations, $i_\xi \Omega$ is a closed 1-form.
If the first cohomology $\H^1 (M)$ of $M$ is trivial, we can write
\beq
i_\xi \Omega = -df, \hskip .3in \xi^\alpha \, \Omega_{\alpha \nu} = - {\del f \over \del q^\nu}
\label{ana6}
\eeq
for some function $f$ on $M$. {\it In other words, to every infinitesimal
canonical transformation, we can associate a function on $M$}.\footnote{
If $\H^1(M)\neq 0$, then there is the possibility that for some
transformations $\xi$, the corresponding $i_\xi \Omega$ in a
nontrivial element of $\H^1(M)$ and hence there is no globally
defined function $f$ for this transformation.
As mentioned before this is related to the possibility
of vacuum angles in the quantum theory.
For the moment, we shall consider the case $\H^1(M)=0$.}
Since $\Omega$ is invertible, we can always
associate a vector field to a function $f$
by the correspondence
\beq
\xi^\mu =  \Omega^{\mu\nu}\partial_\nu f
\label{ana7}
\eeq
What we are saying now is that, for  a canonical transformation, we can go the other way, associating a function with the vector field which gives the canonical transformation, at least when $\H^1 (M) = 0$.
There is a
one-to-one mapping between functions on $M$ and vector fields
corresponding to infinitesimal canonical transformations. 
A vector field corresponding to an infinitesimal canonical trtasnformation
is often referred to as a Hamiltonian vector field. The function $f$
defined by (\ref{ana6}) is called the generating
function for the canonical transformation corresponding to the vector
field.

It is important that for every function $f$ on the phase space $M$, we can associate
a Hamiltonian vector field as in (\ref{ana9}). This means that all observables which are functions
on $M$ generate canonical
transformations.

\subsection{Poisson brackets}

Let $\xi ,~\eta$ be two Hamiltonian vector fields which means that they preserve 
$\Omega$; let their generating functions be $f$ and $g$ respectively.
The Lie bracket or commutator of
$\xi$ and $\eta$ is given in local coordinates by
\beq
[\xi , \eta]^\mu = \xi^\nu \partial_\nu \eta^\mu -
\eta^\nu \del_\nu \xi^\mu
\label{ana8}
\eeq
{\it We can easily verify that the commutator will also preserve $\Omega$.
We must therefore have a function corresponding to $[\xi ,\eta]$}. This is
called
the Poisson bracket of $g$ and $f$ and is denoted by $\{ g,f\}$.
We define the Poisson bracket as
\beqar
\{ f,g\}&=& i_\xi i_\eta \Omega =\eta^\mu\xi^\nu
\Omega_{\mu\nu}\nonumber\\ 
&=& -i_\xi dg =i_\eta df\nonumber\\
&=& \Omega^{\mu\nu}\del_\mu f \del_\nu g
\label{ana9}
\eeqar
Notice that, for the choice $f= q^\mu$, $g = q^\nu$, this reduces to
\beq
\{ q^\mu, \, q^\nu \} = \Omega^{\mu\nu}
\label{ana9a}
\eeq
Because of the antisymmetry of $\Omega_{\mu\nu}$, the Poisson bracket
has the property
\beq
\{ f,g\}= -\{g,f\}
\label{ana11}
\eeq
Further, from the definition, we can write,
using local coordinates,
\beqar
2~ \del_\alpha\{ f,g\}&=& \del_\alpha (\eta\cdot \del f - \xi \cdot \del g)
\nonumber\\
&=& \del_\alpha \eta^\mu \del_\mu f + \eta^\mu (\del_\mu\del_\alpha f)
-\del_\alpha \xi^\mu \del_\mu g -\xi^\mu (\del_\mu \del_\alpha
g)\nonumber\\
&=&\del_\alpha \eta^\mu \del_\mu f -\del_\alpha \xi^\mu \del_\mu g
+\eta\cdot\del (\xi^\mu \Omega_{\alpha\mu}) -\xi\cdot\del (\eta^\mu
\Omega_{\alpha\mu})\nonumber\\
&=& \del_\alpha \eta^\mu \del_\mu f -\del_\alpha \xi^\mu \del_\mu g
+(\xi\cdot \del \eta -\eta\cdot\del \xi)^\mu\Omega_{\mu\alpha}
\nonumber\\
&&~~~~~~~~~~+\eta^{\mu} \xi^\nu (\del_\mu \Omega_{\alpha\nu}+\del_\nu
\Omega_{\mu\alpha})\nonumber\\
&=& [\xi,\eta]^\mu \Omega_{\mu\alpha} +\del_\alpha (\eta^\mu\xi^\nu
\Omega_{\mu\nu}) +\eta^{\mu}\xi^\nu (\del_\mu \Omega_{\alpha\nu}+\del_\nu
\Omega_{\mu\alpha}+\del_\alpha \Omega_{\nu\mu})
\label{ana12}
\eeqar
In local coordinates, the closure of $\Omega$ is the statement $\del_\mu
\Omega_{\alpha\nu}+\del_\nu
\Omega_{\mu\alpha}+\del_\alpha \Omega_{\nu\mu} =0$. We then see that
\beq
-d \{g,f\}= i_{[\xi,\eta]}~\Omega
\label{ana13}
\eeq
which shows the correspondence stated earlier.

Consider now the change in a function $F$ due to a canonical transformation
$q^\mu \rightarrow q^\mu + \xi^\mu$. Let $f$ be the function corresponding to
$\xi^\mu$ via the correspondence (\ref{ana6}).
The change in $F$ is obviously $\xi^\mu \del_\mu F$. We can write this as
\beq
\delta F = \xi^\mu \del_\mu F =  (\Omega^{\mu\alpha} \del_\alpha f )\, \del_\mu F
= \{F, f\}
\label{ana14}
\eeq
Thus the change in a function $F$ due to
$q^\mu \rightarrow q^\mu + \xi^\mu$ is given by the Poisson bracket of $F$ with the
generating function $f$ corresponding to $\xi$.

Another important property of the Poisson bracket is the Jacobi identity
for any three functions $f, g, h$,
\beq
\{f,\{ g, h\}\}+\{h,\{ f, g\}\}+\{g,\{ h, f\}\}=0
\label{ana15}
\eeq
This can be verified by direct computation from the definition of
the Poisson bracket. In fact, if $\xi ,\eta , \rho$ are the Hamiltonian
vector fields corresponding to the functions $f, g, h$, then, by direct computation,
\beq
\{f,\{ g, h\}\}+\{h,\{ f, g\}\}+\{g,\{ h, f\}\} =- i_\xi i_\eta i_\rho
(d\Omega ) \label{ana16}
\eeq
and so the Jacobi identity (\ref{ana15}) follows from the closure of
$\Omega$.

An expression which will be useful later is the
change of the symplectic potential $\A$ under an infinitesimal
canonical transformation; this can be worked out as
\beqar
\delta_\xi \A &=&\A_\mu (q + \xi ) d(q^\mu +\xi^\mu)  - \A_\mu (q) dq^\mu =
\left[ \xi^\alpha \del_\alpha \A_\mu + \A_\alpha {\del \xi^\alpha \over \del q^\mu}
\right] \, dq^\mu \nonumber\\
&=& \left[\xi^\alpha ( \del_\alpha \A_\mu - \del_\mu \A_\alpha ) + \del_\mu ( \xi^\alpha \A_\alpha )
\right] \, dq^\mu\nonumber\\
&=& \left[ \xi^\alpha \Omega_{\alpha \mu} + \del_\mu (\xi^\alpha \A_\alpha ) \right]
\, dq^\mu\nonumber\\
&=& \del_\mu \left( \xi^\alpha \A_\alpha - f \right) \, dq^\mu
\label{ana18}
\eeqar
where we used the definition of $\Omega$ and the fact that $\xi$ is
a Hamiltonian vector field with a corresponding function $f$ defined by
equation (\ref{ana6}). 
Equation (\ref{ana18}) shows that under
a canonical transformation
$\A \rightarrow \A + d \Lambda $, $\Lambda = i_\xi \A -f$.
Evidently $d\A=\Omega$ is unchanged under such a transformation.
This suggests a very useful way of thinking about these structures.
\begin{quotation}
\noindent We may view $\A$ as a $U(1)$ gauge potential
and $\Omega$ as the corresponding field strength. The transformation
$\A \rightarrow \A + d \Lambda $ is thus a gauge transformation. We can use this to construct 
an invariant description, using covariant derivatives and other
properly transforming quantities.
\end{quotation}
A remark which may be useful in comparing the symplectic language we have used to
some other formulations:
If we use the definition of the Poisson bracket, namely (\ref{ana9}), for the
phase space coordinates themselves, we have equation (\ref{ana9a}), 
$\{ q^\mu ,q^\nu \}= \Omega^{\mu\nu}$. This is often interpreted as
saying that the ``basic Poisson brackets" ( i.e., PBs for the phase space coordinates themselves)
are the inverse of the symplectic
structure.

\subsection{Phase volume}

The symplectic two-form can be used to define a volume form
on the phase space $M$ by
\beq
d\sigma (M) = c \,{\Omega \wedge \Omega \wedge \cdots \wedge
\Omega \over (2\pi )^n}
= c\, \sqrt{\det{\left({\Omega \over 2\pi} \right)}}~~d^{2n}q
\label{ana18b}
\eeq
where we take the $n$-fold product of $\Omega$'s for a $2n$-dimensional
phase space. ($c$ is a constant which is undetermined at this stage.)
If the dimension of the phase space is infinite, then a suitable regularized
form of the determinant of
$\Omega_{\mu\nu}$ has to be used. The volume measure defined by equation
(\ref{ana18b}) is called the Liouville measure.

\subsection{Darboux's theorem}

A useful result concerning the symplectic form is Darboux's theorem
which states that in the
neighbourhood of a point on the phase space it is possible to choose coordinates
$p_i, x^i$, $i=1, 2,
\cdots ,n$, (which are functions of the coordinates
$q^\mu$ we started with) such that the symplectic two-form is
\beqar
\Omega &=& dp_i\wedge dx^i = {1\over 2} \, J_{\mu\nu} \, dQ^\mu \wedge dQ^\nu
\nonumber\\
Q^\mu&=& ( p_1, x^1, p_2, x^2, \cdots, p_n, x^n)
\label{darb1}
\eeqar
The tensor $J_{\mu\nu}$ (which is $\Omega_{\mu\nu}$ in these coordinates)
can be expressed in matrix form as
\beq
J_{\mu\nu} = \left[ \begin{matrix}
0&1&0&0&\cdots\\
-1&0&0&0&\cdots\\
0&0&0&1&\cdots\\
0&0&-1&0&\cdots\\
..&..&..&..&\cdots\\
\end{matrix} \right]
\label{darb1a}
\eeq
Evidently from the form of $\Omega$, we see that the Poisson brackets in terms of this 
set of coordinates are
\beqar
\{ x^i, x^j \} &=& 0\nonumber\\
\{ x^i ,p_j \}&=& \delta^i_{\,j} \nonumber\\
\{ p_i ,p_j \} &=& 0\label{darb2}
\eeqar
We will not consider the proof of Darboux's theorem here, although it is not very complicated.
A simple and elegant argument can be found in Arnold's book \cite{arnold}.

\section{ Classical dynamics}

The importance of the symplectic approach is that, classically,
the time-evolution of any quantity is
a particular canonical transformation generated by a function $H$
called
the Hamiltonian. This is the essence of the Hamiltonian formulation
of dynamics. Thus if $F$ is any function on $M$, we then have
\beq
{\del F \over \del t} = \{ F, H \}
\label{ana19}
\eeq
Specifically for the local coordinates $q^\mu$ on $M$ this equation leads to
\beq
{\del q^\mu \over \del t} =  \{ q^\mu ,H\} =
\Omega^{\mu\nu} {\del H \over \del q^\nu}
\label{ana20}
\eeq
Since $\Omega$ is invertible, we can also write this equation as
\beq
\Omega_{\mu\nu}{\del q^\nu \over \del t}=
{\del H\over \del q^\mu}
\label{ana21}
\eeq
If we use the Darboux coordinates $( p_i, x^i)$, these equations
(either (\ref{ana20}) or (\ref{ana21})) become
\beq
{\dot p}_i = - {\del H \over \del x^i}, \hskip .3in
{\dot x}^i = {\del H \over \del p_i}
\label{ana21a}
\eeq
which are more easily recognizable as Hamilton's canonical equations.

We are now in a position to connect the dynamics to an action and a variational
principle. We {\it define} the action as
\beq
\S = \int_{t_i}^{t_f} dt ~\left( \A_\mu {dq^\mu \over dt} ~-~ H \right)
\label{ana22}
\eeq
where $q^\mu (t)$ gives a path on $M$.
Under a general variation of the path $q^\mu (t) \rightarrow q^\mu
(t)
+\xi^\mu (t)$, the action changes by
\beqar
\delta \S &=& \int dt~ \left( {\del \A_\nu \over \del q^\mu}
{d q^\nu \over dt} \xi^\mu +\A_\mu {d \xi^\mu \over dt}
- {\del H \over \del q^\mu }\xi^\mu \right)\nonumber\\
&=& \A_\mu \xi^\mu \Biggr]^{t_f}_{t_i} +\int dt ~\left(
\Omega_{\mu\nu} {d q^\nu \over d t} -{\del H \over \del q^\mu}\right)
\xi^\mu \label{ana23}
\eeqar
The variational principle says that the equations of motion are given by
the extremization of the action, i.e., by $\delta \S =0$,
for resricted set of variations with the boundary
data (initial and final end point data) fixed. From the above variation,
we see that this gives the Hamiltonian equations of motion (\ref{ana21}).
There is a slight catch in this argument because
$q^\mu$ are phase space
coordinates and obey first order equations of motion. So we can only
specify the initial value of $q^\mu$. However, the Darboux theorem tells
us that one can choose coordinates on $M$ such that the canonical one-form
$\A$  is of the form $p_idx^i$. 
The $\xi^\mu$ in the boundary term is just $\delta x^i$.
Therefore, instead of specifying initial data for all
$q^\mu$, we can choose to specify initial and final data for the $x^i$'s.
Since the boundary values are to be kept fixed in the variational principle
$\delta \S =0$, we may set $\delta x^i =0$ at both boundaries
and the equations of motion are indeed just (\ref{ana21}).

We have shown how to define the action if $\Omega$ is given. 
However, going back to the general variations, 
notice that the boundary term resulting form the time-integration
is just the canonical one-form contracted with $\xi^\mu$.
Thus if we start from the action as the given quantity, we can identify
the canonical one-form and hence $\Omega$ from the boundary term which
arises in a general variation. In fact
\beq
\delta \S = i_\xi \A (t_f) - i_\xi \A (t_i) +
\int dt~\left(
\Omega_{\mu\nu} {d q^\nu \over d t} -{\del H \over \del q^\mu}\right)
\xi^\mu \label{ana24}
\eeq
As an example of this, consider a real scalar field theory with the action
\beq
\S = \int d^4x~ \left[ {1\over 2} {\dot \vf}^2 - {1\over 2} ( \nabla \vf )^2
- {1\over 2} m^2 \vf^2  - \alpha \vf^4 \right]
\label{ana24a}
\eeq
The variation of the action leads, upon time-integration, to the boundary term
\beq
\delta \S = \int d^3x ~ {\dot \vf} \, \delta \vf \Bigr]^{t_f}_{t_i} + \int d^4x~ \left[ \cdots \right]
\label{ana24b}
\eeq
The canonical 1-form or the symplectic potential (at a fixed time
$t$) can thus be taken as
\beq
\A = \int d^3x~ {\dot \vf} \, \delta \vf
\label{ana24c}
\eeq
In this analysis, we are at a fixed time, so $\dot \vf$ is a function independent of
$\vf$. The phase space thus consists of the set of functions
$\{ {\dot \vf}, \vf \}$ on the three-dimensional space $\mathbb{R}^3$.
$\A$ in (\ref{ana24c}) is a 1-form on the phase space, interpreting
$\delta \vf$ (which is a functional variation) as the exterior derivative on the space of fields.

If we add a total derivative to the Lagrangian, say, $\S \rightarrow \S + \int dt \, {\dot f}$, it does not affect the equations of motion. However, the new $\A$ obtained from the boundary values has an extra term
$\delta f$. This is the exterior derivative of $f$ and hence the symplectic two-form
$\Omega$ (which is $\delta \A$) is unchanged. We see that the freedom of adding total derivatives to the Lagrangian is thus the freedom of canonical transformations.

An interesting variant for the scalar field theory
 is to consider the light-cone quantization of the same theory.
Introduce light-cone coordinates, corresponding to
a light-cone in the
$z$-direction, as
\beq
u = {1\over \sqrt 2} (t+z), \hskip .3in
v= {1\over \sqrt 2}(t-z)
\label{and5}
\eeq
Instead of considering evolution of the fields in time
$t$, we can consider
evolution in one of the light-cone coordinates, say, $u$.
The analog of `space' is given by the other light-cone
coordinate $v$ and the two coordinates $x^T =x,y$
transverse to the light-cone. They correspond to equal-$u$ hypersurfaces.
The action (\ref{ana24a}) for the real scalar field
$\vf( u,v,x,y)$ can be written in these coordinates as
\begin{equation}
{\cal S} =  \int  du~dv~d^2x^T ~ \left [ \partial_u \vf \partial_v \vf
-\half (\partial_T \vf)^2- \half m^2 \vf^2 - \alpha \vf^4 \right] \label{and6}
\end{equation}
This is first order in the $u$-derivatives, the analog of the time-derivatives.
The time-integration of the variation of this action leads to the boundary term
\beq
\delta \S = \int dv\, d^2x^T \, \del_v \vf \, \delta \vf \Bigr]^{u_f}_{u_i} + {\rm volume ~integral}
\label{and6a}
\eeq
Since $\del_v \vf$ is a spatial derivative now, it is not independent of $\vf$ and so the
phase space is given by field configurations $\vf (v,x^T) $. The symplectic potential is
\beq
\A = \int dv\, d^2x^T ~\del_v \vf \, \delta \vf
\label{and6b}
\eeq

We will consider other cases of determining the symplectic form using this method
when we take up examples.

\section{Geometric quantization}

Quantum theory of any physical system is a unitary irreducible representation of the algebra of observables of the system.
This means that the observables are realized as linear operators on a Hilbert space.
The allowed transformation of
variables are then unitary transformations.
There are thus two key points regarding quantization:
\begin{enumerate}
\item We need a correspondence between canonical transformations and unitary
transformations 
\item We must ensure that the representation of
unitary transformations on the Hilbert space is
irreducible.
\end{enumerate}
Since functions on phase space generate canonical transformations and
hermitian operators generate unitary transformations, the first point is that
we get a
correspondence between
functions on phase space and
operators on the Hilbert space. The algebra of Poisson
brackets will be replaced by the algebra of commutation rules. 
The irreducibility leads to the necessity of choosing a polarization
for the wave functions. Some general references on geometric
quantization are \cite{{geom}, {blau}, {woit}, {reuter}}.

\subsection{Pre-quantization}

We will first consider
the notion of the wave function before discussing how operators act on such wave functions.
In the geometric approach,
the first step is
the so-called prequantum line bundle. 

This is a complex line bundle on the
phase space  with curvature
$\Omega$. Sections of this line bundle form the prequantum Hilbert
space. In less technical terms, we utilize the similarity we mentioned earlier, namely, that
the symplectic potential may be thought of as a $U(1)$ gauge field,
with the transformations $\A \rightarrow \A + d \Lambda$ viewed
as a gauge transformation. We can then
consider complex
functions $\Psi (q)$ defined on open neighborhoods in $M$.
These are like matter fields, they are the sections of the line bundle.
This means that
locally they are complex functions which transform as
\beq
\Psi \rightarrow \Psi ' = \exp(i\, \Lambda )~\Psi
\label{ana25}
\eeq
We can define a covariant derivative
acting on $\Psi (q)$ using $\A$ as 
\beq
\D_\mu \, \Psi \equiv \left( {\del \over \del q^\mu} - i \, \A_\mu \right) \Psi
\label{ana25a}
\eeq
The commutator of two
covariant derivatives gives $-i \,\Omega$, this is the meaning of saying that the curvature
of the line bundle is $\Omega$.

Since canonical transformations correspond to
$\A \rightarrow \A + d \Lambda$, the
 transformation of $\Psi$ as given in (\ref{ana25}) is
equivalent to the requirement of canonical transformations being
implemented as unitary transformations. The transition rules for
the $\Psi$'s from one patch on $M$ to another are likewise given
by exponentiating the transition function for $\A$. The functions
$\Psi$'s so defined form the prequantum Hilbert space with the inner
product
\beq
(1\vert 2)~=~ \int d \sigma (M)~ \Psi_{1}^{*} ~\Psi_{2} 
\label{ana26}
\eeq
where $d \sigma (M)$ is the Liouville measure on the phase space
defined by $\Omega$.

We now turn to
operators (acting on $\Psi$) corresponding
to various functions on the phase space. A function $f(q)$ on the phase
space generates a canonical transformation which leads to the change
$\Lambda = i_\xi \A -f$ in the symplectic potential, see (\ref{ana18}).
The corresponding change in
$\Psi$ is thus
\beqar
\delta \Psi &=& \xi^\mu \del_\mu \Psi ~-~i (~i_\xi\A -f ) \Psi\nonumber\\
&=& \xi^\mu \left( \del_\mu -i \A_\mu \right) \Psi +i f \Psi\nonumber\\
&=& \left( \xi^\mu \D_\mu +i f \right) ~\Psi
\label{ana27}
\eeqar
where the first term on the right hand side in the first line
gives the change in $\Psi$ considered as a
function and the second term compensates for the change of
$\A$. 
The change can be expressed using the covariant derivative as in the last line.
Given (\ref{ana27}), it is natural to define the {\it prequantum operator}
corresponding to $f(q)$ by
\beq
{\cal P}(f) ~=~ -i \bigl( \xi \cdot {\cal D} ~+~ i f \bigr)
\label{ana28}
\eeq
We can easily check that
\beq
\int d^{2n} q\, \sqrt{\det \Omega}~ \Psi_1^* \, \Bigl[\P (f) \, \Psi_2\Bigr]
= \int d^{2n} q\, \sqrt{\det \Omega}~ \Bigl[\P(f) \Psi_1\Bigr]^*  \, \Psi_2 
\label{ana28a}
\eeq
so that $\P(f)$ is a symmetric operator,
which is necessary condition for
a unitary representation. (Strictly speaking, before we can claim a unitary representation, 
we need to consider the completion of the set of such functions and also
make sure the domains and ranges
of operators match; we will not go into this question, since the whole issue has to be
addressed for the true wave functions anyway.)

Now consider the algebra of the prequantum operators.
We have already seen in (\ref{ana13}) that if the Hamiltonian vector fields for
$f$, $g$ are $\xi$ and $\eta$ respectively, then the vector field
corresponding to the Poisson bracket $\{ f,g\}$ is 
$-[\xi ,\eta ]$. Using the definition of the prequantum operator above,
we then find
\beqar
[\P (f) , \P (g) ]&=& \left[ -i\xi \cdot \D +f , -i\eta \cdot \D
+g\right]\nonumber\\
&=& -\left[ \xi^\mu \D_\mu , \eta^\nu \D_\nu \right] -i\xi^\mu [\D_\mu ,g]
+i\eta^\mu [\D_\mu ,f]\nonumber\\
&=& i\xi^\mu \eta^\nu \Omega_{\mu\nu} -(\xi^\mu \del_\mu \eta^\nu )\D_\nu
+ (\eta^\mu \D_\mu \xi^\nu ) \D_\nu -i\xi^\mu \del_\mu g +i \eta^\mu
\del_\mu f \nonumber\\
&=& i \left( - \xi^\mu \eta^\nu \Omega_{\mu\nu} +i [\xi ,\eta ] \cdot \D
\right)\nonumber\\
&=& i \left( -i ~(i_{[\eta ,\xi ]}\D ) + \{ f,g\} \right)\nonumber\\
&=& i\, \P (\{f,g\})\label{ana29}
\eeqar
In other words, the prequantum operators form a representation of
the Poisson bracket algebra of functions on phase space.

\subsection{Polarization}

It seems like we have all the ingredients for the quantum theory, but not quite so.
The prequantum wave functions $\Psi$ depend on all phase space
variables. The representation of the Poisson bracket algebra on
such wave functions, given by the prequantum operators, is
reducible. We will use a simple example to illustrate this point.

Consider a point particle in one dimension, with the symplectic
two-form $\Omega = dp \wedge dx$. We can choose
$\A = p\, dx$. The vector fields corresponding to $x$ and $p$ are
$\xi_x = -{\del /\del p} $ and $\xi_p = \del/\del x$.
The corresponding prequantum operators are
\beq
\P (x) = ~i {\del \over \del p} +x , \hskip .3in
\P (p) = -i {\del \over \del x} 
\label{ana30}
\eeq
which obey the commutation rule
\beq
[\P (x), \P (p) ] = i \label{ana31}
\eeq
We have a representation of the algebra of $\P (x)$, $\P (p)$
in terms of the prequantum wave functions $\Psi (x,p)$. But this is reducible.
For if we consider the subset of functions on the phase space which are
independent of $p$, namely those which obey
the condition
\beq
{\del \Psi \over \del p } =0,
\label{ana32}
\eeq
then the prequantum operators reduce to
\beq
\P (x) = x , \hskip .3in
\P (p) = - i {\del \over \del x}
\label{ana33}
\eeq
which obey the same algebra (\ref{ana31}). Thus we are able to obtain a
representation of the algebra of observables on the smaller space of
$\Psi$'s obeying the constraint ({\ref{ana32}), showing that the previous representation
(\ref{ana30}) is reducible.
\begin{quotation}
\noindent
In order to obtain an irreducible representation, one has to impose
subsidiary conditions which restrict the dependence of the 
prequantum wave functions
to half the number of phase space variables. This is the choice of
polarization and generally leads to an irreducible representation of the
Poisson algebra.
\end{quotation}

If we are talking about ordinary functions $f$ on the phase space $M$, the statement that $f$ is independent of $n$ of the coordinates can be phrased as
\beq
P^\mu_i \, {\del f \over \del q^\mu } = 0
\label{ana33a}
\eeq
where $P_i = P_i^\mu (\del / \del q^\mu )$, $i = 1, 2, \cdots, n$, form $n$ linearly independent
vector fields. An integrability requirement for (\ref{ana33a}) is 
\beq
[ P_i\, ,  P_j ]^\mu\, {\del f \over \del q^\mu} = 0
\label{ana33b}
\eeq
which can be ensured if 
\beq
[ P_i\, , P_j ] = C_{ij}^k \, P_k
\label{ana33c}
\eeq
where the coefficients $C_{ij}^k$ need not be constants.
If we have a set of vector fields $P_i$ obeying (\ref{ana33c}), then they are said to be in involution.
 If this is satisfied, we can integrate, starting from some point on $M$, along these vector fields and obtain, at least locally, a neighborhood of an $n$-dimensional submanifold. (This is ensured by
 Frobenius' theorem.) Such a submanifold is said to be a Lagrangian submanifold
 if we also have the condition
 \beq
 \Omega_{\mu\nu} \, P_i^\mu \, P_j^\nu = 0
 \label{ana33d}
 \eeq
 
 The prequantum wave functions are not functions on $M$, they are sections of a line bundle, i.e.,
 they transform with a phase under $\A \rightarrow \A + d \Lambda$, and so we must impose the covariant version of (\ref{ana33a}). Thus, as the polarization condition we choose
 \beq
 P_i^\mu \, \D_\mu \, \Psi = 0
 \label{ana33e}
 \eeq
 where $P_i$ are $n$ linearly independent vector fields obeying (\ref{ana33c}) and
 (\ref{ana33d}).
 The integrability requirement for (\ref{ana33e}) is automatically satisfied since
 \beqar
 [ P_i^\mu \D_\mu , P_j^\nu \D_\nu ] \, \Psi &=& 
 C_{ij}^k \, P_k^\mu \D_\mu \Psi -i \, \Omega_{\mu\nu} P_i^\mu P_j^\nu \, \Psi
 \nonumber\\
 &=& 0
 \label{ana33f}
 \eeqar
 by virtue of (\ref{ana33e}) and (\ref{ana33d}).
 The prequantum wave functions restricted by the polarization condition
 (\ref{ana33e}) are the true wave functions of the theory.
 There can be different possible choices for the polarization leading to wave functions depending on different subsets of phase space coordinates. For example, the difference between the momentum space wave functions and the coordinate space wave functions familiar from elementary
 quantum mechanics is one of different polarization choices.
 
\subsection{Measure of integration}

The next step is to define an inner product to make
these wave functions into a Hilbert space. Obviously, if the wave functions do not depend on 
half the number of phase space coordinates, it does not make sense to integrate over them
in an inner product. In particular, it would give an undefined or infinite value if those 
directions do not have a finite volume. So one needs to define a volume measure for integration over the coordinates on which the wave functions do depend. The problem is that while the Liouville measure for all of phase space is naturally defined in terms of the symplectic structure, there
is no natural choice of integration measure for the reduced
set of variables, once we impose the polarization requirement.
In many cases, the phase space is the cotangent bundle of some manifold (which is the configuration space $Q$), which means that
it is made of the coordinates and co-vectors. For example, for particle dynamics on
$Q = \mathbb{R}^3$, $M = T^* \mathbb{R}^3$ which has coordinates $x^\mu$ and momenta
$p_\mu$ as the basic coordinates. Then, if we use a polarization given by
$P_i^\mu = (\del /\del p_\mu )$, the wave functions depend on $x$ only. This is the usual coordinate space Schr\"odinger quantum mechanics and one can use the integration just on
$\mathbb{R}^3$ to form the inner product. But generally speaking, unless $M$ is the
cotangent bundle of some manifold, finding a reduced integration measure is not trivial.

However there is one case where there is a natural inner product on
the Hilbert space. This happens when the phase space is also
K\"ahler and $\Omega$ is the K\"ahler form or some multiple thereof.
In this case we can introduce local complex coordinates and write
\beq
\Omega ~=~ \Omega_{a {\bar a}}\, dz^{a}\wedge d\bz^{\bar a}
\label{ana36}
\eeq
$a, {\bar a} ~= 1,2...n$.
The corresponding covariant derivatives are
\beq
{\cal D}_{a} = \partial_{a} ~-~i\A_{a}, \hskip .3in
{\cal D}_{\bar a}
=\partial_{\bar a} ~-~ i\A_{\bar a} \label{ana37}
\eeq
The characteristic of a K\"ahler manifold is the existence of a K\"ahler potential $K$
such that
\beq
\A_a = -{i\over 2}\del_a K , \hskip .3in
\A_{\bar a} =  ~~{i\over 2} \del_{\bar a} K \label{ana38}
\eeq
In this case, one can choose the holomorphic polarization
\beq
\D_{\bar a} \Psi = (\del_{\bar a} +{1\over 2} \del_{\bar a}
K ) \Psi = 0
\label{ana39}
\eeq
The solutions are the polarized wave functions $ \psi$ given by
\beq
\psi = \exp(-\half K ) ~~F\label{ana39a}
\eeq
where $F$ is a holomorphic function on $M$. The wave functions are
thus holomorphic, apart from the prefactor involving the K\"ahler
potential. In this case, $\psi^*$ involves the antiholomorphic functions
$F^*$ and the product depends on all the phase space coordinates.
Integration over all of phase space is acceptable and the
inner product of the prequantum Hilbert 
space can be
retained, may be up to a constant of proportionality,
 as the inner product
of the true Hilbert space; specifically we have
\beq
\la 1 \vert 2\ra = \int d\sigma (M)~ e^{-K} ~~F_1^* F_2
\label{ana40}
\eeq
The cases where $M = T^*Q$ for some manifold $Q$ and the K\"ahler case will 
cover most of the physical situations of interest to us.

\subsection{Representation of operators}

Once the polarized wave functions are defined, the idea is to
represent observables as linear operators on the wave functions as
given by
the prequantum differential operators. Let $\xi$ be the Hamiltonian
vector field corresponding to a function $f(q)$. If 
the commutator of $\xi$ with any polarization
vector field $P_i$ is proportional to $P_i$ itself,
i.e., $[\xi, P_i]= C_i^j P_j$ for some functions $C_i^j$, then,
evidently,
$\xi$ does not change the polarization;
$\xi \Psi$ will obey the same polarization condition as $\Psi$.
In this case the operator corresponding to $f(q)$ is given by
$\P (f)$, but, of course, now acting on the wave functions 
in the chosen polarization.

The situation with operators which do not preserve the polarization is more complicated.
There are many such operators of interest in any physical problem. For example, the
Hamiltonian for a free nonrelativistic particle in one spatial dimension is $H = p^2/2m$, with the vector field
$\xi_H = (p/m)\, (\del / \del x )$. If we choose the polarization which gives
wave functions depending on $x$, namely, choose $P = (\del /\del p)$, then we find
\beq
[ \xi_H , P ] = - {1\over m } {\del \over \del x}
\label{ana40a}
\eeq
We see that $\xi_H$ does not preserve the polarization. 
The solution is also suggested by this example. We can define
$p^2$ trivially by using the momentum-space wave functions, namely, ones corresponding to the
polarization $(\del /\del x)$. It is possible to transform from one type of wave functions to the other, in this case, by Fourier transformation. More generally, there are kernels, known as Blattner-Kostant-Sternberg (BKS) kernels, which map from one polarization to another. Using this, we can define
operators as follows. We carry out a canonical transformation on the wave functions
by the vector field $t \, \xi_f$
where $f$ is the function whose operator version we wish to find and $t$ is a real parameter.
The result is no longer in the same polarization, but we can transform back
using an appropriate BKS kernel. The derivative of the result with respect to
$t$ at $t =0$ will give the action of the operator. Equivalently, we
can work out the form of the operator in a polarization which is preserved 
by the corresponding vector field and then transform to the required 
polarization using an appropriate
BKS kernel. 

\subsection{Comments on the measure of integration, corrected operators, etc.}

The problem of defining the measure of integration in a given polarization has 
implications, which necessitates a certain modified definition for operators.
Fortunately, this will not be an issue for most of what we want to do, but, 
nevertheless, a comment is in order at this stage. (For more detailed analysis, see
\cite{{geom}, {blau}, {woit}, {reuter}}.) To illustrate the problem, consider how we can show that
$\P (f)$ is a symmetric operator, as in (\ref{ana28a}). The relevant partial integration
leads to a discrepancy $\del_\mu \xi^\mu + {\half} \xi^\mu \del_\mu ( \log\det \Omega )$
which is zero by virtue of the closure of $\Omega$ and $\xi$ being a Hamiltonian vector field.
However, the integration measure for the polarized wave functions
is not given by $\Omega$ and hence this argument does not go through.
Consider a real polarization and let the inner product be of the form
\beq
\la 1 \vert 2\ra = \int d^n x \, J \, \psi_1^* \, \psi_2
\label{ana40b}
\eeq
(We do not necessarily mean that $x$ denotes coordinates of some configuration space, it is used as a generic notation here.)
We then find
\beq
\int \psi_1^* \, ( \P (f) \psi_2) - \int (\P (f) \psi_1)^*\, \psi_2 
=
i \int d^n x\, J\, \bigl[ \del\cdot \xi + \xi\cdot \del \log J \bigr]\,
\psi_1^* \, \psi_2
\label{ana40c}
\eeq
Clearly using $\P(f)$ to act on the polarized wave functions will not do.
One strategy is to factorize $J$ as $ {\bar \sigma} \, \sigma$ where $\sigma$ need not be real and consider
$\psi \, \sigma$ in place of the wave function. The quantity $\sigma$ behaves as the square root of the integration measure on the complement of the subspace defined by the polarization vector fields.
For this reason, this way of considering $\psi \, \sigma$ directly, rather than $\psi$ and then the measure of integration separately, is called the half-form quantization.
We then modify the definition of the operator corresponding to $f$ as
\footnote{$L_\xi \sigma$ is again the Lie derivative of $\sigma$.}
\beqar
\P (f) \, \psi \, \sigma &=& \left[ \left( -i \xi\cdot \D + f \right) \psi \right]\, \sigma
- \psi ~ ( i L_\xi \sigma )\nonumber\\
-i L_\xi \sigma &=& - i \xi \cdot \del \sigma - {i \over 2} \del \cdot \xi \, \sigma
\label{ana40d}
\eeqar
With this definition, we can verify that
\beq
\int d^nx~\psi^*{\bar \sigma} ~ \left[ \P (f) \psi \,\sigma \right] = \int d^nx~\left[ \P(f) \psi \,\sigma \right]^* \, \psi \,\sigma
\label{ana40e}
\eeq

It is useful to consider the problem of the integration measure in some more detail.
For this purpose, let us consider the Lagrangian submanifold defined by the polarization 
$\{ P_i \}$. Let $u^i$ denote the local coordinates on this submanifold.
The coordinates $q^\mu$ on the submanifold can be considered as functions 
of $u^i$ and obey equations of the form
\beq
(E^{-1})_i^{~k} \, {\del q^\mu \over \del u^k}  = P_i ^{\,\mu}
\label{ana40e.1}
\eeq
The matrix of functions $E_i^{~k}$ plays the role of frame fields for the subspace and we can define a volume measure of the form
$(\det E )\, d^nu$. 
In the inner product (\ref{ana40b}), the integrand $\psi_1^* \psi_2$ is independent of $u^i$.
So, just as we do in the case of the functional integral for gauge theories, we can introduce a constraint
$\delta^{(n)}(u) \, \det E^{-1} $ and integrate with the full Liouville measure. This will effectively remove the volume element $\det E \, d^n u$ of the Lagrangian submanifold from the Liouville
volume element. But notice that if we make a transformation $P_i \rightarrow C_i^{~j} \, P_j$
on the basis of polarization vectors, $\det E^{-1} \rightarrow (\det C )\, \det E^{-1}$.
Thus the reduced volume must have this transformation property. In order to make this
work for all polarizations, including holomorphic (or partly holomorphic) ones,
we have to factorize $\det E^{-1}$  and hence we must consider quantities transforming as
$(\det C )^{1/2}$.
This may be formalized in terms of what is called the metaplectic structure. We 
consider a vector space $V$ which can eventually be identified as the tangent bundle of $M$.
Let $F \subset V$ be a subspace
of $V$, and further, let $X_a$ be a basis for $F$. (We may think of
$F$ as being defined by the set of polarization vectors with $X_a \sim P_a$.)  We then consider quantities
$\sigma_r (X_a)$ which depend on the basis and which has the property that
$\sigma_r (X_a) \in \mathbb{C}$ and 
\beq
\sigma_r ( N_a^b \, X_b) = (\det N )^{-r }\,
\sigma_r (X_a)\label{ana40f}
\eeq 
If $V$ is the tangent bundle, we can do this at every point on $M$,
thereby defining a bundle which we will designate as
$\delta_r (F)$, sections of which are $\sigma_r$.
If $r=1$ we are effectively talking about the volume element for $F$. 
For $r = {1\over 2}$, $\delta_{\tiny 1/2}$ is a line bundle over $M$;
this is called a metaplectic structure on $M$.
We may think of it as
defining a volume on spinor frames.
The volume for the polarized subspace may be defined using $\delta_{-1/2}(P)$ and
$\delta_{-1/2}({\bar P})$ acting
on $\delta_1(M)$. Let $W = (P \cup {\bar P}) / (P\cap {\bar P}) $.
This can be shown to be a symplectic space with its own volume measure
which will transform as
$\sigma_1$. The general formula for the required integration measure is then
\beq
d \mu = \sigma_{-1/2} (P) \, \sigma_{-1/2}({\bar P})\,  \sigma_1 (W) \, d\sigma (M)
\label{ana40f.1}
\eeq
For a real polarization, $P = {\bar P}$ and $W$ is empty. Thus we get the result
$\sigma_{-1} (P) \, d\sigma (M)$ which is the same as (\ref{ana40b}).
The formula (\ref{ana40f.1}) factors out the effect of the directions defined by $P$.
The two factors $\sigma_{-1/2} (P)$ and $\sigma_{-1/2} ({\bar P})$, which we denoted
by $\sigma$ and ${\bar \sigma}$ in equations (\ref{ana40d}) and (\ref{ana40e}),
are needed for the action of the operators as in (\ref{ana40d})\footnote{By the way,
$\sigma_r ({\bar P}) = \overline{\sigma_r (P)}$.}.
In the integration measure, we can go back to the form $\delta^{(n)}(u) \det E^{-1}$ which
is given by $\sigma_{-1} (P)$.
If we consider holomorphic polarization, then $P \cup {\bar P} = M$
and $P \cap {\bar P} = \emptyset$, so that we get $\sigma_1 (W) = \sigma_1 (M)$, which
gives another factor of $\sqrt{\det \Omega}$. This cancels with the $\sigma_{-1/2}$ factors
retaining $d\sigma (M)$ as the volume.

As we see from (\ref{ana40d}), once we include the half-form factors,
 the definition of operators will have the corrections from
$L_\xi \sigma_{-1/2}$. 
This can give a correction even when the operator 
$\xi_f$ preserves the polarization. If $\xi_f$ preserves the
polarization, then we find
\beq
L_{\xi_f} \, P_i = [ \xi_f, P_i] = C_i^j \, P_j
\label{ana40f.2}
\eeq
Since the polarization is preserved by $\xi_f$, the prequantum operator
with the modification as in (\ref{ana40d}) can be used as the quantum
version of $f$. Thus
\beq
\P (f) =  \left( -i \xi\cdot \D + f \right) - {i \over 2} \Tr\, C
\label{ana40f.3}
\eeq
In comparison with (\ref{ana40f}), $N \approx 1 -i C$.

The condition for the existence of a metaplectic structure is essentially the same as the condition for the existence of spinors on the manifold, namely, the vanishing of the Stiefel-Whitney class; i.e.,
$\H^2 (M, \mathbb{Z}_2) = 0$. (The metaplectic group is the covering group for the symplectic group, and $\delta_{-1/2}$ can be constructed using spinor frames.)
If we have $\H^2 (M, \mathbb{Z}_2) = 0$, then there can still be inequivalent $\delta_{-1/2}$ bundles,
which are classified by $\H^1 (M, \mathbb{Z}_2) $, exactly as for spinors.

The metaplectic structure gives a more formal and better way to address the issue of defining the integration measure for the inner product of the true wave functions and
of having to modify the definition of operators corresponding to $f$ as in (\ref{ana40d}).
We will not go into this in any more detail here. The point is that, overall, while geometric quantization is very beautiful, it must be admitted that defining operators which do not preserve the polarization and defining an integration measure on the space of polarized wave functions are somewhat awkward and cumbersome.
We will be considering mostly the holomorphic polarization which avoids most of these issues.

\section{Topological features of quantization}

There are two aspects of 
the topological features of phase space which have an impact on the quantization.
These are due to
first and second cohomology of the phase space. We will briefly talk about them now.

\subsection{The case of nontrivial $\H^1(M, \mathbb{R})$}

Consider first the case of $\H^1(M, \mathbb{R})\neq 0$, which means that $M$ admits
one-forms which are closed but not exact.
Thus
for a given symplectic two-form $\Omega$, we can have different
symplectic potentials $\A$ and $\A +A$ which lead to the same $\Omega$
since $A$ is closed, i.e., $dA =0$. 
Now if $A$ is exact, there is some globally defined function $h$ on $M$
such that
$A=dh$.
The function $h$ is a canonical transformation and 
physical results will be
unchanged. Thus an exact one-form is equivalent to $A=0$ upon carrying out a
canonical transformation. 
However, if
$A$ is closed but not exact, i.e., it is a nontrivial element of the
cohomology
$\H^1 (M, \mathbb{R})$, then we cannot get rid of it by a canonical
transformation. Locally we can still write $A=d f$ for some $f$, but $f$ will
not be globally defined on $M$. Thus globally we cannot eliminate $A$.

Classical dynamics is defined by the equations of motion as in
(\ref{ana20}) which involves only 
$\Omega$, not the symplectic potential $\A$.
Thus this ambiguity in the choice of the symplectic
potential due to nonzero $\H^1 (M, \mathbb{R})$ will not affect the classical dynamics.
In the quantum theory such $A$'s do make a difference. This can be seen in terms of the action
$\S$;
for a path $C$, parametrized as $q^\mu (t)$
from a point $a$ on $M$ to
a point $b$, the action is
\beq
\S = \int dt~ \left( \A_\mu {dq^\mu \over dt}-H \right)~+~
\int_a^b A_\mu dq^\mu \label{anb1}
\eeq
The action depends on the path but the contribution from 
$A$ is topological. If we change the path slightly from $C$ to $C'$
with the end points fixed, we find, using Stokes' theorem,
\beq
\int_C A -\int_{C'} A =
\oint_{C-C'} A 
= \int_\Sigma dA
= 0\label{anb2}
\eeq
where $C-C'$ is the path where we go from $a$ to $b$ along $C$ and
back from $b$ to $a$ along $C'$. (Since this is the return path, the orientation
is reversed, hence the minus sign.) $\Sigma$ is a surface in $M$ with
$C-C'$ as the boundary. 
The above result shows that the contribution from
$A$ is invariant under small changes of the path, which also explains why it does not contribute to the classical equations of motion viewed as variational equations.
 (The full action $\S$ does depend on the path.) 
 In particular, the value of the integral of $A$
is zero for closed paths so long as they are contractible; for then we
can make a sequence of small deformations of the path (which do not
change the value) and eventually contract the path to zero. If there are
noncontractible loops then there can be nontrivial contributions.
If $\H^1(M, \mathbb{R})\neq 0$, then there are noncontractible loops.
In the quantum theory, it is $e^{i\S}$ which is important, so we need
$e^{i\int A}$. Assume for simplicity that $\H^1(M, \mathbb{R})$ has only one nontrivial element
(say $\alpha$) up to addition of trivial terms and multiplicative factors.
Then there is only one topologically distinct noncontractible loop
apart from multiple traversals of the same. 
Let
$A= \theta\,\alpha$ where $\theta$ is a constant and
$\alpha$ is normalized to unity along the noncontractible loop
for going round once. For all paths which include $n$ traversals
of the loop, we find
\beq
\exp\left({i\oint A }\right) =\exp\left({i\,\theta\oint \alpha}\right)
=\exp\left({i\,\theta\, n} \right)\label{anb3}
\eeq
Notice that a shift $\theta\rightarrow \theta+2\pi$ does not change this
value, so that we may restrict $\theta$ to be in the interval zero to
$2\pi$. Putting this back into the action (\ref{anb2}), we see that, as a
function over all paths, the action has an extra parameter $\theta$.
Thus the ambiguity in the choice of the symplectic potential due to
$\H^1(M, \mathbb{R})\neq 0$ leads to an extra parameter $\theta$ which is
needed to fully characterize the quantum theory. 
Since $\theta$ is in the interval $0$ to $2\pi$, we may regard 
$ A = \theta \, \alpha$ as an element of
$\H^1(M, \mathbb{R})/\H^1(M, \mathbb{Z})$.
If $\H^1(M, \mathbb{R})$ has more than one distinct element,
there are more distinct
paths possible and  there can be more parameters like $\theta$.
Such parameters are generally called vacuum angles.

It is now easy to see these results in terms of wave functions. 
The relevant covariant derivatives are of the form
$\D_\mu \Psi =(\del_\mu -i\A_\mu -iA_\mu )\Psi$. We can write
\beq
\Psi (q) = \exp\left( i \int_0^q A \right) ~~\Phi (q)
\label{anb4}
\eeq
where the lower limit of the integral is some fixed point $a$.
By using this in the covariant derivative, we see that $A$ is removed
from
$\D_\mu$ in terms of action on $\Phi$.
This is like a canonical transformation, except that the relevant
transformation $\exp\left( i \int_0^q A \right)$ is not single valued. As
we go around a closed noncontractible curve, it can give a phase
$e^{i\theta}$. Since $\Psi$ is single-valued, this means that $\Phi$
must have a compensating phase factor; $\Phi$ is not single-valued but
must give a specific phase labelled by $\theta$.
Thus we can get rid of $A$ from the covariant derivatives and hence 
the various operator formulae, but diagonalizing the Hamiltonian
on such $\Phi$'s can give results which depend on the angle $\theta$.

The $\theta$-vacua in a nonabelian gauge theory is an example of this
kind of topological feature. The description of particles of fractional
statistics in two spatial dimensions is another example.

\subsection{The case of nontrivial $\H^2(M, \mathbb{R})$}

We now turn to the second topological feature we mentioned, namely the case of
$\H^2(M, \mathbb{R})\neq 0$. 
This means that there are closed two-forms
on $M$ which are not exact. Correspondingly, there are closed
two-surfaces which are not the boundaries of any three-dimensional
region, i.e., there exists noncontractible closed two-surfaces. In general, elements
of $\H^2(M, \mathbb{R})$ integrated over such noncontractible two-surfaces
will not be zero. If the symplectic two-form
$\Omega$ is some nontrivial element, or it has a part which
is a nontrivial element, of  $\H^2(M, \mathbb{R})$, 
then the
symplectic potential $\A$ cannot be globally defined. 
This can be seen as
follows.
Consider the integral of $\Omega$ over a noncontractible two-surface
$\Sigma$,
\beq
I(\Sigma)= \int_\Sigma ~\Omega \label{anb5}
\eeq
First of all, this is a topological invariant, for if
 $\Sigma '$ is a small deformation of $\Sigma$, then
\beq
I(\Sigma )- I(\Sigma') =\int_{\Sigma -\Sigma'}~\Omega
= \int_V d\Omega =0\label{anb6}
\eeq
where $V$ is a three-dimensional volume with the two surfaces
$\Sigma - \Sigma'$ as the boundary. Thus the integral of $\Omega$ is
 invariant under small deformations of the
surface on which it is integrated. 
If we could write $\Omega$ as
$d\A$ for some $\A$ which is globally defined on $\Sigma$ then
clearly $I(\Sigma )$ is zero by Stokes' theorem. Thus if $I(\Sigma )$ is
nonzero, we must conclude that
there is no potential $\A$ which is globally defined on $\Sigma$. 
We have to use different choices for  $\A$ in different
coordinate patches and have transition functions relating
the $\A$'s in the overlap regions. But we have the same $\Omega$
on a given overlap region whether we use the $\A$ for one patch or the 
$\A$ for the other patch to calculate it.
Thus the transition functions on overlap regions
must be canonical transformations.

As an example, consider a closed noncontractible two-sphere, or any
smooth deformation of it, which may be a subspace of $M$. We can cover it with
two coordinate patches corresponding to the two hemispheres, denoted $N$
and $S$ as usual. The symplectic potential is represented by
$\A_N$ and $\A_S$ respectively. On the equatorial overlap region,
they are connected by 
\beq
\A_N = \A_S +d \Lambda \label{anb7}
\eeq
where $\Lambda$ is a function defined on the overlap region. It gives the
canonical transformation between the two $\A$'s.

The symplectic potential $\A$ is what is needed in setting up the quantum theory.
And since canonical transformations are represented as
unitary transformations on the wave functions, we see that we must also have
a $\Psi_N$ for the patch $N$ and a $\Psi_S$ for the patch $S$.
On the equator they must be related by the canonical transformation,
which from (\ref{ana25}), is given as
\beq
\Psi_N = \exp(i\,\Lambda )~~\Psi_S \label{anb8}
\eeq
Now consider the integral of $d\Lambda$ over the equator $E$, which is
a closed curve being the boundary of either $N$ or $S$. From
(\ref{anb7}) this is given as
\beqar
\Delta \Lambda =\oint_E d\Lambda &=& \int_E \A_N -\int_E \A_S 
=  \int_{\del N}\A_N ~+~ \int_{\del S} \A_S \nonumber\\
&=& \int_N \Omega ~+~ \int_S \Omega
= \int_\Sigma ~\Omega \label{anb9}
\eeqar
(In the second step, we reverse the sign for the $S$-term because $E$
considered as the boundary of $S$ has the opposite orientation compared
to it being the boundary of $N$.)
The above equation shows that the change of $\Lambda$
as we go around the equator once, namely 
$\Delta\Lambda$, is nonzero if $I(\Sigma )$ is nonzero;
$\Lambda$ is not single-valued on the equator. But the wave function must
be single-valued. From (\ref{anb8}), we see that this can be achieved if 
$\exp (i\Delta \Lambda )=1$ or if $\Delta\Lambda =2\pi n$ for some
integer $n$. Combining with (\ref{anb9}), this can be stated as 
a topological quantization rule implied by the
single-valuedness of wave functions in the quantum theory,
\beq
\int_\Sigma ~\Omega = 2\,\pi \,n \label{anb10}
\eeq
The integral of the symplectic two-form $\Omega$ on closed noncontractible
two-surfaces must be quantized as $2\pi$ times an integer. (Or we may say that 
$\Omega$ must belong to an integral cohomology class of $M$.)
We have given
the argument for surfaces which are deformations of a two-sphere, but a
similar argument can be made for general noncontractible two-surfaces.
The result (\ref{anb10}) is quite general.

The quintessential example of this kind of topological feature is the motion of
a charged particle in the field of a magnetic monopole. The condition
(\ref{anb10}) is then the famous Dirac quantization condition.
The Wess-Zumino terms occuring in many field
theories are another example.

\section{Summary of holomorphic polarization and quantization}

Since we will be using geometric quantization with holomorphic polarization in some of the examples
later, this is a good point to summarize the
key features of
quantization using the
holomorphic polarization. 
\begin{enumerate}
\item We need a phase space which is also K\"ahler; the symplectic two-form
must be a multiple of the K\"ahler form.
\item The prequantum
wave functions are sections of a bundle which is the product of the
holomorphic line bundle with curvature equal to the symplectic form
and a half-form bundle. 
(The existence of the half-form bundle requires the vanishing of the Stiefel-Whitney class
as mentioned earlier.)
\item The true wave functions are obtained by
imposing the polarization condition,
which, for the holomorphic polarization is
${\cal D}_{\bar a}\,\Psi~=~0$.
\item The inner product of the prequantum Hilbert space, which is
essentially square integrability on the phase space with the 
Liouville measure of integration, is
retained  as the inner product on the true Hilbert space in the
holomorphic polarization.
\item The operator corresponding to an observable $f(q)$ which preserves the
chosen polarization is given by the prequantum operator $\P (f)$ acting on
the true (polarized) wave functions.
The half-form part of the wave functions, while not important for the integration measure
in the holomorphic polarization, can modify the operators as in (\ref{ana40d})
or (\ref{ana40f.3}).
\item For observables which do not preserve the polarization,
one has to construct infinitesimal unitary transformations whose classical
limits are the required canonical transformations.
\item If $\H^1(M, \mathbb{R})$ is not zero, then there are inequivalent
$\A$'s for the same $\Omega$ and we need extra angular parameters
to specify the quantum theory completely.
\item If the phase space $M$ has noncontractible two-surfaces, then the
integral of $\Omega$ over any of these surfaces must be quantized in
units of $2\pi$.
\end{enumerate}

\section{Coherent states, the two-sphere and $G/H$ spaces}

\subsection{Coherent states}

We will start with the simplest case of coherent states for a one-dimensional quantum system
to illustrate how these ideas take concrete form.
In one spatial dimension, $\Omega = dp\wedge dx =
i dz\wedge d\bz$, where $(p\pm i x)/\sqrt{2} =z, \bz $.
Choose 
\beq
\A = {i\over 2} (z ~d\bz - \bz ~dz )\label{ana41}
\eeq
The space is K\"ahler, with $K = \bz z$.
The covariant derivatives corresponding to
(\ref{ana41}) are ${\cal D}_z= \del_z -\half \bz $ and ${\cal D}_\bz=\del_\bz +\half
z$. Holomorphic polarization corresponds to $P= {\del /\del \bz}$, so that
the polarization condition on the prequantum wave functions is
\beq
{\cal D}_\bz \Psi = (\del_\bz +\half z )\Psi =0
\label{ana42}
\eeq
The solutions are of the form
\beq
\Psi = e^{-\half z\bz } ~\vf (z)
\label{ana43}
\eeq
where $\vf (z)$ is holomorphic in $z$. 
The Hamiltonian vector fields corresponding to $z, \bz$ are
\beq
z \longleftrightarrow  -i{\del \over \del \bz} , \hskip .3in
\bz \longleftrightarrow  ~~i {\del \over \del z}\label{ana44}
\eeq
These commute with $P= {\del /\del \bz}$ and so are
polarization-preserving. The prequantum operators corresponding to these
are
\beqar
\P (z)&=& -i (-i)\left( {\del \over \del \bz} +\half z \right) +z =
-{\del \over \del \bz} +\half z\nonumber\\
\P (\bz )&=& -i (~i) ~\left( {\del \over \del z} -\half \bz \right) +\bz
= ~~{\del \over \del z}+\half \bz\label{ana45}
\eeqar
In terms of their action on the functions $\vf (z)$ in (\ref{ana43}),
corresponding to
$\Psi$'s 
obeying the polarization condition, we define the operator versions of
$z$ and $\bz$ by
\beq
\P (z ) \, \Psi = e^{- {\half} \bz z} \, {\cal O}(z) \vf (z), \hskip .3in 
\P (\bz ) \, \Psi = e^{- {\half} \bz z} \, {\cal O}(\bz) \vf (z)
\label{ana45a}
\eeq
so that
\beqar
{\cal O} (z)\, \vf (z) &=& z ~\vf (z)\nonumber\\
{\cal O} (\bz ) \, \vf (z) &=& {\del \vf \over \del z}\label{ana46}
\eeqar
The inner product for the $\vf (z)$'s is
\beq
\la 1\vert 2\ra = \int i{ dz \wedge d\bz \over 2\pi} 
~e^{-z\bz}~~ \vf_1^*~ \vf_2 \label{ana47}
\eeq
What we have obtained is the standard coherent state (or Bargman)
realization of the Heisenberg algebra. 

It is somewhat illuminating to consider the quantization of the function
$\bz z$. The vector field corresponding to this is
$\xi = i ( z \del_z - \bz \del_\bz )$. The prequantum operator for this is easily seen to 
be $z \del_z$ acting on $\vf (z)$. For the polarization we have chosen,
\beq
[ \xi, \del_\bz ] = i \, \del_\bz
\label{ana47a}
\eeq
Thus $\xi$ preserves polarization and we can identify $C = i$ in comparing with
(\ref{ana40f.2}). The operator corresponding to $\bz z$, including the metaplectic correction,
is thus
\beq
{\cal O} (\bz z) = z {\del \over \del z} + {1\over 2}
\label{ana47b}
\eeq

\subsection{Quantizing the two-sphere}

We now consider the example of the phase space being a two-sphere $S^2$.
This space can be considered as $\mathbb{CP}^1$, the complex projective space
in one (complex) dimension. It is
a K\"ahler manifold. We may also regard $S^2$ as $SU(2)/ U(1)$, a point of view which is
useful for generalization later. We will consider quantization of the two-sphere
first in local coordinates, then using homogeneous coordinates for $\mathbb{CP}^1$, and then
from the group theory point of view.

\noindent{\underline{\it Quantization using local coordinates}}

We introduce local complex coordinates for $\mathbb{CP}^1$ as
$z = x+iy, \bz =x-iy$, the standard K\"ahler two-form is given by
\beq
\omega = i ~{dz \wedge d\bz \over (1+z\bz )^2}\label{ana48}
\eeq
These coordinates can be related to an embedding of $S^2$ in $\mathbb{R}^3$ via
\beq
X_1 = {z + \bz  \over (1 + z \bz )}, \hskip .2in
X_2 = {i (z - \bz )  \over (1 + z \bz )}, \hskip .2in X_3 = {1- z \bz \over (1 + z \bz )}
\label{ana48a}
\eeq
so that we may view $z, \bz$ as the coordinates of a plane onto which
the sphere is stereographically projected.
The metric is given by $ds^2 = e^1 e^1 +e^2 e^2$ where the frame fields
are
\beq
e^1 = {dx \over 1+r^2} ~,\hskip 1.2in e^2= {dy\over 1+r^2}
\label{ana49}
\eeq
The Riemannian curvature is $ R^1_{\,2 } = 4\, e^1 \wedge e^2$ giving the Euler
number
\beq
\chi = \int {R_{12}\over 2\pi} =2
\label{ana50}
\eeq
The phase space has nonzero $\H^2(M, \mathbb{R})$ with its generating element
given by the K\"ahler two-form, which is also proportional to the volume form for
$S^2$.
As the discussion which led to
(\ref{anb10}) showed, the symplectic two-form must belong to an integral
cohomology class of $M$ to be able to quantize properly. So we consider
the symplectic form
\beq
\Omega = n~\omega = i~n ~{dz \wedge d\bz \over (1+z\bz )^2}
= i \, \del \, {\bar \del} \, K, \hskip .3in
K = n \log (1 + z \bz )
\label{ana51}
\eeq
where $n$ is an integer; 
In this case, $\int_M \Omega = 2\pi n$ as required
by the quantization condition.
$K$ is the K\"ahler potential for $\Omega$.
Classically the Poisson bracket of two functions $F$ and $G$ on the phase
space is given by
\beqar
\{ F,G\}&=& \Omega^{\mu\nu}\, \del_\mu F \,\del_\nu G\nonumber\\
&=& {i\over n}(1+z\bz )^2 \left( {\del F \over \del z}
{\del G \over \del \bz} ~-~{\del F \over \del \bz}{\del G \over \del z}
\right)\label{ana60}
\eeqar

Turning to the quantization, first of all,
the symplectic potential corresponding to the $\Omega$ in
(\ref{ana51}) can be taken as
\beq
\A = {in\over 2} \left[ {z~d\bz - \bz ~dz \over (1+z\bz )}\right]
\label{ana52}
\eeq
The covariant derivatives are given by $\del -i \A$. The holomorphic
polarization condition is 
\beq
(\del_\bz -i \A_\bz ) \Psi = \left[ \del_\bz +{n\over 2} {z\over (1+z\bz
)}\right] ~\Psi =0
\label{ana53}
\eeq
This can be solved as
\beq
\Psi =\exp\left ( -{n\over 2} \log (1+z\bz )\right)~f(z) 
\label{ana54}
\eeq
Notice that we have a factor $\exp ({-\half K} )$ as expected.
The inner product is given by
\beq
\la 1\vert 2\ra = i\,c \int { dz \wedge d\bz \over 2\pi (1+z\bz )^{n+2}}
~{f_1}^* f_2
\label{ana55}
\eeq
Here $c$ is an overall constant, which can be absorbed into the
normalization factors for the wave functions.
Since $f(z)$ in (\ref{ana54}) is holomorphic, we can see that a
basis of
nonsingular wave functions is given by
$f(z) =1, ~z, ~z^2,\cdots , ~z^n$; higher powers of $z$ will not have
finite norm.
The dimension of the Hilbert space is thus $(n+1)$. We could have seen
that this dimension would be finite from the semiclassical estimate of
the number of states as the phase volume. Since the phase volume is
finite for $M= S^2$, the dimension of the Hilbert space should be finite. \\
\hrule

It is interesting to see this dimension in another way. The polarization
condition (\ref{ana53}) is giving the ${\bar \del}$-closure of
$\Psi$ with a $U(1)$ gauge field $\A$ and on a space of Riemannian
curvature $R_{12}$.
The number of normalizable solutions to (\ref{ana53}) is thus given by
the index theorem for the twisted Dolbeault complex, i.e.,
\beq
{\rm index} ({\bar \del}_V )= \int_M {\rm td}(M) \wedge {\rm ch} (V)
\label{ana56}
\eeq
where, for our two-dimensional case, the Todd class ${\rm td}(M)$ is
$R/4\pi$ and the Chern character ${\rm ch}(V) = \Tr (e^{F/2\pi})$
is $\int \Omega /2\pi$ for us \cite{eguchi}. We thus have
\beq
{\rm index} ({\bar \del}_V ) =  \int_M {\Omega \over 2\pi} +\int_M {R\over
4\pi}
= n+1
\label{ana57}
\eeq
Notice that, semiclassically, we should expect the number of states
to be $\int \Omega /2\pi =n$. The extra one comes from the Euler number
in this case. (The semiclassical counting is supposed to apply only for
large $n$, so this is all consistent with expectations.)\\
\hrule
An orthonormal basis for the wave functions may be taken to be
\beq
f_k (z)= \left[ {n! \over  k!~(n-k)!}\right]^{1\over 2} ~z^k
\label{ana58}
\eeq
with the inner product
\beq
\la 1\vert 2\ra = i(n+1) \int { dz \wedge d\bz \over 2\pi (1+z\bz )^{n+2}}
~{f_1}^* f_2
\label{ana59}
\eeq
Here we have chosen the parameter $c$ in (\ref{ana55}) such that
the trace of the identity operator is $n+1$.

Consider now the vector fields
\beq
\xi_+ = i \left( {\del\over \del \bz} +z^2{\del \over \del
z}\right), \hskip .2in
\xi_- = i \left( {\del\over \del z} +\bz^2{\del \over \del
\bz}\right), \hskip .2in
\xi_3 = i \left( z{\del \over \del z} - \bz {\del \over \del
\bz}\right) \label{ana61}
\eeq
It is easily verified that these are the standard $SU(2)$ isometries of
the sphere. The Lie commutator of the $\xi$'s give the $SU(2)$ algebra.
Further, these
are Hamiltonian vector fields corresponding to the functions 
\beq
J_+ = -n ~{z\over 1+z\bz } ,\hskip .2in
J_- = -n ~{\bz \over 1+z\bz}, \hskip .2in
J_3 = -{n\over 2} \left( {1-z\bz \over 1+z\bz }\right)
\label{ana62}
\eeq
The prequantum operators $-i\xi\cdot\D +J$ corresponding to these
functions are
\beqar
\P (J_+)&=& ~\left( z^2\del_z - {n z\over 2}\, {2 + \bz z \over 1+ \bz z}\right) ~~-i \xi_+^\bz \D_\bz \nonumber\\
\P (J_{-})&=& \left(-\del_z -{n\over 2}{\bz \over 1+z\bz}\right)
-i\xi_-^\bz \D_\bz \nonumber\\
\P (J_3)&=& \left( z\del_z -{n\over 2} {1\over 1+z\bz}\right) -i\xi_3^\bz
\D_\bz \label{ana63}
\eeqar
Acting on the polarized wave functions, $\D_\bz $ in these expressions
will give zero. Writing
$\Psi$ as in (\ref{ana54}), we can work out the action of the operators
on the holomorphic wave functions $f(z)$, by moving the derivatives through the
$e^{- \half K}$ factor. We then find 
\beqar
{\hat J}_+ \, f &=& (z^2 \del_z -n ~z ) \, f\nonumber\\
{\hat J}_-  \, f&=& ( -\del_z)\, f \nonumber\\
{\hat J}_3 \, f&=& ( z\del_z -\half ~n ) \, f
\label{ana64}
\eeqar
If we define $j=n/2$, which is therefore half-integral, we see that
the operators given above correspond to a unitary irreducible
representation of $SU(2)$ with $J^2 = j(j+1)$ and dimension $n+1=2j+1$.
Notice that there is only one representation here and it is fixed by the
choice of the symplectic form $\Omega$.
In other words, the quantization of the two-sphere with the symplectic form
(\ref{ana51}) gives one unitary irreducible representation of $SU(2)$ with
$j = n/2$.

\noindent\underline{\it Quantization using homogeneous coordinates}

The complex coordinates we used are only local coordinates valid in a coordinate patch around
$z =0 $; strictly speaking we need at least another
coordinate patch with a different choice of coordinates to describe the sphere
in a nonsingular way. This would be valid around $z = \infty$; it did not matter too much
 in what we did, because the potential coordinate singularity is basically a point with
zero measure.

A more
global approach is to use the homogeneous coordinates of the 
sphere viewed as $\mathbb{CP}^1$. 
Recall that the complex projective space
$\mathbb{CP}^k$ is defined by $(k+1)$ complex coordinates
$( u_1, u_2, \cdots, u_{k+1}) \in \mathbb{C}^{k+1}$ with the identification
$( u_1, u_2, \cdots, u_{k+1})  \sim \lambda\, ( u_1, u_2, \cdots, u_{k+1}) $,
for any complex nonzero $\lambda$, $\lambda \in \mathbb{C} - \{ 0 \}$.
Thus, for $\mathbb{CP}^1$, we wil need two $u$'s which we may think of as a
two-component spinor $u_\alpha$, $\alpha =1, 2$,
with the identification $u_\alpha \sim \lambda u_\alpha$.
We also define $\bu_1 = u_2^*,~ \bu_2 =-u_1^*$
or $\bu_\alpha =\epsilon_{\alpha\beta}u_\beta^*$, where
$\epsilon_{\alpha\beta}=-\epsilon_{\beta\alpha}$, $\epsilon_{12}=1$.
The symplectic form is
\beq
\Omega = -i\, n \left[ {du \cdot d\bu \over \bu \cdot u} -{\bu\cdot du
~u\cdot d\bu\over (\bu \cdot u)^2}\right]
\label{ana69}
\eeq
where the notation is $u \cdot v = u_\alpha v_\beta \epsilon_{\alpha
\beta}$. This means that
$\bu \cdot v = u^\dagger v = u_1^* v_1 + u_2^* v_2$.
It is easily checked that $\Omega (\lambda u )=\Omega (u)$;
it is invariant under $u\rightarrow \lambda u$ and hence is properly 
defined on $\mathbb{CP}^1$ rather than $\mathbb{C}^2$. 
The choice of
$u_2/u_1 =z$ leads to the previous local parametrization; this is valid around 
$u_1 \neq 0$. We can use another coordinate patch with the local coordinates
$w = u_1 / u_2$. These two patches will correspond to the north and south hemispheres
of the sphere, in a stereographic projection.

The
symplectic potential corresponding to (\ref{ana69}) is
\beq
\A =-i {n\over 2} \left[ {u\cdot d\bu ~-~ du \cdot \bu \over \bu \cdot
u}\right] \label{ana70}
\eeq 
Directly from the above expression we see that
\beq
\A (\lambda u )= \A (u) ~+~ d \left( i {n\over 2} \log ({\bar
\lambda}/\lambda ) \right)\label{ana71}
\eeq
This means that $\A$ cannot be written as a globally defined form on
$\mathbb{CP}^1$ since it is not invariant under the needed identification
$u_\alpha \sim \lambda u_\alpha$.
 This is to be expected because $\int \Omega \neq 0$
and hence we cannot have a globally defined potential on $\mathbb{CP}^1$.
From the transformation law (\ref{ana71}) and (\ref{ana25}), we see that the prequantum wave functions must
transform as
\beq
\Psi (\lambda u , {\bar\lambda}\bu )= \Psi (u, \bu ) ~\exp\left[{n\over
2}\log(\lambda /{\bar \lambda})\right]\label{ana72}
\eeq 
The polarization condition for the wave functions becomes
\beq
\left[ {\del \over \del \bu_\alpha} -{n\over 2} {u_\beta\,
\epsilon_{\beta\alpha}\over \bu\cdot u}\right] ~\Psi =0
\label{ana73}
\eeq
The solution to this condition is
\beq
\Psi = \exp \left(- {n\over 2} \log (\bu \cdot u)\right)~ f(u)
\label{ana74}
\eeq
Combining this with (\ref{ana72}), we see that the holomorphic functions
$f(u)$ should behave as
\beq
f (\lambda u) = \lambda^n ~f (u)
\label{ana75}
\eeq
$f(u)$ must thus have $n$ $u$'s and hence is of the form
\beq
f (u) = \sum_{\alpha 's} ~ C^{\alpha_1 \cdots \alpha_n}~~
u_{\alpha_1}\cdots u_{\alpha_n} \label{ana76}
\eeq
Because of the symmetry of the indices, there are $n+1$ independent
functions, as before. There is a natural linear action of $SU(2)$ 
on the $u ,~\bu$ given by
\beq
u'_\alpha = U_{\alpha \beta} ~u_{\beta},\hskip 1.2in
\bu'_\alpha = U_{\alpha\beta} ~\bu_\beta \label{ana77}
\eeq
where $U_{\alpha\beta}$ form a $(2\times 2)$ $SU(2)$ matrix.
The corresponding generators are the $J_a$ we have constructed
in (\ref{ana63}, \ref{ana64}).
We have thus recovered all the previous results in a more global way.

\noindent\underline{\it Group theoretic version}

Equation (\ref{ana22}) relating the action and the symplectic potential $\A$ shows that
the potential of interest to us, namely,
(\ref{ana52}) can be obtained from the action
\beq
\S = i {n\over 2} \int dt~ {z{\dot \bz} - \bz {\dot z} \over 1+z\bz}
\label{ana65}
\eeq
where the overdot denotes differentiation with respect to time.
This action may be written as
\beq
\S = i {n\over 2} \int dt ~~\Tr (\sigma_3 \,g^{-1}{\dot g})
\label{ana66}
\eeq
where $g$ is an element of $SU(2)$ written as a $(2\times 2)$-matrix,
$g =\exp (i\,(\sigma_i/ 2)\theta_i)$ and $\sigma_i$, $i=1,2,3$, are the Pauli
matrices. In this action, the dynamical variable is thus an element of
$SU(2)$.
There are many ways to parametrize the group element, corresponding to
local coordinates on the group viewed as a Riemannian manifold.
One convenient parametrization  is given by
\beq
g= {1\over \sqrt{1+z\bz}}\left(\begin{matrix}
1&z\\
-\bz &1\\
\end{matrix} \right)~\left[
\begin{matrix} e^{i\theta} &0\\ 0&e^{-i\theta}\\ \end{matrix} \right]
\label{ana68}
\eeq
If this is used in (\ref{ana66}), we get (\ref{ana65}).

In the action (\ref{ana66}), if we make a transformation $g\rightarrow g~h$,
$h=\exp(i\sigma_3 \vf )$, we get
\beq
\S \rightarrow \S -n \int dt~ {\dot \vf}\label{ana67}
\eeq
The extra term is a boundary term and does not affect the equations of
motion. (It is for this same reason that $\theta$ in (\ref{ana68}) does not appear in
(\ref{ana65}).)
Since equations of motion do not depend on $\theta$,
we see that classically the dynamics is actually restricted to
$SU(2)/U(1) =S^2$. 

Even though the classical dynamics is restricted to $SU(2)/U(1)$, 
the boundary term  in (\ref{ana67}) does have an effect in the quantum
theory.
Consider choosing $\vf (t)$ such that $\vf (-\infty )=0$ and
$\vf (\infty )=2\pi$. In this case $h(-\infty )=h (\infty )=1$
giving a closed loop in the $U(1)$ subgroup of $SU(2)$ defined by the
$\sigma_3$-direction. For this choice of $h(t)$, the action changes by
$-2\pi n$. $e^{i\S}$ remains single-valued and, even in the quantum theory,
the extra $U(1)$ degree of freedom is consistently removed. If the
coefficient were not an integer, this would not be the case and we would
have inconsistencies in the quantum theory. The quantization of the
coefficient to an integral value is seen again from a slightly different
point of view.

We can now move ahead and complete the quantization. 
The canonical one-form is obtained from $\S$ as
\beq
\A = i {n \over 2} \Tr (\sigma_3 \, g^{-1} d g )
\label{ ana67a}
\eeq
The prequantum wave functions
are sections of a bundle on $SU(2)/ U(1)$. Let us start with functions on $SU(2)$. A function on $SU(2)$ may be written as a linear combination of the representation matrices
$\D^{(j)}_{ab}(g)$ as
\beq
\Psi = \sum_j \sum_{a, b} \, C^{(j)}_{ab} \, \D^{(j)}_{ab}(g)
= \sum_j \sum_{a, b} \, C^{(j)}_{ab} \, \la a\vert e^{i {\hat J}_i \theta_i } \, \vert b\ra
\label{ana67b}
\eeq
where ${\hat J}_i$ is the angular momentum or $SU(2)$ generator in an arbitrary representation.
(The matrices $\D^{(j)}_{ab}(g)$ are also known as the Wigner $\D$-functions.)
Consider the transformation $g \rightarrow g \, h$, $h = \exp (-i {\sigma_3\over 2}  \theta )$;
the change in $\A$ is given by $\A \rightarrow \A + (n/2)\, d \theta$. Since $\sigma_3/2$ corresponds to
${\hat J}_3$ in an arbitrary representation, this implies that the
wave functions must obey
\beq
\Psi \left( g\, e^{-i {\hat J}_3 \theta } \right) =  \Psi (g) ~ \exp \left( { i \,n\over 2} \theta  \right)
\label{ana67c}
\eeq
This identifies the $J_3$-eigenvalue of the state corresponding to $b$ in (\ref{ana67b}) as
$- n/2$, so that $\vert b\ra = \vert j, -{n\over 2}\ra$.

We have considered translations of $g$ on the right by $h \in U(1)$.
The remaining generators for the right action are
$R_\pm = R_1 \pm i R_2$, where $R_i$ is defined by
\beq
R_i \, g = g \, {\sigma_a \over 2}
\label{ana67d}
\eeq
The combinations $R_\pm$ are complex and conjugate to each other. We can take $R_-$ as the polarization condition, requiring the wave functions to obey
\beq
R_- \, \Psi = R_- \sum_j \sum_{a, b} \, C^{(j)}_{ab} \, \la a\vert e^{i {\hat J}_i \theta_i } \, \vert b\ra
= \sum_j \sum_{a, b} \, C^{(j)}_{ab} \, \la a\vert e^{i {\hat J}_i \theta_i } \, {\hat J}_- \vert b\ra
= 0
\label{ana67e}
\eeq
This is a holomorphicity condition and upon using the parametrization
(\ref{ana68}) will be seen to be identical to the condition (\ref{ana53}), namely, 
$\D_\bz \Psi =0$.
From the group theory point of view, (\ref{ana67e}) means that
the state $\vert b\ra$ must also be the lowest weight state.
A lowest weight state with $J_3 = -n/2$ means that $j = n/2$. Thus only
one representation in (\ref{ana67b}) will have nonzero coefficients, identifying the general 
wave function as
\beq
\Psi = \sum_{a} \, C^{({n\over 2})}_{a, -{n\over 2}} \, \D^{({n \over 2})}_{a, -{n \over 2}}(g)
\label{ana67f}
\eeq
A general state is a linear combination of $\D^{({n \over 2})}_{a, -{n \over 2}}(g)$; since
$a$ takes $2 j +1$ values, we see that the Hilbert space corresponds to a unitary irreducible
representation of $SU(2)$ with $j = n/2$.
The operators $J_i$ given in (\ref{ana63}) or (\ref{ana64}) correspond to the left action
on $g$, i.e.,
\beq
J_i \, \Psi (g) = \sum_{a} \, C^{({n\over 2})}_{a, -{n\over 2}} \, \D^{({n \over 2})}_{a, -{n \over 2}}({\sigma_i\over 2}\,g)
= \sum_{a, c} \, C^{({n\over 2})}_{a, -{n\over 2}} \, (J_i)_{ac} \D^{({n \over 2})}_{c, -{n \over 2}}(g)
\label{ana67g}
\eeq
Here $(J_i)_{ac}$ is the matrix version of ${\hat J}_i$ in the representation with
$j = n/2$.
We have thus reproduced the previous results from a purely group theoretic point of view.

Before we consider the generalization of this to arbitrary groups, it is useful to mention
some examples where these results turn up. 
We may regard the $\Omega =n \omega$ as a $U(1)$ magnetic field which is constant
(in the appropriate coordinates) on the sphere.
This point of view is further supported by looking at $R_\pm$. These are translation operators
on the sphere, but their commutator is given by
$[R_+ , R_- ] \, \Psi = 2 \, R_3 \Psi  = n\, \Psi$. The commutator of derivatives is the gauge field, so we can identify a magnetic field for this case as $2 \, B = n$.
(We set the electric charge to be $1$; also we took the sphere to have radius equal to $1$, otherwise this would read
$2 \, B R^2 = n$ where $R$ is the radius.)
The quantization of the magnetic flux is the Dirac quantization condition again.
Thus the states (\ref{ana67f}) we find are
the angular part of the wave functions
 for a charged particle in the presence of a magnetic monopole \cite{bal}.
 Also they can be thought of as the lowest Landau levels for a constant magnetic field on the 
 sphere \cite{{haldane}, {KN1}}.
 The left action of the $J_i$ as in (\ref{ana67g}) correspond to the so-called magnetic translations
 for the Landau levels. So quantum Hall effect on the sphere can be discussed using these
 wave functions.
 
 This can also appear as part of the dynamics of a particle with spin; we get one UIR of $SU(2)$, so we have exactly what is needed for spin. It can also be thought of as describing the internal symmetry structures, such as the color degrees of freedom for a particle with nonabelian charges
 for the case of the color group being $SU(2)$.

\subsection{K\"ahler spaces of the $G/H$-type}

The two-sphere $S^2 =SU(2)/U(1)$ is an example of a group coset which is
a K\"ahler manifold. There are many K\"ahler manifolds which are
of the form $G/H$ where $H$ is a subgroup of a compact Lie group $G$.
In particular $G/H$ is a K\"ahler manifold for any compact Lie group if
$H$ is its maximal torus. Another set of K\"ahler
spaces of this type is given by
$\mathbb{CP}^k = SU(k+1) /U(k)$. 
There are also examples of this type corresponding to noncompact groups.
For example, the Lobachevskian space $SL(2, \mathbb{R})/ U(1)$ is also a K\"ahler manifold,
although its volume defined by the K\"ahler two-form is infinite.
There are many other cases as well \cite{BWB}.

In these cases, one can consider theories
where the symplectic form is proportional to the K\"ahler form 
or is a combination of the generators of $\H^2(M, \mathbb{R})$ for these manifolds
and quantize as we have done for the case of $S^2$.
The general result is that 
they lead to one unitary irreducible representation (UIR) of the
group $G$, the specific representation being determined
by the choice of $\Omega$.

\noindent\underline{\it Quantizing $\mathbb{CP}^2$}

In most of these cases with $G/H$ structure, it is rather simple and straightforward to
construct the K\"ahler form for these spaces. We will consider in some detail
another example, namely, the quantization of
$\mathbb{CP}^2 =SU(3)/U(2)$.
A general element of $SU(3)$ can be represented as a unitary 
$(3\times 3)$-matrix. This is of the form
$g = \exp (i t_a \theta^a )$, where the generators $\{ t_a \}$ in the 
$3 \times 3$ matrix 
representation can be chosen as
\beqar
    &&
   t_1 = {1\over 2}
    \left(
      \begin{matrix}
        0 & 1 & 0 \\
        1 & 0 & 0 \\
        0 & 0 & 0 \\
      \end{matrix}
    \right)~~~
     t_2 = {1\over 2}
    \left(
      \begin{matrix}
        0 & -i & 0 \\
        i & 0 & 0 \\
        0 & 0 & 0 \\
      \end{matrix}
    \right)~~~
   t_3 = {1\over 2}
    \left(
      \begin{matrix}
        1 & 0 & 0 \\
        0 & -1 & 0 \\
        0 & 0 & 0 \\
      \end{matrix}
    \right)~~~    t_4 = {1\over 2}
    \left(
      \begin{matrix}
        0 & 0 & 1 \\
        0 & 0 & 0 \\
        1 & 0 & 0 \\
      \end{matrix}
    \right)
    \nonumber \\
    &&~\label{ana77}\\
    &&
     t_5 = {1\over 2}
    \left(
      \begin{matrix}
        0 & 0 & -i \\
        0 & 0 & 0 \\
        i & 0 & 0 \\
      \end{matrix}
    \right)~~~
    t_6 = {1\over 2}
    \left(
      \begin{matrix}
        0 & 0 & 0 \\
        0 & 0 & 1 \\
        0 & 1 & 0 \\
      \end{matrix}
    \right)
~~~
    t_7 = {1\over 2}
    \left(
      \begin{matrix}
        0 & 0 & 0 \\
        0 & 0 & -i \\
        0 & i & 0 \\
      \end{matrix}
    \right)~~~
     t_8 = \frac{1}{\sqrt{12}}
    \left(
      \begin{matrix}
        1 & 0 & 0 \\
        0 & 1 & 0 \\
        0 & 0 & -2 \\
      \end{matrix}
    \right)
\nonumber
\eeqar
(Our $t_a$ are normalized so that $\Tr (t_a t_b ) = \half \delta_{ab}$.)
We define a $U(1)$ subgroup by elements
of the form $U= \exp(i t_8 \theta^8 )$ and we
can also define an $SU(2)$
subgroup which commutes with this $U(1)$ subgroup; the latter has
elements of the form $U= \exp (i t_a \theta^a )$ for $ a = 1, 2, 3$.
These two subgroups together form the $U(2)$ subgroup of $SU(3)$.\footnote{
Strictly speaking there is an identification of certain elements involved.
There is a common $\mathbb{Z}_2$ subgroup
for the factors in $SU(2)\times U(1)$
defined by $\mathbb {Z}_2 = \{ 1, h_Z \}$, $h_Z = (h_2, h_1)$ with
$h_Z^2 =1$ and
\beq
h_2= \left( \begin{matrix}
 -1_{2\times 2} &0\\
  0&1\\
\end{matrix} \right), \hskip .3in h_1 = \exp (i t_8 \sqrt{12} \, \pi ) =
\left( \begin{matrix}
 -1_{2\times 2} &0\\
  0&1\\
\end{matrix} \right)
\label{ana79a}
\eeq
The $U(2)$ subgroup is given by $SU(2) \times U(1) / \mathbb{Z}_2$.}
Consider now the one-form
\beq
\A (g) = i\, w ~\Tr (t_8\, g^{-1}dg ) = - i \, w\, {\sqrt{3}\over 2}\, u_\alpha \, du^*_\alpha
\label{ana80}
\eeq
where $g$ is an element of the group $SU(3)$ and $w$ is a
numerical constant; $u^*_\alpha = g_{\alpha 3}$.
If $h$ is an element of
$U(2)\subset SU(3)$ of the form $h = U \, \exp ( i \,t_8 \,\theta )$, we find
\beq
\A (g\,h) = \A (g) - {w\over 2}\, d \theta \label{ana81}
\eeq
We see that $\A$ changes by a total
differential under the $U(2)$-transformations. The two-form
$d\A$ is therefore independent of $\theta$ or it is invariant under
$U(2)$ transformations; it is a two-form defined on
the coset space $SU(3)/U(2)$. 
Evidently it is closed ($d \,d \A = 0$ since $d^2 = 0$), but it is not exact since the
corresponding one-form is not globally defined on $SU(3)/U(2)$, but only
on $G= SU(3)$. Thus $d\A$ is a nontrivial element of $\H^2( SU(3)/U(2), \mathbb{R})=
\H^2(\mathbb{CP}^2, \mathbb{R})$.
There will be quantization conditions on $w$ and the lowest possible choice,
with our choice of normalization for $t_8$, will
be $ 2/\sqrt{3}$. The K\"ahler 2-form for $SU(3)/U(2)$ is
\beq
\omega = d \, \left( i {2\over \sqrt{ 3}} \,\, \Tr ( t_8 \, g^{-1} dg ) \right)
\label{ana81a}
\eeq
The connection with the complex projective space is clarified by introducing
$Z_\alpha = \rho\, u_\alpha $, where $\rho$ is an arbitrary complex number, not
equal to zero. We can then consider
\beq
\A = - i \, w\,{\sqrt{3}\over 2}\, {Z \cdot d{\bar Z}\over {\bar Z}\cdot Z}
=  -i \, w\,{\sqrt{3}\over 2}\, \left[ u_\alpha \, du^*_\alpha + d \log\rho \right]
\label{ana81b}
\eeq
This $\A$ differs from (\ref{ana80}) by a total derivative and hence 
$d \,\A$ will be the same for both
$\A$'s. We thus see that we can write $\omega$ as
\beq
\omega = -i \left[ {dZ \cdot d {\bar Z} \over (Z \cdot {\bar Z})}
- {dZ \cdot {\bar Z}\, Z\cdot d {\bar Z} \over (Z \cdot {\bar Z})^2} \right]
\label{ana81c}
\eeq
which is the expected K\"ahler form on $\mathbb{CP}^2$.

As the symplectic form for quantization, we can consider
any integral multiple of $\omega$; we need an integral multiple, since
the integrals of $\Omega = d \A$ over
nontrivial two-cycles on
$\mathbb{CP}^2$ will have to be integers.
Thus the possible choices for $w$ are of the form
$w = 2 \, n / \sqrt{3}$, $n \in \mathbb{Z}$.\footnote{
It should be kept in mind that different choices of $w$
correspond to different theories and different physics.}
Therefore we will consider the symplectic two-form
\beqar
\Omega &=& - i { 2\, n \over \sqrt{3}} \, \Tr \left( t_8 \,g^{-1} dg \wedge g^{-1} dg \right)
\nonumber\\
&=&n \, \omega = - i\, n \left[ {dZ \cdot d {\bar Z} \over (Z \cdot {\bar Z})}
- {dZ \cdot {\bar Z}\, Z\cdot d {\bar Z} \over (Z \cdot {\bar Z})^2} \right]
\label{ana81d}
\eeqar
The action which leads to the chosen $\A$ and the $\Omega$ in (\ref{ana81d})
is
\beq
\S =  i {2\, n \over \sqrt{3}} \int dt ~ \Tr ( t_8 \, g^{-1} {\dot g} )
\label{ana81e}
\eeq
Again, for $e^{i\S}$ to be well defined on $\mathbb{CP}^2$, the values of
$w$ will have to be restricted to the form given above, namely,
$ w= 2 \,n /\sqrt{3}$, $n \in \mathbb{Z}$.
The wave functions are functions on $SU(3)$ subject
to the restrictions given by the action of $SU(2)$ and $U(1)$ and a holomorphicity condition.
In other words, we can write, using the Wigner ${\cal D}$-functions for $SU(3)$
\beq
\Psi \sim {\cal D}^{(r)}_{AB} (g) = 
\la r, A \vert\, {\hat g}\, \vert r, B\ra
\label{ana81f}
\eeq
Here $(r )$ is a set of indices which
labels the representation, $A,B$ label the states within a representation.
Only the finite-dimensional (and hence unitary) representations can occur here, since they form a complete set for functions on $SU(3)$.

The groups involved in the
quotient can be taken as the right action on $g$. The transformation law for
$\A$ then tells us that $\Psi$ must transform as
\beq
\Psi (g\,h) = \Psi (g) ~\exp \left(-i\,{n\over\sqrt{3}} \, \theta \right) 
\label{ana81g}
\eeq
This shows that the wave functions must be singlets under the $SU(2)$ subgroup and
carry a definite charge $n/\sqrt{3}$ under the $U(1)$ subgroup generated by $t_8$. This restricts the choice of
values for the state $\vert r, B \ra$ in (\ref{ana81f}).
Further $w = 2 \, n /\sqrt{3}$ must also be quantized so that it can be one of the allowed values in
the representations of $SU(3)$ in (\ref{ana81f}).
This is the same as what we already found, namely, that $n$ must be an integer.

One has to choose a polarization condition as well. The generators of
$SU(3)$ can be divided into those of the $SU(2)$ and $U(1)$ subgroups, and
the coset ones which correspond to $t_i$ with $i = 4, 5, 6, 7$.
These can be grouped into $t_a, \, t_{\bar a}$ and
corresponding to $= t_4 +i t_5, t_6+i t_7$ and their conjugates, $a, {\bar a} = 1, 2$.
Correspondingly, we can define the right translation operators
\beq
R_a \, g = g \, t_a , \hskip .3in
R_{\bar a} \, g = g \, t_{\bar a}
\label{ana81h}
\eeq
As the holomorhic polarization condition, we choose
\beq
R_{\bar a} \, \Psi (g) = 0
\label{ana81i}
\eeq
This requires the state $\vert r, B\ra$ to be a highest weight state. This requirement, along with
the earlier statement that 
$\vert r, B\ra$  should be an $SU(2)$ singlet with eigenvalue $n/\sqrt{3}$ for $t_8$ transformation,
completely fixes the representation $r$ and the state $\vert r, B\ra$.
The left index $A$ is, however, free, running over the possible states in 
the representation $r$.
Thus the
result of the quantization is to yield a Hilbert space which is one
unitary irreducible representation (UIR) of the group
$SU(3)$.

\noindent\underline{\it Quantizing general $G/H$ spaces}

More generally, with a view of obtaining UIRs of a compact Lie group $G$,
one can take 
\beq
\A (g) = i \sum_a w_a \Tr ( t_a \,g^{-1} dg )\label{ana82}
\eeq
where $t_a$ are diagonal elements of the Lie algebra of $G$ and
$w_a$ are a set of numbers. $H$ will be the subgroup commuting with
$\sum_a w_a t^a$; if $w_a$ are such that all the diagonal elements
of $\sum_a w_a t^a$ are distinct, then $H$ will be the maximal torus of
$G$. $\A$ will change by a total differential under $g\rightarrow gh$,
$h\in H$ and $d\A$ will be a closed nonexact form on $G/H$.
If some of the eigenvalues of $\sum_a w_a t^a$ are equal, $H$ can be
larger than the maximal torus. Upon quantization, for suitably chosen
$w_a$, we will get one unitary irreducible representation of $G$ and
$w_a$ will be related to the highest
weights defining the representation \cite{BWB}.

There is another way to think about this problem. Let us say that we want to construct
a UIR of a group $G$. We ask the question: Is there a classical action which upon quantization
gives exactly one UIR of the group $G$?
Recall that if we quantize the rigid rotor we get all UIR's of the angular momentum group
$SO(3)$. That is not what we want, we want one and only one representation.
The answer to this is the action 
\beq
\S = i \sum_a w_a \int dt~ \Tr ( t_a \,g^{-1} {\dot g}  )\label{ana82a}
\eeq
with the choice of $\{ w_a\}$ determined by which representation we wish to obtain
upon quantization.

One can use an action similar to (\ref{ana82a}) for noncompact groups as well. 
The key here is that, since we are quantizing the system, the representation
we obtain
is unitary.
Thus if one carries out the quantization of $SL(2, \mathbb{R})/ U(1)$,
we will get a UIR of $SL(2, \mathbb{R})$. Such representations are infinite dimensional
since $SL(2, \mathbb{R})$ is noncompact. The representation obtained will be one of the
series needed for the completeness relation for functions on $SL(2, \mathbb{R})$.
The infinite dimensionality is also in agreement with the semiclassical counting
of the dimension of the Hilbert space since the phase volume
($=$ the volume of $SL(2, \mathbb{R})/U(1)$ as measured by its K\"ahler form) is
infinite.

\noindent\underline{\it A short historical note}

Historically, geometric quantization arose out of representation theory for groups.
The construction of UIR's of a compact group using the K\"ahler two-form
on $G/T$ where $T$ is the maximal torus was carried out in the 1950s. It goes by the name of
Borel-Weil-Bott theory. 
Geometric quantization was developed in the 1970s (by Kostant, Souriau, Kirillov and others)
as an attempt to generalize this 
to arbitrary symplectic manifolds.
The use of actions of the form (\ref{ana82a}) for various physical problems was
pursued in the 1970s by Balachandran and others. This action (\ref{ana82a}) may also 
be viewed as the
prototypical Wess-Zumino term.
The usual Wess-Zumino term was introduced in the context of meson physics by
Wess and Zumino in 1971 as an effective action for anomalies \cite{WZ}.
 It was developed and its full import was realized in the work of
Witten \cite{witten}.
(In this context, Novikov's work on the Wess-Zumino term in a (2+1)-dimensional
setting should be mentioned, although
the physics implications were not fully evident \cite{novikov}.
There were also a few other earlier papers which focused on certain aspects of the Wess-Zumino term.)

\section{The Chern-Simons theory in 2+1
dimensions} 

The Chern-Simons (CS) theory is a gauge theory in two
space (and one time) dimensions \cite{CS1}. The action is given by
\beqar
\S&=& -{k\over 4\pi}\int_{\Sigma\times[t_i,t_f]} \Tr \left[ A \wedge dA +{2\over
3} A\wedge A\wedge A\right]\nonumber\\
&=&-{k\over 4\pi}\int_{\Sigma\times[t_i,t_f]}d^3x~\epsilon^{\mu\nu\alpha}
~\Tr\left[ A_\mu\partial_\nu
A_\alpha+{2\over3} A_\mu A_\nu A_\alpha\right]\label{cs0a}
\eeqar
\noindent Here $A_\mu$ is the Lie algebra valued gauge potential,
$A_\mu=-i\,t_a A_\mu^a$, corresponding to a compact Lie group
$G$. $t_a$ are hermitian matrices forming a basis of
the Lie algebra in the fundamental representation of the gauge group.
We shall the gauge group to be
$G = SU(N)$ in what follows and normalize the $t^a$ as
$\Tr (t_a t_b) =\half \delta_{ab}$.
Thus, for example, for the case of the gauge group being $SU(3)$, the set of
matrices $t_a$ can be taken as
the ones given in (\ref{ana77}).
In addition to the choice of the group, the theory has one parameter $k$,
which is a real constant whose precise value we do not need to specify
at this stage. We shall consider the spatial manifold to
be some Riemann surface $\Sigma$ and
we shall be using complex
coordinates \cite{{CS1}, {CS2}}. The classical equations of motion for the theory are
\beq
F_{\mu\nu}=0\label{cs0b}
\eeq

We shall now consider the quantization of the theory in the formalism we have developed.
For this purpose, the theory is best analyzed in the
gauge where $A_0$ is set to zero. 
In the $A_0=0$ gauge, the action becomes
\beq
\S=-{ik\over\pi}\int dt d\mu_\Sigma ~\Tr (A_{\bar z}\partial_0 A_z)
\label{cs1}
\eeq
Taking the variation of the action, we see that the boundary term which results from the time-integration is
\beq
\delta \S~=~ -{ik\over{\pi}}\int_\Sigma ~\Tr (A_\bz \delta A_z )
\Biggr]^{t_f}_{t_i}\label{cs2}
\eeq
As in (\ref{ana24}), this identifies the
symplectic potential as
\beq
\A~=~ -{ik\over{\pi}}\int_{\Sigma}\Tr\bigl(A_\bz\delta A_z\bigr)~+~\delta
\rho [A]\label{cs3}
\eeq
where $\rho [A]$ is an arbitrary functional of $A$. The freedom of adding
$\delta \rho$ is the freedom of canonical transformations. 
We have written $A$ as a one-form on space; we also need one-forms on the space of the fields
$A$; to avoid confusion, we use $\delta$ to denote exterior derivatives on the space of fields.
We may define the space of fields as follows. The potential $A_i$ is a map from
$\Sigma$ to Lie algebra valued one-forms on $\Sigma$.
\beq
\mathfrak{F} = \{ {\rm Set ~of ~all ~gauge ~potentials}~ A_i\}
\label{cs3a}
\eeq
This is not the true space of physical field configurations, since potentials which differ by a gauge transformation are physically equivalent. We define
\beq
\mathfrak{G}_*= \{ {\rm Set ~of ~all ~maps} ~g(x): \Sigma \rightarrow G, ~ g \neq \, {\rm constant}\}
\eeq
(The subscript is to emphasize that we exclude constant $g$'s.)
The true configuration space should be $\mathfrak{C} = \mathfrak{F} / \mathfrak{G}_*$.

\subsection{Analysis on $S^2 \times \mathbb{R}$}

We will now consider a simple special case, namely, $\Sigma = S^2$.
We will carry out the analysis on the space of potentials, imposing the condition to
eliminate gauge freedom later.
Notice that (\ref{cs3}) is defined on $\mathfrak{F}$; this is the
phase space of the theory before reduction by the action of gauge 
symmetries. The
symplectic two-form $\Omega$ is given by $\delta \A$, i.e.,
\beqar
\Omega&=& -{ik\over{\pi}}~\int_{\Sigma}d\mu_\Sigma~\Tr\bigl(\delta
A_\bz\delta A_z\bigr)\nonumber\\
&=& {ik\over 2\pi} \int_{\Sigma}d\mu_\Sigma~ \delta A_\bz^a \delta A_z^a
\label{cs4}
\eeqar
(We do not write the wedge sign for exterior products on the field
space from now on since it
is clear from the context.)

The complex structure on $\Sigma$ induces a complex structure on 
$\mathfrak{F}$. We may take
$A_{z},~A_{\bar z}$ as the local complex coordinates on
$\mathfrak{F}$. Indeed we have a K\"ahler structure on 
$\mathfrak{F}$, 
$\Omega /k$
being the K\"ahler form with the K\"ahler potential 
\beq
K~=~ {k\over{2\pi}}
\int_{\Sigma} A_{\bar z}^{a} A_{z}^{a}\label{cs5}
\eeq
The Hamiltonian vector fields corresponding to $A_{z}$ and $A_{\bar z}$ 
are
\beq
A^a_z (z) \longrightarrow - {2\pi \over i k} {\delta \over \delta A^a_{\bz} },
\hskip .3in
A^a_\bz (z) \longrightarrow  {2\pi \over i k} {\delta \over \delta A^a_{z} }
\label{cs5a}
\eeq
The
Poisson brackets for $A_\bz$ , $A_z$ are
obtained using the general formula (\ref{ana9}) as
\beqar
\{A_{z}^a(z), A_{w}^b(w)\}&=&0\nonumber\\
\{A_{\bar z}^a(z), A_{\bar
w}^b(w)\}&=&0\nonumber\\
\{A_{z}^{a}(z), A_{\bar w}^{b}(w)\}&=& -{2\pi i\over k} \delta^{ab}
\delta^{(2)}(z-w)
\label{cs6}
\eeqar
These become commutation rules upon quantization.
Gauge transformations are given by
\beq
A^{g} = g\,A\,g^{-1} - dg\, g^{-1}\label{cs7}
\eeq
The infinitesimal version of this (for $g \approx 1 -i t_a \theta^a$)
is generated by the vector field
\beq
\xi~=~ -\int_{\Sigma} ~ \bigl( (D_{z}\theta)^{a} {\delta\over{\delta
A_{z}^a}}~+~ (D_{\bar z}\theta)^{a} {\delta\over{\delta A_{\bar
z}^a}}\bigr)\label{cs8}
\eeq
where $D_{z}$ and $D_{\bar z}$ denote the corresponding gauge covariant
derivatives. By contracting this with $\Omega$ we get 
\beq
i_\xi\Omega ~=~ -\delta \left[{ik\over{2\pi}} \int_{\Sigma}F^a_{z\bz}
\theta^a\right]\label{cs9}
\eeq
which shows that the generator of infinitesimal gauge
transformations is 
\beq
G^a= {ik\over{2\pi}}F^a_{z\bz}\label{cs10.1}
\eeq
Reduction of the phase space to gauge-invariant degrees of freedom 
can thus be performed by setting $F_{z \bz}$ to zero.
This takes us from $\mathfrak{F}$ to
$\mathfrak{C} = \mathfrak{F}/ \mathfrak{G}_*$.
The constraint $G^a = 0$ is just the Gauss law of the Chern-Simons
gauge theory. 
Notice also that in the $A_0 =0$ gauge, the
equations of motion (\ref{cs0b}) tell us that $A_z$, $A_{\bar z}$
are independent of time, but must satisfy the constraint
\beq
F_{{\bar z}z}\equiv\partial_{\bar z}A_z-
\partial_z A_{\bar z}+[A_{\bar z},A_z]=0\label{cs0c}
\eeq
This is the equation of motion for the $A_0^a$ component.
We see that the constraint which reduces the phase space
to the physical degrees of freedom is just the equation of motion for $A_0^a$.
Notice also that, for finite transformations, we get
\beqar
\Omega (A^{g})~-~\Omega(A)&=& \delta
\left[ {ik\over{\pi}}\int_\Sigma \Tr(g^{-1}\delta g
~F_{z\bz})\right]\nonumber\\ 
&=& \delta
\left[ -{k\over{2\pi}}\int_\Sigma \Tr(g^{-1}\delta g
~F)\right]
\label{cs10.2}
\eeqar
(In the second term $F$ is the two-form $dA+A\wedge A$.)

The construction of the wave functions proceeds as follows. One has to
consider a line bundle on the phase space with curvature $\Omega$.
Sections of this bundle give the prequantum Hilbert space. In other words
we consider functionals $\Phi[A_{z},A_{\bar z}]$ with the condition that
under the canonical transformation $\A\rightarrow \A~+~ \delta \Lambda,~~
\Phi\rightarrow e^{i\,\Lambda}~\Phi $. The inner product on the prequantum
Hilbert space is given by
\beq
(1\vert 2)~=~ \int~d \mu (A_{z}, A_{\bar z})~{\Phi}_{1}^{*}[A_{z}, A_{\bar
z}] ~{\Phi}_{2} [A_{z},  A_{\bar z}] \label {cs10.3}
\eeq
where $d\mu(A_{z}, A_{\bar z})$ is the Liouville measure associated with
$\Omega$. Given the K\"ahler structure $\Omega /k$, this is just the volume
$[dA_{z}dA_{\bar z}]$ associated with the metric $\vert\vert \delta
A\vert\vert ^{2} ~=~ \int_{\Sigma}~ \delta A_{\bar z}\delta A_{z}$.

The wave functions so constructed depend on all phase space variables. 
We must now choose the polarization conditions on the $\Phi$'s so that they
depend only on half the number of phase space variables. 
This reduction of the prequantum Hilbert space
leads to the Hilbert space of the quantum theory. Given the K\"ahler
structure of the phase space, the most appropriate choice is the Bargmann
polarization which can be implemented as follows. With a specific choice
of $\rho[A]$ in (\ref{cs3}), the symplectic potential can be taken as
\beq
\A~=~ -{ik\over{2\pi}} \int_{\Sigma} ~\Tr\bigl(A_{\bar z} \delta
A_{z}~-~ A_{z}\delta A_{\bar z}\bigr)~=~ {ik\over{4\pi}}\int_{\Sigma}
\bigl(A_{\bar z}^{a}
\delta A_{z}^{a}-A_{z}^{a} \delta A_{\bar z}^{a}\bigr)\label{cs11}
\eeq
The covariant derivatives with $\A$ as the potential are
\beq
\nabla~=~ \bigl( {\delta \over{\delta A_{z}^{a}}}+{k\over{4\pi}} A_{\bar z}^{a}
\bigr),~~~~~~~~ {\overline\nabla} ~=~ \bigl({\delta\over{\delta A_{\bar
z}^{a}}} -{k\over{4\pi}} A_{z}^{a}\bigr)\label{cs12}
\eeq
The holomorphic (or Bargmann) polarization condition is
\beq
\nabla ~\Phi =0
\eeq
which has solutions of the form
\beq
\Phi ~=~ \exp\left( -{k\over{4\pi}}\int A_{\bar z}^{a} A_{z}^{a}\right)
~~\psi [A_{\bar z}^{a}]~=~ e^{-\half K}~\psi [A_{\bar
z}^{a}]\label{cs13}
\eeq
where $K$ is the K\"ahler potential of (\ref{cs5}). The
states are represented by wave functionals $\psi [A_{\bar z}^{a}]$ which
are holomorphic in $A_{\bar z}^{a}$. Further, the
prequantum inner product can be retained as the inner product of the Hilbert
space. Rewriting (\ref{cs10.3}) using (\ref{cs13}) we get the inner
product as 
\beq
\big< 1\vert 2\big>~=~ \int [dA_{\bar z}^{a}\,d A_{z}^{a}]~ e^{-K(A_{\bar
z}^{a}, A_{z}^{a})}~\psi_{1}^{*}~\psi_{2}\label{cs14}
\eeq
On the holomorphic wave functionals, 
\beq
A_{z}^{a} ~\psi [A_{\bar z}^{a}]=
{2\pi\over k}{\delta \over{\delta A_{\bar z}^{a}}}~\psi [A_{\bar
z}^{a}]\label{cs14a}
\eeq

As we have mentioned before, one has to make a reduction of the Hilbert
space by imposing gauge invariance on the states, i.e., by setting the
generator $F_{z \bar z}^{a}$ to zero on the wave functionals. This amounts to
\beq
\left(D_{\bar z}~{\delta \over{\delta A_{\bar z}^{a}}}~-~ {k\over
{2\pi}} \partial_{z}A_{\bar z}^{a}\right)~\psi [A_{\bar
z}^{a}]~=0.\label{cs15}
\eeq
Consistent implementation of gauge invariance can lead to quantization
requirements on the coupling constant $k$. For nonabelian groups $G$ this
is essentially the requirement of integrality of $k$ based on the invariance
of $e^{iS}$ under homotopically nontrivial gauge transformations. It is the same as the Dirac
quantization conditon (\ref{anb10}). Also, once we impose the gauge invariance condition,
the integration in (\ref{cs14}) must be restricted to the gauge-invariant volume.

\subsection{Argument for quantization of $k$}

We will now work out how the quantization of $k$ arises, in some detail,
staying within the geometric quantization framework.
Since we are on $S^2$, the 
group of gauge transformations 
$\mathfrak{G}_* ~=~ \{ g(x): ~S^{2}\rightarrow G , ~ g \neq ~{\rm constant}\}$. We have chosen $G = SU(N)$,
so obviously 
\beq
\Pi_{0}(\mathfrak{G}_*) = \Pi_{2}(G) =0, \hskip .3in
\Pi_{1}(\mathfrak{G}_*) = \Pi_{3}(G) = \mathbb{Z}
\label{cs15a}
\eeq
Correspondingly one has 
\beq
\Pi_{1}(\mathfrak{F}/ \mathfrak{G}_*) = 0, \hskip .3in
\Pi_{2}(\mathfrak{F}/ \mathfrak{G}_*)  = \mathbb{Z}
\label{cs15b}
\eeq
The nontriviality of  $\Pi_{2}(\mathfrak{F}/ \mathfrak{G}_*) $ arises
from the nontrivial elements of $\Pi_{1}(\mathfrak{G}_*)$. Therefore consider
a noncontractible
loop $C$ of gauge transformations,
\beq
C=  g(x,\alpha), \hskip .2in 0\leq \alpha \leq 1, \hskip .1in {\rm with} \hskip .1in
g(x,0)= g(x,1) =1\label{cs16}
\eeq
With the boundary condition given, $g(x, \alpha )$ may be considered as
a map from $S^3$ to $G$. Such elements fall into homotopy classes corresponding to
$\Pi_3 (G) = \mathbb{Z}$.
We can now use this $g (x, \alpha )$ to construct a noncontractible two-surface in 
the gauge -invariant space $\mathfrak{F}/ \mathfrak{G}_*$.
We start with a square in the space of gauge potentials
parmetrized by 
$0\leq\alpha, \sigma\leq 1$ with the potentials given by
\beq
A(x,\alpha, \sigma) = (g\,A\,g^{-1}-~dg\,g^{-1}) \,\sigma~+~ (1-\sigma) A
\label{cs19}
\eeq
We can simplify this even further by taking $A =0$, so that
\beq
A(x,\alpha, \sigma) = - \sigma \,dg \, g^{-1}
\label{cs19a}
\eeq
This potential goes to zero on the boundaries $\alpha =0$ and $\alpha =1$
and also on $\sigma =0$. $A$ goes to the pure gauge $-dgg^{-1}$ at
$\sigma =1$, which is gauge-equivalent to
$A =0$. Thus the boundary corresponds to a single point
on the quotient $\mathfrak{F}/ \mathfrak{G}_*$ and we have a closed two-surface.
This surface is noncontractible if we take $g(x, \alpha )$ to be a
nontrivial element of $\Pi_3 (G) = \mathbb{Z}$ since the contraction of the two-surface would constitute a homotopy mapping $g$ to the identity; this is impossible if $g$ belongs to
a nontrivial element of $\Pi_3 (G)$.
Using this set of configurations in $\Omega$ and carrying out the integration
over $\sigma$ we get 
\beq
\int \Omega = 2 \pi \, k\, Q[g]\label{cs20}
\eeq
where 
\beq
Q[g]~=~ {1\over{24\pi^{2}}}\int \Tr(dgg^{-1})^{3}\label{cs20a}
\eeq
$Q[g]$ is the winding
number (which is an integer) characterizing the class in $\Pi_{1}(\mathfrak{G}_*)
=\Pi_{3}(G)$ to which
$g$ belongs. From (\ref{anb10}) we know that the integral of $\Omega$
over any closed noncontractible two-surface in the phase space must be an
integer. Thus (\ref{cs20}) and (\ref{cs20a}) lead to the
requirement that $k$ has to be an integer.
The parameter $k$ is referred to as the {\it level number}
of the Chern-Simons theory.
(Even though we
presented the arguments for quantization of the coefficient of the action
for $ \Sigma =S^2$, similar arguments and results hold more generally.)

\subsection{The ground state wave function}

We now turn to the solution of (\ref{cs15}).
For this we introduce the Wess-Zumino-Witten action given by \cite{{witten2}, {novikov}}
\beqar
{\cal S}_{WZW} &=& {1\over 8\pi} \int_{\Sigma} d^2x\sqrt{g} ~ g^{ab}\,\Tr
(\del_a K
\del_b K^{-1}) ~+~\Gamma [K]
\nonumber\\
\Gamma [K] &=&  {i\over 12\pi} \int_{{\cal M}^3} \Tr ( K^{-1} d K )^3
\label{an2.38}
\eeqar
The fields are matrices $K$ which can generally belong to
$GL(N, \mathbb{C})$.
Also $\Sigma$ is the two-dimensional space on which the
fields are defined. Since it can in general be a curved manifold,
we use the two-dimensional metric
tensor $g_{ab}$. ($g^{ab}$ is the inverse metric
and $g$ denotes the determinant of $g_{ab}$ as a matrix.)
(This model can be defined and used 
for fields on $\mathbb{R}^2$ as well, by choosing the 
boundary condition $K \rightarrow 1$ 
(or some fixed value independent of
directions)
as $\vert \vx\vert
\rightarrow \infty$; topologically,
such fields are equivalent to fields on
the closed manifold $S^2$.)

The second term in the action,
$\Gamma [K]$, is the so-called Wess-Zumino term.
It is defined in terms of integration over a three-dimensional space
${\cal M}^3$ which has
$\Sigma$ as its boundary.
The integrand  does not require
metrical factors for the integration since it is a differential
three-form.
However, it requires an extension of the field $K$ to the three-space
${\cal M}^3$.
There can be many spaces ${\cal M}^3$ with the same boundary 
$\Sigma$,
or equivalently, there can be many different ways to 
extend the fields to
the three-space ${\cal M}^3$. The physical results of the theory
are independent of how this extension is chosen, if we consider actions
of the form $k \, S_{WZW}$ where $k$ is an integer. By direct calculation, we can verify 
the Polyakov-Wiegmann identity
\beq
\S_{WZW} [ K \, h] = \S_{WZW}[K] + \S_{WZW}[h] - {1\over \pi} \int_\Sigma
\Tr ( K^{-1} \del_\bz K \, \del_z h \, h^{-1} )
\label{cs20b}
\eeq
where we have used local complex coordinates.
Now, in two dimensions, we can parametrize a nonabelian gauge potential
as
\beq
A_z = - \del_z M \, M^{-1}, \hskip .3in
A_\bz = M^{\dagger -1} \del_\bz M^\dagger
\label{cs20c}
\eeq
where $M$ is a complex matrix which may be taken to be in
$SL(N, \mathbb{C})$ for gauge fields corresponding to the
gauge group $SU(N)$. The identity (\ref{cs20b}) shows that
\beq
\delta \S_{WZW} = \S_{WZW}[ M^\dagger \, (1+ \theta )]
- \S_{WZW}[M^\dagger ] =
{1\over \pi} \int \Tr ( \del_z A_\bz \, \theta )
\label{cs20d}
\eeq
With $D_\bz$ denoting the covariant derivative with respect to
$A_\bz$, we have the identity
\beq
\del_z A_\bz = D_\bz ( M^{\dagger -1} \del_z M^\dagger )
\label{cs20e}
\eeq
Notice that, since $\delta M^\dagger = M^ \dagger \, \theta$, we may write
$\theta = M^{\dagger -1} \delta M^\dagger$; further, from (\ref{cs20c}), $\delta A_\bz = D_\bz ( M^{\dagger -1} \delta M^\dagger ) = D_\bz \theta$. Combining these relations with
(\ref{cs20e}), we can simplify (\ref{cs20d}) as
\beq
D_\bz {\delta \S_{WZW}\over \delta A^a_\bz } = {1\over 2 \pi } \del_z {A^a_\bz}
\label{cs20f}
\eeq
where we have also evaluated the trace in terms of the components.
Comparing this with (\ref{cs15}), we see that we can solve it as
\beq
\psi ( A_\bz ) = {\cal N} \, \exp \left( k \, \S_{WZW} [ M^\dagger ]\right)
\label{cs20g}
\eeq
The normalization factor ${\cal N}$ is to be fixed by using the inner product
(\ref{cs14}).
There is only one state for this theory. On $S^2$, there are no degrees of freedom left
for the Chern-Simons theory after one reduces to the physical configuration space.
Thus there is only the vacuum state of the theory. What we have found is the
expression for the ground state wave function in terms of the variables on
$\mathfrak{F}$.
If we consider higher genus Riemann surfaces, or two-manifolds with a boundary, 
then the Chern-Simons theory will have nontrivial degrees of freedom.

\subsection{Abelian theory on the torus}

We will now consider an Abelian Chern-Simons theory, with
$G=U(1)$ and with $\Sigma$ being a torus
$S^{1}\times S^{1}$. This will illustrate some of the topological features we mentioned.
The torus can be described by $z =\xi_1+\tau \xi_2$,
where $\xi_1$, $\xi_2$ are real and have periodicity of
$\xi_i \rightarrow \xi_i +$ integer, and $\tau$, which is a complex number,
is the modular parameter of the torus. The metric on the torus is
$ds^2 = \vert d\xi_1 +\tau d\xi_2\vert^2$. The two
basic noncontractible cycles of the torus are usually labelled as the
$\alpha$ and $\beta$ cycles. Further the torus has a holomorphic one-form
$\omega$ with
\beq
\int_\alpha ~\omega =1,\hskip .3in \int_\beta ~\omega =\tau
\label{cs21}
\eeq
Since $\omega$ is a zero mode of $\del_\bz$, we can parametrize $A_\bz$ as
\beq
A_\bz = \del_\bz \chi ~+~ i\,  {\pi\, {\bar \omega}\over {\rm Im}\tau}~a
\label{cs22}
\eeq
where $\chi$ is a complex function and 
$a$ is a complex number corresponding to the value of $A_\bz$ along
the zero mode of $\del_z$. 

For this space
$\Pi_{0}(\mathfrak{G}_*)= \mathbb{Z} \times \mathbb{Z}$, because the gauge
transformations
$g_{m,n}$ can have nontrivial winding numbers $m,n$ around the two cycles.
Consider one connected component of $\mathfrak{G}_*$, say $\mathfrak{G}_{m,n}$. 
A homotopically nontrivial
$U(1)$ transformation can be written as $g_{m,n}=
e^{i\alpha }~e^{i\theta_{m,n}}$, where $\alpha (z,\bz )$ is a
homotopically trivial gauge transformation and
\beq
\theta_{m,n}~=~ {i\,\pi\over{{\rm Im} \tau}} \left[ m\int^{z} {\bar
\omega}-\omega ~+~ n\int^{z} \tau\, {\bar\omega} -{\bar\tau}\,\omega
\right], \hskip .3in m, n \in \mathbb{Z}
\label{cs23}
\eeq
With the parametrization of $A_\bz$ as in (\ref{cs22}), the effect of
this gauge transformation can be represented as
\beq
\chi \rightarrow \chi ~+~ \alpha , \hskip .3in
a \rightarrow a ~+~ m ~+~n \tau \label{cs24}
\eeq
The real part of $\chi$ can be set to zero by an appropriate choice of
$\alpha$. (The imaginary part also vanishes when we impose the condition
$F_{z\bz}=0$.) The physical subspace (which has only the zero modes left after reduction)
 is given by the
values of $a$ modulo the transformation (\ref{cs24}), or in other words,
\beq
{\rm Physical ~space~for ~zero ~modes} \equiv \mathfrak{C}
=
{\mathbb{C}
\over \mathbb{Z} + \tau \mathbb{Z}}\label{cs25}
\eeq
This space is known as the Jacobian variety of the torus. It is
also a torus and therefore we see that the phase space $\mathfrak{C}$ has
nontrivial $\Pi_1$ and $\H^2$.
In particular, $\Pi_1 (\mathfrak{C}) = \mathbb{Z} \times \mathbb{Z} $ and this leads
to two angular parameters $\varphi_\alpha$ and $\varphi_\beta$ which are
the phases the wave functions acquire under the gauge transformation
$g_{1,1}$. 
The symplectic two-form  can be written as
\beqar
\Omega &=& {k\over 4\pi} \int \bdel \delta \chi \wedge \del \delta {\bar
\chi} ~+~ {k\pi \over 4} {d{\bar a}\wedge da\over {\rm Im}\tau}
\int_\Sigma {{\bar\omega}\wedge\omega\over {\rm Im}\tau}\nonumber\\
&=& \Omega_\chi ~-~ i{k\pi \over 2} {d{\bar a}\wedge da\over {\rm
Im}\tau}\label{cs26}
\eeqar
Integrating the zero mode part over the physical space of zero modes
$\mathfrak{C}$, we get
\beq
\int_\mathfrak{C} \Omega = k\, \pi \label{cs27}
\eeq
showing that $k$ must be quantized as an even integer for $U(1)$ fields on
the torus due to (\ref{anb10}).
\footnote{Since there has been some argument over this point in the literature, a comment might be in order.
In geometric quantization, we are considering the wave functions as sections of  a line bundle.
This means that each quantum state has a wave function which is a complex number.
One can avoid the quantization condition on $k$ for the Abelian theory if one is willing to
go beyond this and allow for multicomponent wave functions (for each state). However, the interpretation of such wave functions is not very clear.}

The modular parameter of the torus is subject to the so-called
modular transformations which are homotopically nontrivial
diffeomorphisms of the torus.
The vacuum angles change under 
such transformations and can eventually
be set to zero. To continue with the quantization,
we focus
on the zero modes for which the
symplectic potential can be written as
\beq
\A = -{\pi \,k\over 4}\, {({\bar a} -a) (\tau\, d{\bar a} - {\bar \tau}\, da )\over
{\rm Im}\tau }\label{cs28}
\eeq
The polarization condition then becomes
\beq
\left[ {\del \over \del {\bar a}} +i{\pi k \over 4} {({\bar a} -a) \tau \over ({\rm
Im}\tau )^2} \right] ~\Psi  =0
\label{cs29}
\eeq
with the solution
\beq
\Psi = \exp \left[ -i {\pi k\over 8} {({\bar a}-a)^2 \tau \over ({\rm Im}\tau )^2}
\right] ~~f(a) \label{cs30}
\eeq
where $f(a)$ is holomorphic in $a$. Under the gauge transformation (\ref{cs24})
we find
\beq
\Psi (a+m +n\tau ) = \exp\left[ -i{\pi k ({\bar a} -a) \over 8 ({\rm Im}\tau )^2}
~- {\pi k n ({\bar a}-a)\tau \over {\rm Im}\tau} +i \pi k \tau n^2 \right]~
f(a+m+n\tau )
\label{cs31}
\eeq
Under this gauge transformation $\A$ changes by $d \Lambda_{m,n}$ where
\beq
\Lambda_{m,n}= i{\pi\, k\, n\, (\tau {\bar a} - {\bar \tau }a ) \over 2~{\rm Im}\tau}
\label{cs32}
\eeq
The change in $\Psi$ should thus be given by $\exp (i\Lambda_{m,n}) \Psi$; requiring
the transformation (\ref{cs31}) to be equal to this, we get
\beq
f(a+m+n\tau )= \exp\left[ -i {\pi\, k\, n^2\, \tau \over 2} -{2\pi i\, k \,n \,a\over 2}
\right] ~f(a) \label{cs33}
\eeq
This transformation rule shows that $f(a)$ is a Jacobi $\Theta$-function. On these functions $f(a)$, ${\bar a}$ is
realized as
$(2~{\rm Im}\tau / k \pi )(\del / \del a ) +a$. The inner product for the wave
functions of the zero modes is
\beq
\la f \vert g \ra = \int \exp\left[ - {\pi k \,{\bar a} a\over 2~{\rm Im}\tau }
+{\pi k\, {\bar a}^2 \over 4~{\rm Im}\tau} +{\pi k \,a^2 \over 4~{\rm Im}\tau}
\right] ~{\bar f} g \label{cs34}
\eeq
It is then convenient to absorb the holomorphic part of the
exponent into the wave function defining 
the new set of holomorphic wave functions
\beq
\Phi \equiv \exp \left[ {\pi k a^2 \over 4~{\rm Im}\tau}\right] ~f (a)
= \exp \left[ {\pi k a^2 \over 4~{\rm Im}\tau}\right] ~\Theta (a) \label{cs35}
\eeq
On these functions, ${\bar a}$ acts as
\beq
{\bar a} = {2~{\rm Im}\tau \over \pi  k }~{ \del \over \del a }
\label{cs36}
\eeq
The key point we wanted to illustrate here is the use of the homotopically nontrivial gauge transformations. 

\section{$\theta$-vacua in a nonabelian gauge
theory}

Consider a nonabelian gauge theory in four spacetime dimensions,
the gauge group is some compact Lie group $G$. We can choose the gauge
where $A_0=0$ so that there are only the three spatial components of the
gauge potential, namely, $A_i$, considered as an antihermitian
Lie algebra valued vector field.. The choice
$A_0=0$ does not completely fix the gauge, one can still do gauge
transformations which are independent of time. These are given by
\beq
A_i \rightarrow A_i' = g A_i g^{-1} -\del_i g ~g^{-1}
\label{th1}
\eeq
The Yang-Mills action gives the symplectic two-form as
\beq
\Omega = \int d^3x~ \delta E^a_i ~\delta A^a_i
= -2\int d^3x~ \Tr \left( \delta E_i ~\delta A_i\right)
\label{th2}
\eeq
where $E^a_i$ is the electric field $\del_0 A_i^a$, along the Lie algebra
direction labelled by $a$. The gauge transformation of $E_i$ is 
$E_i \rightarrow g E_i g^{-1}$. By combining this with the transformation
(\ref{th1}), we identify the vector field generating
infinitesimal gauge transformations, with $g \approx 1+ \vf$, as
\beq
\xi = -\int d^3x~\left[(D_i\vf)^a {\delta \over \delta A_i^a}
+[E_i,\vf ]^a {\delta \over \delta E_i^a}\right]\label{th3}
\eeq
This leads to
\beq
i_\xi \Omega = -\delta \int d^3x~ \left[ -(D_i\vf )^a
E_i^a\right]\label{th4}
\eeq
The generator of time-independent gauge transformations is thus
\beq
G (\vf )= -\int d^3x~ (D_i\vf )^a E_i^a \label{th5}
\eeq
For transformations which go to the identity at spatial infinity,
$G^a = (D_iE_i)^a$. This is Gauss law, one of the Yang-Mills equations
of motion. This is to be viewed as a condition on the allowed
initial data and enforces a reduction of the phase space to gauge
invariant variables.
We again define the space of fields and gauge transformations as
\begin{eqnarray}
\mathfrak{F}&=& \Bigl\{ {\rm space~of~gauge~potentials~}A_i
\Bigr\}\label{th6}\\
\mathfrak{G}_*&=& \left\{ \begin{matrix}
{\rm space~of ~gauge~transformations~} g(\vx ): \mathbb{R}^3\rightarrow
G\\
{\rm such~that}~g \rightarrow 1~{\rm as}~\vert\vx\vert \rightarrow \infty\\
\end{matrix}\right\}
\label{th7}
\end{eqnarray}
The transformations $g(\vx )$ which go to a constant
element $g_\infty \neq 1$ act as a Noether symmetry. The states fall into unitary
irreducible representations of such transformations, which are isomorphic to
the gauge group $G$, upto $\mathfrak{G}_*$-transformations. The true gauge freedom
is only $\mathfrak{G}_*$. The physical
configuration space of the theory is thus
$\mathfrak{C} = \mathfrak{F}/ \mathfrak{G}_*$.\footnote{For further elaboration on this question, see \cite{nair} and references therein.}

With the boundary condition on the $g$'s, the
gauge functions are equivalent to a map from $S^3$ to $G$, and hence
there are homotopically distinct transformations corresponding to the
fact that $\Pi_3 (G) = \mathbb{Z}$. (In other words,
$\Pi_0 ( \mathfrak{G}_*) = \mathbb{Z}$.) These can be labelled by the winding
number $Q[g]$ given in (\ref{cs20a}).
We can write $\mathfrak{G}_*$ as the sum of different components, each of which is
connected and is characterized by the winding number $Q$. i.e.,
\begin{equation}
\mathfrak{G}_* = \sum_{Q=-\infty}^{+\infty} \!\!\oplus\, \,\mathfrak{G}_{Q}\label{th8}
\end{equation}
where 
each $\mathfrak{G}_{Q}$ consists of all maps with winding number $Q$.
$\mathfrak{G}_{Q}$ and $\mathfrak{G}_{Q'}$ are disconnected from each other
for $Q\neq Q'$, since if they are connected, $g_Q\in \mathfrak{G}_{Q}$ and
$g_{Q'}\in \mathfrak{G}_{Q'}$ should be deformable to each other and this is
impossible since $Q\neq Q'$.
One can easily check that $Q[g\,g']= Q[g]+Q[g']$
and hence this structure is isomorphic to the additive group of
integers $\mathbb{Z} $.
The space of gauge potentials $\mathfrak{F}$ is an affine space and is
topologically trivial. Combining these facts, we see that
the configuration space
has noncontractible loops, with $\Pi_1 (\mathfrak{C}) =\Pi_3(G) = \mathbb{Z}$.

An example of a noncontractible loop in $\mathfrak{C}$ is as follows.
Let $g_1(x)$ be a gauge transformation with winding number $1$ and
consider the line in
$\A$ given by 
\begin{equation}
A_i(x, \tau )= A_i(x) (1-\tau )+A_i^{g_1}\tau \label{th9}
\end{equation} 
for $0\leq \tau \leq 1$ or more generally
\beq
A_i(x, \tau )\hskip .1in 
{\rm with}\hskip .1in A_i(x,0)= A_i(x), \hskip .1in
A_i(x,1)= A_i^{g_1} (x)
\label{th10}
\eeq
where
$A_i^{g_1}$ is the gauge transform of $A_i$ by $g_1(x)$.
This is an open path in $\mathfrak{F}$. But since $A_i^{g_1}$ is the gauge transform
of $A_i$, both configurations $A_i$ and $A_i^{g_1}$ represent the same point
in $\mathfrak{C} = \mathfrak{F}/ \mathfrak{G}_*$. Thus $A_i(x, \tau )$ describes a closed loop in
$\mathfrak{C}$. If this loop is
contractible, we can deform the trajectory to a curve purely along the gauge flow
directions which connects $g=1$ to $g_1(x)$. This would imply that
$g_1(x)$ is smoothly deformable to the identity. But this is impossible
from our discussion of the structure of $\mathfrak{G}_*$. 
In turn this implies that $A_i(x,\tau)$ of (\ref{th10}) is a noncontractible
loop.
By considering other values of the winding number, we can easily establish that
$\Pi_1 (\mathfrak{C})= \mathbb{Z} $. Our general discussion shows that there must be an
angle $\theta$ which appears in the quantum theory.
We can see how this emerges by writing the symplectic potential.

We will first construct a flat potential on the space of fields.
For this we start with the instanton number which is
given, for a four-dimensional potential, by
\beqar
\nu [A]&=& -{1\over 8\pi^2} \int d^4x~ \Tr \left( F_{\mu\nu} F_{\alpha\beta}\right)
\epsilon^{\mu\nu\alpha\beta}\nonumber\\
&=& {1\over 16\pi^2} \int d^4x~ E^a_i F^a_{jk} \epsilon^{ijk}\label{th11}
\eeqar
The density in the above integral is a total derivative in terms of the potential
$A$, but it cannot be written as a total derivative in terms of gauge-invariant
quantities. $\nu [A]$ is an integer for any field configuration which is
nonsingular up to gauge transformations. It is possible to construct configurations
which are have nonzero value of $\nu$ which are nonsingular; 
these are instantons in a general sense.
\footnote{There is a more specific sense in which the word
instanton is used; it applies to self-dual solutions of the Yang-Mills equations
which further have $\nu [ A] \neq 0$.}
An example of a $\nu =1$ configuration, for $G=SU(2)$, is
\beq
A_\mu (x) = {x^2 \over x^2 + \alpha^2} \, \omega^{-1} \del_\mu \omega,
\hskip .3in
\omega ={ x_4 + i\, {\vec\sigma}\cdot \vx \over \sqrt{x^2}}
\label{th11a}
\eeq
For our purpose, we can transform this to the gauge with $A_0 = A_4 = 0$ obtaining
\begin{eqnarray}
A_i &=& U \left( {i x^2\over x^2+\alpha^2}\omega^{-1}\partial_i \omega \right) U^{-1}
~-~ \partial_i U ~U^{-1}, \hskip .62in
U= \exp \left( i{\vec \sigma}\cdot {\hat x}\, \rho\right) \nonumber\\
\rho &=& {\vert\vx \vert \over \sqrt{\vert\vx\vert^2+\alpha^2}}\left[ \arctan\left(
{x_4\over \sqrt{\vert\vx\vert^2+\alpha^2}}\right) -{\pi\over 2}\right],
\hskip .2in {\partial \rho \over \partial x_4} = {\vert\vx\vert \over x^2+\alpha^2}\label{th15}
\end{eqnarray}
Here $\sigma_i$ are the Pauli matrices and the path is parametrized by $x_4$,
$-\infty \leq x_4 \leq \infty$. $x^2 =\vx^2 +x_4^2$.
Since $x_4$ parametrizes the path, we see that $\nu [A]$ can be written as
\beqar
\nu [A] &=& \oint K[A]\nonumber\\
K[A] &=& \int d^3x~F^a_{jk} \delta A^a_i \epsilon^{ijk}\label{th16}
\eeqar
The integral of the one-form $K$ around a closed curve is the instanton number $\nu$
and is nonzero, in particular, for the loop corresponding to (\ref{th15}).
We can also see that the one-form $K[A]$ on $\mathfrak{C}$ is closed 
in the following way.
\beqar
\delta K[A] &=& -2\int d^3x~\, \delta \Bigl( \Tr \left(F_{jk}\delta
A_i\right) \Bigr) \epsilon^{ijk}
= -4 \int d^3x~\Tr \left( (D_j \delta A_k ) \,\delta
A_i\right) \epsilon^{ijk}\nonumber\\
&=&-4 \int d^3x~\Tr \left( \del_j \delta A_k ~\delta A_i +[A_j,\delta A_k] \delta A_i
\right) \epsilon^{ijk} \nonumber\\
&=&0\label{th17}
\eeqar
In the last step we have used the antisymmetry of the expression under permutation of
$\delta$'s, cyclicity of the trace and have done a partial integration.
We see from the above discussion that $K[A]$ is a closed one-form which is not exact
since its integral around the closed curves can be nonzero. 

With this flat potential
on $\mathfrak{C}$, we can construct a general solution
for the symplectic potential
corresponding to the $\Omega$ in(\ref{th2}) as
\beq
\A = \int d^3x~ E^a_i \delta A^a_i +\theta ~K[A]
\label{th18}
\eeq
Use of this potential will lead to a quantum theory where we need the parameter
$\theta$, in addition to other parameters such as the coupling constant, to
characterize the theory. The potential $\A$ in (\ref{th18}) is obtained from an
action
\beq
\S = -{1\over 4} \int d^4x~ F^a_{\mu\nu} F^{a\mu\nu} ~+~ \theta ~\nu [A]
\label{th19}
\eeq
This shows that
the effect of using (\ref{th18}) can be reproduced in the functional integral approach by
using the action (\ref{th19}). Since it is $\exp (i\S)$ which is imporatnt, we see
that $\theta$ is an angle with values $0\leq \theta < 2\pi$.
Alternatively, we can see that one can formally eliminate the $\theta$-term
in $\A$ by making a transformation $\Psi \rightarrow \exp(i \theta\Lambda ) \Psi$
where
\beq
\Lambda = -{1\over 8\pi^2} \int \Tr \left(A\wedge dA +{2\over 3} A\wedge A\wedge
A\right) \label{th20}
\eeq
Notice that $2\pi \Lambda$ is the Chern-Simons action (\ref{cs0a}) for $k=1$.
$\Lambda$ is not invariant under homotopically nontrivial transformations. 
The wave
functions get a phase equal to $e^{i\theta Q}$ under the winding number
$Q$-transformation, showing that $\theta$ can be restricted to the interval
indicated above. This is in agreement with our discussion after equation (\ref{anb4}).

\section{Fractional statistics in quantum Hall effect}

Excitations in the fractional quantum Hall effect will provide another example of how
the nontrivial connectivity of the configuration space can affect the physics.
We will discuss this briefly now.
The simplest and best description we have so far for
the fractional quantum Hall effect is in terms of the many-electron wave functions
\cite{fqhe}.
For example, for the states with filling fractions $\nu = 1/(2 p +1)$ where $p$ is an integer,
the $N$-electron wave function is usually taken as the Laughlin function
\beq
\Psi_{Laughlin} = {\cal N} \exp \left( - \half \sum_{i=1}^{N} \bz_i z_i \right)
\prod_{1 \le i < j \le N} ( z_i - z_j )^{2p+1}
\label{fqhe1}
\eeq
where $z = x_1 +i x_2$. and the subscript refers to the particle.
This leads to an electric current of the form
\beq
\la J_i \ra= - \nu \frac{e^2}{2 \pi} \epsilon_{ij} E_j, \hskip .3in
\nu = {1\over 2 p +1}
\label{fqhe2}
\eeq  
This corresponds to the observed Hall conductivity, quantized as the reciprocals of
odd integers. Among the excited states of the system are hole-like excitations
with a wave function of the form
\beqar
\Psi_{hole} &=& \prod_{i=1}^{N} (z_i - w ) \Psi_{Laughlin}\nonumber\\
&=& \prod_{i=1}^{N} ( z_i - w )  ~{\cal N} \exp \left( - \half \sum_{i=1}^{N} \bz_i z_i \right)
 \prod_{1 \le i < j \le N} ( z_i - z_j )^{2p+1}
\label{fqhe3}
\eeqar
where $w$ is the position of the hole.
We want to briefly consider the statistics of holes in fractional quantum Hall effect.
We can do this in an effective description with an action of the form
\beq
	S = \int d^3 x \left[ \frac{k}{4\pi} \epsilon^{\mu \nu \alpha} a_\mu \del_\nu a_\alpha
	+ a_\mu \left( j^\mu - \frac{e}{2 \pi}\epsilon^{\mu \nu \alpha} \del_{\nu} A_\alpha   \right) \right]
	\label{fqhe4}
\eeq
where $a^\mu$ ($\mu = 0,1,2$) is a new auxiliary field and $j^\mu$ denotes the hole current.
The value of the constant $k$ will be specified shortly.
$A_\mu$ is the electromagnetic vector potential.
(We are using a three-dimensional covariant notation now. $B^0 =
\epsilon^{0ij} \del_i A_j$ is the magnetic field
along the $x_3$-axis.)
The variation of the action with  respect to $A_\alpha$ identifies the electromagnetic current
as
\beq
J^\alpha = - \frac{e}{2 \pi} \epsilon^{\alpha \mu \nu} \del_\mu a_\nu
\label{fqhe5}
\eeq
The equation of motion for the auxiliary field $a_\mu$ is
\beq
\frac{k}{2\pi} \epsilon^{\mu\nu\alpha} \del_\nu a_\alpha + j^\mu - \frac{e}{2 \pi} 
\epsilon^{\mu\nu\alpha} \del_\nu A_\alpha = 0
\label{fqhe6}
\eeq
From (\ref{fqhe5}) and this equation, we see that
\beq
J^\mu 
= \frac{e}{k}\, j^\mu - \frac{e^2}{2 \pi k} \epsilon^{\mu\nu\alpha} \,\del_\nu A_\alpha \, .
\label{fqhe7}
\eeq
Choosing $k = 2 p +1$ we see that we can reproduce the Hall conductivity
correctly in the absence of holes.
The first term then shows that the charge per hole is $e/k$.

For a pair of well-separated holes we can take
\beq
j^\mu = {\dot w_1}^\mu\,\delta^{(2)}(x- w_1 ) + {\dot w_2}^\mu \,\delta^{(2)}(x- w_2)
\label{fqhe8}
\eeq
Leaving the electromagnetic field and focusing on the holes, the action becomes
\beq
S_{hole} = \frac{k}{4\pi}  \int d^3 x~ \epsilon^{\mu\nu\alpha} a_\mu \del_\nu a_\alpha
+ \int dt\, \left( a_\mu (w_1) \dot{w_1}^\mu + a_\mu (w_2) \dot{w_2}^\mu 
+ \frac{m \dot{w_1}^2}{2} +  \frac{m \dot{w_2}^2}{2} \right) 
\label{fqhe9}
\eeq
where we have also added a regular kinetic energy term for the holes.
(The specific form of this will not be important for our purpose.)
The time-component of the equation of motion for $a_\mu$, namely (\ref{fqhe6}), can be 
simplified as
\beq
\del_z \a_\bz - \del_\bz a_z = -i \, {\pi \over k} \left( \delta^{(2)}(x-w_1) 
+ \delta^{(2)}(x-w_2)  \right)
\label{fqhe10}
\eeq
Using
\beq
\del_z\, {1\over \bz - {\bar w}} = \del_\bz \, {1\over z- w} = \pi \, \delta^{(2)}(x-w)
\label{fqhe11}
\eeq
the solution to (\ref{fqhe10}) can be worked out as
\beq
a_\bz = - \frac{i}{2k} \left( \frac{1}{\bz - \bw_1} + \frac{1}{\bz - \bw_2} \right) , \hskip .2in
a_z = \frac{i}{2k} \left( \frac{1}{z - w_1} + \frac{1}{z - w_2} \right) 
\label{fqhe12}
\eeq
The coincident point $w_1 = w_2$ has to be excluded for consistency.
We will also use the $a_0 = 0 $ gauge so that
the action (\ref{fqhe9}) for the holes simplifies to
\beq
S = \int dt \left[  {m\over 2} (\dot{\bw}_1 \dot{w}_1 +  \dot{\bw}_2 \dot{w}_2)
+ a_{w_1} \dot{w}_1 + a_{\bw_1} \dot{\bw}_1
+ a_{w_2} \dot{w}_2 + a_{\bw_2} \dot{\bw}_2\right]
\label{fqhe13}
\eeq
where we have removed the singularities at the poles; in (\ref{fqhe13}),
\begin{align}
a_{w_1} = {i \over 2 k} \,{1\over w_1 - w_2}, & \hskip .1in
a_{{\bar w}_1} = -{i \over 2 k}\, {1 \over \bw_1 - \bw_2}\nonumber\\
a_{w_2} = {i \over 2 k} \,{1\over w_2 - w_1}, & \hskip .1in
a_{{\bar w}_2} = -{i \over 2 k}\, {1 \over \bw_2 - \bw_1}
\label{fqhe14}
\end{align}
The two equations (\ref{fqhe13}) and (\ref{fqhe14}) suffice for our consideration of
the statistics of holes.

Because the coincident point $w_1 = w_2$ has been excluded, the closed path of one hole going around the other is not smoothly deformable to zero. Thus $\Pi_1$ of the configuration space
is nonzero, equal to $\mathbb{Z}$. In fact, with $w_2$ fixed,
\beq
a_{w_1} dw_1 + a_{\bw_1} d \bw_1 =
d \left[ {i \over 2 k}\, \log \left( {w_1 - w_2 \over \bw_1 - \bw_2 }\right)\right]
\label{fqhe15}
\eeq
This is evidently closed, but cannot be considered exact since
\beq
\oint_C a = - {2 \pi \over k} \neq 0
\label{fqhe16}
\eeq
where $C$ is a contour enclosing $w_2$. 

The Hamiltonian corresponding to the action (\ref{fqhe13}) is
\beq
H = {1\over 2} m \left( {\dot \bw_1} {\dot w_1} + {\dot \bw_2 } {\dot w_2} \right)
\label{fqhe17}
\eeq
From the action we also identify the operators
\begin{align}
&m \dot{w}_1 = -i \frac{\del}{\del \bw_1} - a_{\bw_1}  , \hskip .3in
m \dot{\bw}_1 = -i \frac{\del}{\del w_1} - a_{w_1}  \nonumber\\
&m \dot{w}_2 = -i \frac{\del}{\del \bw_2} - a_{\bw_2} , \hskip .3in
m \dot{\bw}_2 = -i \frac{\del}{\del w_2} - a_{w_2} 
\label{fqhe18}
\end{align}
Written as a differential operator, the Hamiltonian will involve the $a$'s. Because of this, it is convenient
to write the wave function as
\beq
\Psi (x_1, x_2) = \exp \left[ {1\over 2 k} \log \left( {\bw_1 - \bw_2 \over w_1 - w_2}\right)
\right] ~ \Phi (x_1, x_2)
\label{fqhe19}
\eeq
The action of $H$ on $\Phi$ is then the usual one,
\beq
H \, \Phi = - {1\over 2 m} \left( {\del \over \del w_1 } {\del \over \del \bw_1 }
+{\del \over \del w_2 } {\del \over \del \bw_2 } \right) \, \Phi
\label{fqhe20}
\eeq

We can now consider the exchange of the two holes as due to a rotation of the two points
by $\pi$ followed by a translation to bring them back to the same points.
We take $\Phi$ to be symmetric under exchange.
As for the phase factor in (\ref{fqhe19}), 
the translation does not change it.
The $\pi$-rotation leads to
\beq
\Psi (x_2, x_1)  = e^{- i \pi /k } \, \Psi (x_1, x_2)
\label{fqhe21}
\eeq
With $k = 2 p +1$, we see that the two holes do display fractional statistics.
The origin of this can be traced to
the closed but not exact one-form (\ref{fqhe15}).

In two spatial dimensions, it is also possible to have fractional values for the spin
for a particle \cite{anyon} . The usual argument for the
quantization for spin in three spatial dimensions
relies on the fact that the components of the
angular momentum operators do not commute among themselves
and because we need a unitary representation.
In two spatial dimensions, there is only one rotation and 
fractional values for spin are possible.
This is true even in a Lorentz-invariant theory, because of the
noncompact nature of the Lorentz group.
There is a spin-statistics theorem in two spatial dimensions as well.
In accordance with this, the result we have shown
implies that
the holes have fractional spins or that they are ``anyons" \cite{anyon}.

\section{Fluid dynamics}

We now turn to considerations about how an action for fluid dynamics can be constructed
using the results on quantization of $G/H$ spaces and how anomalies can be incorporated into fluid
dynamics \cite{review}. We start with the well known formulation of classical nonrelativistic fluid dynamics.

\subsection{The Lagrange formulation}

The so-called Lagrange formulation of fluid dynamics, developed more than two centuries ago by Euler and Lagrange, is an elegant method of obtaining the equations
of fluid dynamics starting from Newton's equations for point-particles.
Here one considers a collection of, say, $N$ particles obeying the equations of motion
\beq
 {d \over dt} {\dot X^i_\lambda} = - {\del V \over \del X_{i \lambda}}
\label{fluid1}
\eeq
where $X^i_\lambda$ denote the position of the $\lambda$-th particle, $\lambda = 1, 2, \cdots, N$.
For simplicity, we have taken all particles to have the same mass $m$, with units adjusted so that
$ m = 1$.
We can label the particles by their positions at time $t=0$, assuming that there is no overlap of particles. 
In the limit of a large number of particles, this means that we may take
$\lambda$ to be continuous; it is a three-vector corresponding to the initial position vector.
Let $\rho_0 (\lambda )$ be the number density of particles.
Then we sum equation (\ref{fluid1}) over a small range of $\lambda$ and go to the continuous
$\lambda$-limit to obtain
\beq
 \rho_0 (\lambda )\, d^3\lambda~ {d \over dt} {\dot X^i(t, \lambda)}  = - \rho_0 (\lambda ) 
 \, d^3\lambda ~ 
{\del V \over \del X_i (t,\lambda )}
\label{fluid2}
\eeq
Now, the particle at position ${\vec \lambda}$ at $t=0$ moves to $X^i (t, \lambda )$ at time
$t$. This is a continuous transformation of the $\lambda^i$ into $X^i$, which is invertible
at least for small $t$. We can therefore solve for $\lambda^i$ as a function of $X^i$ and $t$
and write various quantities as functions of $t, X^i$.
Since the number of particles is conserved, we should have
$\rho_0 (\lambda )\, d^3\lambda = \rho (t,X)\, d^3X$. 
This shows that we can define the density of particles in terms of $X$ as
\beq
\rho (t, X) = {\rho_0 (\lambda ) \over \det (\del X /\del \lambda)}
\label{fluid4}
\eeq
The density $\rho$ so defined obeys an equation of continuity.
By direct differentiation with respect to time, we find
\beqar
{\del \rho \over \del t} + {\del \rho \over \del X^i} {\dot X^i} &=& \rho_0 (\lambda)
{d \over dt } {1 \over \det (\del X /\del \lambda)}
= - {\rho_0 (\lambda) \over \det (\del X / \del \lambda)} {d\over dt} \left(\log \det (\del X/\del \lambda)\right)\nonumber\\
&=& - \rho ~ {\del \lambda^i \over \del X^k} {\del {\dot X^k}\over \del \lambda^i}
= - \rho ~ \nabla_k {\dot X^k}
\label{fluid5}
\eeqar
We now define the velocity at a point $X$ as 
\beq
v^i (t, X) = {\dot X}^i(t, \lambda)\Bigr]_{\lambda = \lambda (t, X)}
\label{fluid6}
\eeq
Equation (\ref{fluid5}) then reduces to the continuity equation
\beq
{\del \rho \over \del t} + \nabla_k ( \rho v^k ) =0
\label{fluid7}
\eeq
The equation of motion (\ref{fluid2}) involves the time-derivative of the velocity ${\dot X}^i (t, \lambda )$. We can substitute for $\lambda$ in terms of $X$ in this equation.
Thus the quantity on the left hand side involves
\beq
\left. {d \over dt}{\dot X^i} (t,\lambda)\right]_{\lambda = \lambda (t,X)} =
\left. {d \over dt}{\dot X^i} (t,\lambda (t,X))\right]_{\lambda ~fixed}
\label{fluid8}
\eeq
In other words, we should substitute $\lambda = \lambda (t, X)$ after the second time-derivative has been evaluated.
Since $\lambda^i$ do not depend on time, being initial data, we get the identity
\beq
0 = {\del \lambda^i  (t,X) \over \del t} + {\del \lambda^i (t,X) \over \del X^k } {\dot X}^k
\label{fluid10}
\eeq
Simplifying (\ref{fluid8}) using (\ref{fluid10}), the equation of motion (\ref{fluid2}) becomes
\beq
\rho \left[ {\del v^i \over \del t} +  v^k \nabla_k v^i \right] 
= - {\rho } ~{\del V \over \del X_i (t,\lambda)}
\label{fluid12}
\eeq
This equation, along with the continuity equation (\ref{fluid7}), defines perfect fluid dynamics.
The right hand side of (\ref{fluid12}) can be expressed in terms of the gradient
of the pressure, but we will not need that for now.

In modern physics, a point-particle is defined as a unitary irreducible representation of the Poincar\'e group.
In addition, we may want to consider particles with
 internal symmetries such as nonabelian
color charges, the latter being also described by an appropriate representation
of the symmetry group. So we may ask:
\begin{quotation}
\noindent
Can we do a Lagrange trick and describe fluid dynamics in terms of group theory, with each particle corresponding to a unitary irreducible representation
of the symmetry group (Poincar\'e $\otimes$ internal symmetry group)?
\end{quotation}
We can also equally well ask the counter question: Why would this be interesting beyond the
pure mathematical joy of showing that it can be done? There are some good reasons.
In such a formalism, symmetry would be really foundational and this would facilitate
the inclusion of nonabelian internal symmetries
and spin in magnetohydrodynamics and also incorporate anomalous symmetries as well.
These have all become issues of interest in recent research partly because of the deconfined
fluid phase of quarks and gluons.

We start with a simple case of a nonrelativistic particle which carries an internal symmetry, say,
$SU(2)$ to see how this can all work out.
(This internal symmetry could be ``color" or spin or something else depending on the
physical context.)
The action for such a particle coupled to an $SU(2)$ gauge field is given by
\begin{align}
\S &= \int dt \left[ {1\over 2 } m {\dot x}^2 - A^a_i Q^a {\dot x}_i 
- i\,{n \over 2}\,\Tr (\sigma_3 \, g^{-1} {\dot g} ) \right]\nonumber\\
&= \int dt \left[ {1\over 2 } m {\dot x}^2 
- i\,{n\over 2}\, \Tr (\sigma_3 \, g^{-1} {D_0\, g} ) \right]
\label{fluid13}
\end{align}
where $Q^a = {n \over 4} \Tr ( \sigma_3\, g^{-1} \sigma^a g )$
and $D_0 = \del_0 + A^a_i {\dot x}_i (-i \sigma^a /2)$.
$D_0$ is the covariant derivative of $g$ with respect to the $SU(2)$ gauge field
evaluated on the trajectory of the particle.
This action was proposed in the 1970s by Balachandran and collaborators \cite{bal}; the equations of motion
corresponding to this action were written down earlier, in 1971, by Wong \cite{wong}.
The last term in (\ref{fluid13}), apart from the gauge field term,
 is familiar to us as the action
(\ref{ana66}) for $G/H = SU(2)/U(1)$.
The quantization of the action is also familiar. The usual kinetic term
$\half m {\dot x}^2$ will lead to the usual point particle dynamics, with a minimal
coupling to the gauge field via the charge operator
$Q^a$. The degrees of freedom represented by $g$ will lead to a unitary representation of
$SU(2)$, with $j = n/2$. This part will describe the dynamics of the internal symmetry and how it influences and is influenced by the kinetic motion of the particle and the external field.

We can now see how to generalize to fluids. We will focus on the last term
in (\ref{fluid13}) as it is the key term for obtaining UIRs of the group after quantization.
 We consider a large number of particles, using a variable $\lambda$ to label them. As with the Lagrangian approach to fluids, we will eventually take
$\lambda$ to be continuous and to correspond to
a three-volume. For the last term in (\ref{fluid13}) we get
\beq
\S = - i\, {n \over 2}\int dt~ \Tr (\sigma_3 \, g^{-1} {\dot g} )
\longrightarrow \S = -{i \over 2}  \int dt \sum_\lambda  n_\lambda  \Tr (\sigma_3\, g_\lambda ^{-1} {\dot g}_\lambda )
\label{fluid14}
\eeq
We can take the continuum limit by
$\sum_\lambda  \rightarrow \int  J\,d^3x /v$, where $J$ is the Jacobian of the transformation
$\vec \lambda \rightarrow \vx$, $J = \vert \del \lambda /\del x\vert$ and $v$ indicates a small volume
over which the dynamics is coarse-grained. Defining a density
by $j^0 = n \, J / v$, we get
\beq
\S = -i \int d^4x~ j^0\, \Tr (t_3 g^{-1} \del_0 g ), \hskip .3in t_3 = {\sigma_3 \over 2}
\label{fluid15}
\eeq
where $g(t, \lambda ) = g (t, \vx )$ is to be considered as a spacetime-dependent
group element. The form of the action
(\ref{fluid15}) also suggests a natural relativistic generalization
\beq
\S = -i \int j^\mu~ \Tr (t_3\, g^{-1} \del_\mu g )
\label{fluid16}
\eeq
The remaining terms in the action can be added on at this stage, but before doing that, we pause to consider what happens with the Poincar\'e group. If we follow the same strategy we should consider
the analog of the term $\Tr (t_3 \,g^{-1} {\dot g})$ for the Poincar\'e group,
which has the translational parameters $x^\mu$ and the rotational and Lorentz boost parameters;
the latter set of parameters may be gathered into a Lorentz group element $\Lambda$.
The action is then given by
\beq
S = - \int d\tau~ p_\mu \, {\dot x}^\mu + i \,{n \over 4} \int d \tau~\Tr (\Sigma_3\, \Lambda^{-1} \, {\dot\Lambda} )
\hskip .3in \Sigma_3 = \left[ \begin{matrix} \sigma_3 &0 \\  0&\sigma_3\\ \end{matrix} \right]
\label{fluid17}
\eeq
where we have chosen to display the term involving the Lorentz group element
$\Lambda$ in terms of the usual spinor representation.\footnote{This is like using the
$2\times 2$-matrix  version of $g$  to display the action (\ref{fluid13}). It 
does not imply that there is anything special about this representation.}
This is almost what we want, but the first term in the action (\ref{fluid17}) is
a bit awkward.
In going over to a fluid description, the position variables $x^\mu$ are a bit awkward.
First of all, there should only be three independent $x$'s or corresponding velocities.
For the point-particle, this is naturally implemented by a mass-shell type constraint.
It is not clear how to do this for fluids.
Secondly, the role of diffeomorphisms versus translations is not clear
in this language.
So we will first deal with this problem before returning to the main line of development.

\subsection{Clebsch variables and the general form of action}

We return to the usual approach to fluids briefly.
It has been known for a long time that fluid dynamics can be described as a Poisson bracket
system. This means that the equations of motion are derived from a Hamiltonian
\beq
H= \int d^3x\left[ {1\over 2} \rho~ v^2 ~+~ V(\rho )\right]
\label{fluid18}
\eeq
by using the Poisson brackets
\beqar
[ \rho (x) , \rho (y) ] &=& 0\nonumber\\
{~}[ v_i (x) , \rho (y) ] &=& \del_{xi}\delta^{(3)} (x-y)\nonumber\\
{~}[ v_i (x), v_j (y) ] &=& - {\omega_{ij} \over \rho }~ \delta^{(3)}(x-y)
\label{fluid19}
\eeqar
The pressure is related to $V(\rho )$ as $p = \rho {\del V \over \del \rho} - V$.
The Poisson brackets can be summarized for arbitrary functions $F$, $G$ of the 
fluid variables as
\beq
[F, G] = \int \biggl[ {\delta F \over \delta \rho}  \del_i \left( {\delta G \over \delta v_i}\right)
- {\delta G \over \delta \rho}  \del_i \left( {\delta F \over \delta v_i}\right)
- { \omega_{ij}\over \rho} {\delta F \over\delta v_i} {\delta G \over \delta v_j}\biggr]
\label{fluid20}
\eeq
It is then easy to check that any local observable $F$ will Poisson commute with the helicity
which is defined as
\beq
C = {1\over 12 \pi^2} \int \epsilon^{ijk} ~v_i\, \del_j v_k 
\label{fluid21}
\eeq
where we take the velocity to vanish at the boundary of the spatial region of integration.
Denoting the variables $\rho, \, v_i$ collectively as $q^\mu$, and writing the
Poisson brackets as $\{ q^\mu , \, q^\nu \} = K^{\mu\nu}$, we can check from
(\ref{fluid20}) that $\delta C /\delta v_i$ is a zero mode for $K^{\mu\nu}$.
This means that $K^{\mu\nu}$ is not invertible. Comparing with
(\ref{ana9a}) we see that we have a problem.
If $K^{\mu\nu}$ has an inverse, that would be the symplectic structure $\Omega_{\mu\nu}$ and 
we can construct an action. But that is not possible because $K^{\mu\nu}$ has a zero mode.
This is a problem, but the way to a solution 
 is also clear. Since $C$ Poisson commutes with any local observable, it must be superselected. We must fix its value and then consider only those velocities which
keep the value unchanged. Such a parametrization is given by
the Clebsch variables which expresses the velocity as
\beq
v_i = \del_i \theta ~+~ \alpha ~\del_i \beta
\label{fluid22}
\eeq
where $\theta$, $\alpha$ and $\beta$ are 3 independent fields.
One can easily check that the integrand of $C$ is a total derivative with
this parametrization and gives zero upon integration.
(We can also accommodate other values of $C$, see below.)
A suitable action which gives the fluid equations is then
\beq
\S =\int d^4x~\left[ \rho ~{\dot \theta} + \rho \,\alpha ~{\dot \beta}\right] - \int d^4x~ \left[ {1 \over 2} \rho\, v^2 + V\right]
\label{fluid23}
\eeq
We can also write this as
\beq
\S = \int d^4x~\left[ j^\mu \, \left( \del_\mu \theta + \alpha \, \del_\mu \beta \right)\right]~
-~ \int d^4x~\left[ j^0~-~ {j^i j^i \over 2 \, \rho } + V \right]
\label{fluid24}
\eeq
where $j^0 = \rho$ and we introduce an auxiliary field ${\vec j}$. Elimination of
${\vec j}$ takes us back to (\ref{fluid23}).
This is easily generalized to the relativistic case as
\beq
S =  \int d^4x~ \bigl[ j^\mu \, \left( \del_\mu \theta + \alpha \, \del_\mu \beta \right)~
-~   F(n) \bigr]
\label{fluid25}
\eeq
where $F(n) =  n + V(n) $ and $n^2 = j^2 = (j^0)^2 - j^i j^i$. Notice that
$n = \sqrt{(j^0)^2 - {\vec j} \cdot {\vec j}} \approx j^0 - (j^i j^i / 2 j^0 ) + \cdots$,
so that (\ref{fluid24}) is recovered in the nonrelativistic case.

The Clebsch parametrization can also be written in a more group-theoretic form.
For this purpose, we can use either $SU(1,1)$ or $SU(2)$. We parametrize an element of the
group as\footnote{Whether we should choose $SU(1,1)$ or $SU(2)$ depends on the 
vorticity which is given as $d \alpha \, d\beta$. The group $SU(2)$ would describe situations with quantized vorticity, $SU(1,1)$ would give no quantization condition on vorticity.}
\beq
g =  {1\over \sqrt{1\mp {\bu} u}} \left( \begin{matrix} 
      1 & u \\
      {\pm\bu} & 1 \\
   \end{matrix}\right)~\left( \begin{matrix} e^{i\theta}&0\\ 0&e^{-i\theta}\\ \end{matrix}
   \right),
\label{fluid26}
\eeq
We can easily check that
\beq
-i~ \Tr \left({\sigma_3\over 2}\,  g^{-1} dg \right) = d \theta + \alpha ~d\beta, \hskip .2in   
   \alpha ={ {\bu}u \over (1\mp {\bu} u)}, \hskip .1in
    \beta = \mp\, {i\over 2} \log (u/ {\bu})\label{fluid26a}
\eeq
where the upper sign applies to $SU(1,1)$ and the lower to $SU(2)$.\footnote{By the way, we are also saying that ordinary fluid dynamics can display an $SU(1,1)$ or $SU(2)$ symmetry,
which is effectively replacing the diffeomorphism symmetry. This is a point worth further exploration.}
We can now write the usual ordinary fluid dynamics action as
\beq
\S =  \int d^4x~ \left[ -i  j^\mu \, \Tr (\sigma_3 \, g^{-1}\del_\mu g )
-~  F(n) \right]
\label{fluid27}
\eeq
We have thus brought the action, even for the usual fluid dynamics, to a form
consistent with the group-theoretic approach. We can now see how the Poincar\'e group can be accommodated.
For the translational part we use the Clebsch way of writing the action. For the rest of it, we can use the usual group-theoretic way which we have already discussed.
Thus our action for a general fluid dynamics is given by
\beqar
\S \!\!&=&\!\!\! \int d^4x~ \Bigl[ 
j^\mu\, (\del_\mu \theta + \alpha\, \del_\mu \beta ) 
- { i \over 4}\, j^\mu_{(s)} \, \Tr ( \Sigma_3 \,\Lambda^{-1} \del_\mu \Lambda )
+ i \sum_a\, j^\mu_{(a)} \Tr (q_a \, g^{-1} D_\mu \, g) \nonumber\\
&&\hskip .7in
- F( \{ n \} ) )\Bigr] ~+~ S(A) \label{fluid28}
\eeqar
We use $q_a$ to denote the diagonal generators of the internal symmetry group
$G$ with $g \in G$.
The currents $j^\mu$, $j^\mu_{(s)}$, $j^\mu_{(a)}$ correspond to the transport of
mass, spin and internal quantum numbers, respectively.
Generally, we must have different currents $j^\mu$, $j^\mu_{(s)}$, $j^\mu_{(a)}$ for
mass flow, spin flow and the transport of other quantum numbers, since they are independent.
For example, we may have a cluster of particles of zero total spin moving off in some direction, giving mass transport but no spin transport; we can have a similar situation with internal symmetry groups as well. Generally these currents are independent; any relations among them must be viewed as
``constitutive relations" characteristic of the physical system.
The coupling of the system to 
gauge fields follows from covariant derivatives on the group elements. The function
$F( \{ n \} )$ depends on all the invariants such as
$n = \sqrt {j^\mu\, j_\mu}$, ~
$n_a = \sqrt{j^\mu_{(a)}\, j_{\mu \, (a)}}$, etc., which we can make from the currents
and the gauge fields. We have explicitly indicated the action for the gauge fields.
The group-valued fields are related to flow velocities and currents and are given by
the equations of motion,
\beqar
{1\over n} {\del F \over \del n} \,\, j_\mu &=& \del_\mu \theta + \alpha\, \del_\mu \beta\nonumber\\
{1\over n_a} {\del F \over \del n_a}\,\, j_{\mu \, (a)}&=& i \, \Tr \, ( q_a \, g^{-1} D_\mu \, g ),
\hskip .2in {\rm etc.} \label{fluid29}
\eeqar

\subsection{Assorted comments}

Many new concepts (or at least concepts which may not be very familiar)
 have been introduced, so a few clarifying remarks
are in order at this point.

\noindent\underline{Helicity}

In terms of the group-valued variables, the
helicity is given by the topological invariant
\beq
C = {1\over 24 \pi^2} \, \int \Tr ( g^{-1} \, dg )^3
\label{fluid30}
\eeq
This shows how we may generalize the Clebsch parametrization to situations with
nonzero value of $C$. We choose a particular $g$, say $g_1$ which gives the
desired value $C$. Then we write use  $g_1 \, g$ in place of $g$ in
(\ref{fluid26a}) to get the parametrization for velocities. $g$ is taken to have zero $C$.
It is easy to check that $C[ g_1 \,g] = C[g_1] + C[g] = C[g_1]$.

\noindent\underline{The action and the density matrix}

The idea of using an action of the form
$\int j^0 \, \Tr (\sigma_3 g^{-1} {\dot g} )$ may be seen from another more general point of view as well.
The full quantum dynamics for a state with density matrix $\rho$ is given by the action
\beq
S = \int dt~ \Tr \left[ \rho_0 \, \left( U^\dagger i \, {\del U \over \del t}  - U^\dagger \, H \, U
\right) \right]
\label{fluid31}
\eeq
 The variational equation for this is
\beq
i {\del \rho \over \del t} = H \, \rho - \rho \, H, \hskip .3in \rho = U \, \rho_0 \, U^\dagger
\label{fluid32}
\eeq
which is the expected equation for the time-evolution of the density matrix.
The canonical one-form corresponding to this action is
\beq
{\cal A} = i \, \Tr ( \rho_0 \, U^\dagger \, \delta U )
\label{fluid33}
\eeq
where $\delta U$ includes all possible observables.
Consider a subset of transformations (such as symmetry transformations which can survive into the hydrodynamic regime). We write $\delta U$ as
\beq
\delta U = -i\, ( t_A U ) \,\delta \theta^A ~+~ {\rm other ~transformations}
\label{fluid34}
\eeq
where the other transformations correspond to other observables, beyond the subset we are interested in. Neglecting those, we find that $\A$ restricted to the variables of interest is
\beq
{\cal A} = \Tr (U \, \rho_0 \, U^\dagger \, t_A ) \, \delta \theta^A =
T_A \, \delta \theta^A, \hskip .3in T_A (\theta ) = \Tr\,( \rho\, t_A)  = \la t_A \ra
\label{fluid35}
\eeq
$T_A$ will have appropriate group composition/commutation properties.
$\theta$'s are essentially collective variables for the theory.
We can then ask the question: What is the action 
(at the level of the $\theta$'s) which gives this
${\cal A}$? This is evidently the co-adjoint orbit action of the form
we have been using.

\noindent\underline{Diffeomorphisms and Clebsch variables}

Finally, we can think of the Clebsch variables in another way as well.\footnote{
In principle, we can use the action (\ref{fluid17}) with the translational degrees of freedom
$x^\mu$ even in the fluid case.
If we keep ${\dot x}^\mu$ as fluid velocity, then we do get the correct fluid
equations, but with no pressure.}
We start by looking at the diffeomorphism algebra,
\beq
[ M(\xi ) , M(\xi')] = M (\xi \times \xi' ), \hskip .3in (\xi \times \xi' )^i
= \xi^k \del_k \xi^{'i} - \xi^{'k} \del_k \xi^{i}
\label{fluid36}
\eeq
where $M$ is the generator of spatial diffeomorphisms, given by
$T_{0i}$ where $T_{\mu\nu}$ is the energy-momentum tensor.
The algebra (\ref{fluid36}) can be realized by
\beq
J_i = \pi_1 \,\del_i \vf_1 + \pi_2\, \del_i \vf_2 + \cdots \label{fluid37}
\eeq
for any number of canonical pairs of variables $(\pi_i , \vf_i )$. 
 We need two such pairs for a complete characterization in 3 spatial dimensions.
Hence, we can see that diffeomorphism symmetry can be traded for
an $SU(1,1)$ or $SU(2)$ symmetry for the pairs $\pi_i, \vf_i$.
The redesignation of variables as $\pi_1 =\rho$, $\pi_2 = \rho\, \alpha$,
$\vf_1 = \theta$, $\vf_2 = \beta$ takes us back to the usual Clebsch form.

We can also view
$\pi_1, \vf_1$ as the modulus and phase
of  a complex field
$\psi, \, \psi^*$. But then how do we interpret the extra fields $\alpha$, $\beta$ ?
These are what we need to get vorticity.
We may observe that
for vorticity, we need to compare the velocities of nearby particles.
Thus in attributing some nonzero vorticity to each local coarse-graining unit,
we see that inside each such
unit (around, say, $\vx$), we must have distinct fields
representing these particles whose velocities are to be compared. This means that
$\psi (x)$ and $\psi( x+ \epsilon )$ must be counted as independent fields since we want to replace them by fields at a single point $\vx$ upon coarse-graining.
This gives some understanding of how the $SU(1,1)$ or $SU(2)$ group emerges.

\subsection{Examples}

\subsubsection{Nonabelian magnetohydrodynamics}

We will briefly mention a few examples before going on to the question of anomalies.
Our first example is about nonabelian magnetohydrodynamics, say with $SU(2)$ as the
internal symmetry \cite{bistro}.
Picking out the relevant terms in the general action
(\ref{fluid28}), we see that we can take the action for this case as
\beqar
\S &=& \int J_m^\mu \, \left( \del_\mu \theta + \alpha\, \del_\mu \beta
\right) -i \int j^\mu~ \Tr (\sigma_3\, g^{-1} D_\mu g )  - \int F(n)  ~+~ S_{YM}
\label{fluid38}\\
D_\mu g &=& \del_\mu g + A_\mu \,g \hskip .5in A_\mu = -i\, t^a\, A^a_\mu ,~~~~t^a = \half \sigma^a
\nonumber\\
J_m^\mu &=& n_m \, U^\mu , \hskip .9in U^2 = 1\nonumber\\
j^\mu &=& n ~u^\mu , \hskip 1.03in u^2 =1\nonumber
\eeqar
$J^\mu_m$ denotes the mass current, while
$j^\mu$ corresponds to the current for the diagonal generator of the internal symmetry.
 We have also defined the flow velocities
$U^\mu$ and $u^\mu$ in terms of the currents.
The current which couples to the gauge field may be obtained as
\beq
J_a^{\mu} =  - {\delta \S \over \delta A^a_\mu}=
 \Tr (\sigma_3 \,g^{-1} t_a g )  ~j^\mu = Q_a~ u^\mu ,
\hskip .3in  Q_a = n \,\Tr (\sigma_3 \,g^{-1} t_a g ) 
\label{fluid39}
\eeq
Notice that the current factorizes into a charge density $Q_a$ and a flow
velocity $u^\mu$. This is known as the Eckart factorization.
The equations of motion may be derived from the action (\ref{fluid38}) by varying
all the fields. We show some of the equations here:
\beqar
\del_\mu j^\mu &=&0\nonumber\\
(D_\mu J^\mu )_a &=& 0\nonumber\\
n\,u^\mu \del_\mu  (u_\nu F' ) - n\, \del_\nu F' &=&  \Tr ( J^\mu F_{\mu\nu})
\label{fluid40}
\eeqar
The first two are conservation laws, while the last one is the Euler equation for
the (nonabelian) charge transport.\footnote{There is another equation for mass transport which we are not displaying.
Here we are zeroing in on just the ``new" equations, namely, those beyond the usual ones.}
The first two equations in (\ref{fluid40}) also give
\beq
u^\mu  (D_\mu Q)_a =  (D_0 Q)_a + {\vec u} \cdot ({\vec D} Q)_a = 0 
\label{fludi41}
\eeq
This may be viewed as the fluid version of the Wong equations for the transport of
nonabelian charge by a point-particle. We also have 
$\del_\mu T^{\mu\nu} = \Tr\, ( J^\mu F_{\mu\nu})$ where the energy-momentum tensor
$T_{\mu\nu}$ has the perfect fluid form.

The group element $g$ may be given a nice physical interpretation.
The nonabelian charge density $\rho = \rho_a\, t_a$ (which is the time-component of
$J_a^{\mu}$) transforms, under gauge transformations, as
\beq
\rho \rightarrow \rho' = h^{-1} \rho ~h, \hskip .5in h \in SU(2)
\label{fluid42}
\eeq
Thus we can diagonalize $\rho$ at each point by an
$(\vx ,t)$-dependent transformation $g$.
Then we can write $\rho = g\, \rho_{diag} \,g^{-1}$, with
$\rho_{diag} = \rho_0 \sigma_3 $.
In other words,
\beq
\rho_a = \rho_0 ~\Tr ( g\,\sigma_3\, g^{-1}\, t_a) = j^0\, \Tr (g\, \sigma_3\, g^{-1}\, t_a)
\label{fluid43}
\eeq
The group element $g$ diagonalizes the charge density at each point.
The eigenvalues are gauge-invariant and are represented by $n$.
We may thus view $g$ as describing
 the degrees of freedom corresponding to the orientation of the local charge density
 in color space.
Under a gauge transformation, $g \rightarrow h^{-1} ~g$.

The Poisson brackets involving the charge densities are
\beqar
\{ j^0 (\vx ), j^0 (\vy )\} &=& 0\nonumber\\
\{ j^0 (\vx ) , ~g (\vy )\} &=& -i \, g (\vx ) ~\left({\sigma_3 \over 2}\right) ~\delta (x-y)\nonumber\\
\{ \rho_a (\vx ), \rho_b (\vy )\} &=& f_{abc}\, \rho_c (\vx ) \,\delta (x-y)\nonumber\\
\{ \rho_a( \vx ) , ~g (\vy )\} &=& - i \left( {\sigma_a \over 2}\right) \, g (\vx ) ~ \delta (x- y)
\label{fluid44}
\eeqar
Notice that $\rho_a$ generates left transformations on $g$, while
$j^0$ generates right transformations  along the $\sigma_3$-direction.

\subsubsection{Spin and fluids}

Another example we will briefly quote is for fluids with spin \cite{KN2}.
Consider a special case where mass transport and charge transport
are described by the same flow velocity. In other words, impose a ``constitutive relation"
$J^\mu_m = j^\mu_e$. Such a relation is reasonable
when we have one species of particles with the same charge.
Further, for dilute systems, if we neglect the possibility of
spin-singlets forming (and moving independently), we can take
spin flow velocity $\approx$ charge flow velocity, so that we can further impose
$J^\mu_m = = j^\mu_s $.
In this case, the action
(\ref{fluid28}) simplifies as
\beq
\S =  S(A) + \int d^4x~
 \Bigl[j^\mu\, (\del_\mu \theta + \alpha \del_\mu \beta + e A_\mu ) 
- { i \over 4}  j^\mu\, \Tr ( \Sigma_3 \,\Lambda^{-1} \del_\mu \Lambda )
- F( n, \sigma ) \Bigr]
\label{fluid45}
\eeq
The Lorentz group element
$\Lambda$ may be written as $\Lambda = B\, R$, where $B$ is a specific boost transformation taking us from a rest frame to a moving frame and $R$ is a spatial rotation.
Explicitly,
\beq
B(u) =  {1\over \sqrt{2  (u^0 +1 )}} \left[
\begin{matrix} u^0 + 1 & {\vec\sigma}\cdot\vu \\
{\vec\sigma}\cdot\vu & u^0 + 1 \\
\end{matrix}
\right]
\label{fluid46}
\eeq
The statement that $J^\mu_m = = j^\mu_s $ means that $B$ contains the same velocity
$u^\mu$ as for the mass transport,
 as in $j^\mu = n \, u^\mu$.
$F$ depends on $n$ and $\sigma = S^{\mu\nu} \, F_{\mu\nu}$, where
$S^{\mu\nu}$ is the spin density,
\beq
S^{\mu\nu} = {1\over 2} \, \Tr \,(\Sigma_3\, \Lambda^{-1}\, J^{\mu\nu} \, \Lambda ),
\hskip .3in J^{\mu\nu} = {i \over 4} [ \gamma^\mu , \gamma^\nu ]
\label{fluid47}
\eeq

One interesting feature which emerges from this analysis, and the equations of motion for the action
(\ref{fluid45}), is that
the spin density is subject to precession effects
due to pressure gradient terms in addition to the expected precession due to the 
magnetic field.
This is seen explicitly from the equations of motion
\beqar
u^\alpha \del_\alpha ( F' \, u_\nu ) - \del_\nu F' &=&
e \, u^\lambda \, F_{\lambda \nu} - 
{16 e\over F'}\, \del_\nu S^{\lambda \beta} ( S\, F \,S - F\, S \,S )_{\lambda \beta} ~+~ 
\cdots\nonumber\\
u^\alpha \del_\alpha S_{\mu\nu}&=& {e\over F' } \left[S_{\mu}^{~\lambda}  F_{\lambda \nu} 
- S_{\nu}^{~\lambda}  F_{\lambda \mu} \right]
+  \left[ S_\mu^{~\lambda} \,f_{\lambda \nu} 
- S_\nu^{~\lambda} \, f_{\lambda \mu} \right] \nonumber\\
&&\hskip .2in - {16 \, e\over F^{' 2}} ( u_\mu S_{\nu}^{~\lambda} - u_\nu S_{\mu}^{~\lambda} )
\del_\lambda S^{\rho \beta} ( S\, F \,S - F\, S\, S)_{\rho \beta} ~ + ~\cdots
\label{fluid48}
\eeqar
where $F' = ( \del F /\del n)$ and
\beqar
f_{\lambda \nu} &=& {1\over F'}\left[  u_\lambda \, \del_\nu F' - u_\nu \, \del_\lambda F' 
\right] \nonumber\\
(S\, F\, S - F\, S\, S )_{\lambda \beta} 
&=& S_{\lambda}^{~\rho}\, F_{\rho \tau}\, S^{\tau}_{~\beta} -
F_{\lambda}^{~\rho} \, S_{\rho \tau} \, S^{\tau}_{~\beta} 
\label{fluid49}
\eeqar
The first equation in (\ref{fluid48}) is the expected Lorentz force formula for fluids, with
corrections depending on the gradient of the spin density. The second describes the precession of the spin density in the electromagnetic field. The term $S_\mu^{~\lambda} \,f_{\lambda \nu} 
- S_\nu^{~\lambda} \, f_{\lambda \mu}$ describes a spin precession effect due to 
pressure gradient terms which can exist even in the absence of external fields.
This is a bit unusual and somewhat unexpected.

\subsection{Anomalies in fluid dynamics}

We will now consider how anomalies can affect fluid dynamics. 
Anomalies arise in the quantum
theory because of the need to regularize the theory. This involves a cut-off on the integrations
over loop momenta in various Feynman diagrams. If a situation arises that one cannot find a regulator which preserves all the classical symmetries, then we have to ensure that the regulator we choose
preserves gauge symmetries (for consistency reasons). This may mean that we have to give up
some of the other non-gauge symmetries. We say that those symmetries are anomalous.

Even though anomalies arise out of ultraviolet regulators, they have a deeper topological origin and one consequence of this aspect of the anomalies is that they are not renormalized.
Further, they can also be reproduced from infrared physics. As a result, we can expect them to be
relevant in the hydrodynamical regime as well.

\subsubsection{Anomalous electrodynamics}

First of all we will consider a very simple case, that of an Abelian $U(1)$ theory which has anomalies.
We may think of this as electromagnetism. The basic equations we need are
the conservation laws,
\beqar
\del_\mu T^\mu_{~\nu} &=& F_{\lambda \mu}  \, J^\mu \nonumber\\
\del_\mu J^\mu &=& - {c \over 8} \epsilon^{\mu\nu\alpha\beta} 
F_{\mu\nu} F_{\alpha\beta}
\label{fluid50}
\eeqar
The first equation is the expected relation for the divergence of the energy-momentum tensor.
The second one is the conservation law for charge which is anomalous, with the anomaly
as given on the right hand side. Here $c$ is a constant, the anomaly coefficient, which can be calculated
from the underlying quantum physics. The lack of conservation for the electric current
will, of course, lead to inconsistencies, so we must really regard this system as
describing a subsystem which is anomalous, with another subsystem which will cancel this anomaly
for the full system, thus avoiding any inconsistencies.
These two equations (\ref{fluid50}) are to be supplemented by the form of $T^\mu_{~\nu}$ and $J^\mu$, given by
\beqar
T^\mu_{~\nu} &=& \mu \, n \, U^\mu\, U_\nu + \delta^\mu_{~\nu} \, P\nonumber\\
J^\mu &=& n\, U^\mu + \epsilon^{\mu\nu\alpha\beta}  \left[{c \over 6}\, \mu\, U_\nu \,\del_\alpha
(\mu \, U_\beta ) + {c \over 2}\,  \mu \, U_\nu\, \del_\alpha A_\beta \right]
\label{fluid51}
\eeqar
where $\mu$ is the chemical potential corresponding to the particle number and $P$ is the pressure.
Notice that $T^\mu_{~\nu}$ has the perfect fluid form.
These equations (\ref{fluid50}) and (\ref{fluid51}) were written down by Son and Surowka
as a minimal way to incorporate anomalies \cite{son}.
We may then ask the question: Can we find an action which leads to these equations?
We may expect such an action in terms of the formalism we have developed.
Indeed such an action can be found, it is given by
\beq
\S = \int d^4x\, \left[ j^\mu ( V_\mu + A_\mu )
+ {c \over 6} \epsilon^{\mu\nu\alpha\beta} \left( A_\mu \, V_\nu \del_\alpha V_\beta
+ V_\mu \, A_\nu \del_\alpha A_\beta\right) - \mu \, \sqrt{- j^2}
+ P (\mu ) \right]
\label{fluid52}
\eeq
where $V_\mu = \del_\mu \theta + \alpha\, \del_\mu \beta$ and the flow velocity
$U^\mu$ is related to $V_\mu$ by
\beq
(V+A)_\mu = - \mu \, U_\mu\label{fluid53}
\eeq
It is not difficult to see why the action is of the form (\ref{fluid52}). The terms representing the 
anomaly must be independent of the metric, and hence it must be a differential four-form.
The only one-forms available are the electromagnetic gauge potential
$A = A_\mu \, dx^\mu$ and the velocity of the fluid for which we can use the Clebsch form,
$ V = V_\mu \, dx^\mu = d\theta + \alpha \, d\beta$. Thus we can take a linear combination
of $A \, V \, dV$ and $V\, A \, dA$. The coefficients can be fixed by comparison with
(\ref{fluid50}) and (\ref{fluid51}). This leads to the action (\ref{fluid52}).
The equations which follow from this action
have been analyzed in more detail in \cite{MAN}.

\subsubsection{Anomalies in the fluid phase of the standard model}

A more interesting scenario is where there are no gauge anomalies and we ask the question of how we can include the anomalies for the non-gauge symmetries. The most physical realization of this would be the standard model, so we will phrase our arguments in terms of it.
We may regard the fluid we are talking about as
 the quark-gluon
plasma phase for three flavors of quarks, say, $u,d,s$. In other words,
we consider a phase with thermalized $u,d,s$ quarks, so that they
must be described by fluid variables while the heavier quarks are
described by the field corresponding to each species. 
We will also neglect the
quark masses so that we have the full flavor symmetry $U(3)_{L}\times U(3)_{R}$.
Thus the group $G$ to be used in (\ref{fluid28}) is 
\begin{equation}
G=SU(3)_{c}\times U(3)_{L}\times U(3)_{R}\label{fluid54}
\end{equation}
with individual flows corresponding to the charges.
Here we want to focus on the flavor transport, as this is the sector with anomalies, so we will drop the color group $SU(3)_{c}$
from the equations to follow. 

The flavor symmetry is not
fully preserved even in the absence of masses; this is because of the anomalies. 
It may be useful at this point to recall the argument 
why we expect a term in the effective action which reproduces anomalies \cite{'tHooft}.
We set up a {\it gedanken} argument, where we consider all
 flavor symmetries to be gauged with their anomalies canceled by
an extra set of fermions; the latter will not play any role in the dynamics except for
the anomalies, so they are referred to as
spectator fermions. The full theory is nonanomalous. The usual argument is that
if, instead of the quarks, we consider the confined phase with mesons and baryons as the basic degrees of freedom, the theory will continue to remain nonanomalous.
Even though the confined phase is obtained only at low energy, anomalies, because of their
topological origin, are unaffected. Thus in the effective action for baryons and mesons,
we should be able to find
a term which reproduces the original anomalies, thereby ensuring cancellation with the
spectator fermions.
This is the Wess-Zumino term written in terms of the pseudoscalar meson fields.
Clearly, we can expect a similar reasoning for the
fluid phase where $u,d,s$ are
replaced by fluid variables. We must then have a term in the fluid
action which can reproduce the anomalies so that the cancellation
with spectator fermions still remains valid. How do we write this term?
Since we have formulated fluid dynamics in terms of
group-valued variables, the solution is almost
trivial. We can simply use the usual Wess-Zumino term, but interpret the group-valued
variables in it, not in terms of mesons, but as describing the fluid flow velocities for various flavor quantum numbers.

Adapting (\ref{fluid28}) to the case at hand with $U(3)_L \times U(3)_R$ symmetry,
the action for fluid phase of the standard model is \cite{NRR}
\beqar
\S &=& -i \int \Biggl[ j^\mu_3\, \Tr \left( {\lambda_3 \over 2} \, g_L^{-1} D_\mu \, g_L\right)
+  j^\mu_8 \,\Tr \left( {\lambda_8 \over 2} \, g_L^{-1} D_\mu \, g_L\right)
+  k^\mu_3\, \Tr \left( {\lambda_3 \over 2} \, g_R^{-1} D_\mu \, g_R\right)\nonumber\\
&&\hskip .4in + \, k^\mu_8\, \Tr \left( {\lambda_8 \over 2} \, g_R^{-1} D_\mu \, g_R\right)
+  j^\mu_0 \,\Tr \left( g_L^{-1} D_\mu \, g_L\right)
+\, k^\mu_0 \,\Tr \left( g_R^{-1} D_\mu \, g_R\right)
\Biggr]\label{fluid55}\\
&&~\hskip .4in - F(n_3, n_8, n_0, m_3, m_8,  m_0) ~+~ S_{YM} (A) ~+~ \Gamma_{WZ}(A_L, A_R, g_L\,g_R^\dagger ) - \Gamma_{WZ}(A_{L},A_{R}, \, {\mathbb 1}) \nonumber
\eeqar
The three diagonal generators correspond to
the $t_3$, $t_8$ and the identity for $U(3)_L$ and $U(3)_R$, with the corresponding
currents $j^\mu_3$, $j^\mu_8$, $j^\mu_0$ and 
$k^\mu_3$, $k^\mu_8$, $k^\mu_0$.
We have also defined
$n_{l}^{2}=j_{l}^{2}$, $m_{l}^{2}=k_{l}^{2}$ with $l=0,3,8$.
$g_L \in U(3)_L$ and $g_R \in U(3)_R$ will describe the various flow velocities; their relation to
the currents is seen upon eliminating the latter by the equations of motion.
Further,
$\Gamma_{WZ}(A_L, A_R, g_L\,g_R^\dagger )$ is the standard Wess-Zumino term
$\Gamma_{WZ}(A_L, A_R, U )$ with $U \Longrightarrow
g_L \, g_R^\dagger$.
We have also subtracted $\Gamma_{WZ}(A_{L},A_{R}, \, {\mathbb 1}) $ which is necessary to bring the analysis to the so-called Bardeen form of the anomalies \cite{KRS}. The Bardeen form is the
one which not only preserves the vector gauge symmetries, but also gives a manifestly
vector-gauge-invariant form to the remaining axial anomalies. This form is what is appropriate 
for the fluid phase.
The explicit expression for $\Gamma_{WZ}(A_L, A_R, g_L\,g_R^\dagger )$
is 
\beqar
\Gamma_{WZ} &=& -{iN\over 240\pi^2} \int_D \Tr ( dU~U^{-1})^5
-{iN \over 48\pi^2} \int_{\cal M}
\Tr [ (A_L \,dA_L +dA_L\, A_L +A^3_L)\,dU U^{-1}]\nonumber\\
&&-{iN \over 48\pi^2} \int_{\cal M}
\Tr [(A_R \,dA_R + dA_R\, A_R +A^3_R )\, U^{-1} dU]\nonumber\\
&&+{iN \over 96\pi^2} \int_{\cal M}
\Tr [ A_L \,dU U^{-1} A_L \,dU U^{-1} - A_R\, U^{-1} dU \,A_R \,U^{-1}dU]\nonumber\\
&&+{iN \over 48\pi^2} \int_{\cal M}\Tr [ A_L (dU U^{-1})^3 \!\!+\!
A_R (U^{-1} dU)^3 +dA_L\, dU\, A_R \,U^{-1}\!\! - dA_R\,
d(U^{-1})\, A_L\, U]
\nonumber\\
&&+{iN \over 48\pi^2} \int_{\cal M} \Tr [ A_R\, U^{-1}\, A_L\, U (U^{-1}dU)^2
-A_L \,U\, A_R\, U^{-1} (dU U^{-1})^2 ]\label{fluid56}\\
&&-{iN \over 48\pi^2} \int_{\cal M}
\Tr [ (dA_R \,A_R + A_R\, dA_R )\,U^{-1} \,A_L\, U
-(dA_L \,A_L + A_L \,dA_L )\,U \,A_R \,U^{-1}]\nonumber\\
&&-{iN \over 48\pi^2} \int_{\cal M}
\Tr [ A_L \,U \,A_R\, U^{-1}\,A_L \,dU U^{-1} +
A_R \,U^{-1}\, A_L \,U \,A_R \,U^{-1} dU]\nonumber\\
&&-{iN \over 48\pi^2} \int_{\cal M}
\Tr [ A^3_R\,\, U^{-1}\,A_L\, U - A^3_L\, \,U\, A_R\, U^{-1} 
+{\half }
U\, A_R \,U^{-1}\, A_L \,U\, A_R\, U^{-1} \,A_L]\nonumber
\eeqar
with $U \Longrightarrow g_L \, g_R^\dagger$. ($N$ is the number of colors, $= 3$ for us.)
This is evidently a very complicated expression and we will need to pick out some pieces to 
highlight some physical effects. The most relevant of such effects is the chiral magnetic effect.

\subsubsection{The chiral magnetic effect}

The chiral magnetic effect corresponds to the following.
In the quark-gluon plasma, in the presence of a magnetic field, there is charge separation
and a chiral induction which may be displayed as
\beq
J_0 = {e^2 \over 2 \pi^2}\, \nabla \theta \cdot {\vec B},
\hskip .3in
J_i = - {e^2 \over 2 \pi^2}\, {\dot \theta} \, B_i
\label{fluid57}
\eeq
Here $\theta$ is an axial $U(1)$ field, similar to the $\eta'$-meson.
In the plasma, we can replace ${\dot \theta}$ by the difference of the chemical
potentials corresponding to the $U(1)_L$ and $U(1)_R$ subgroups
of $U(3)_L \times U(3)_R$ as ${\dot \theta} \rightarrow \half ( \mu_L - \mu_R) $. In this case, we find
\beq
J_i = - {e^2 \over 4 \pi^2} \, ( \mu_L - \mu_R ) \, B_i
\label{fluid58}
\eeq
We see that the chiral asymmetry of chemical potentials can lead to an electromagnetic current
in the direction of the magnetic field \cite{CME}.
In the experiment with colliding heavy nuclei which produces this fluid phase, if the collision is slightly off-center, the two nuclei constitute a current which produces, for a very short time, an intense magnetic field of the order of $10^{17} \, G$.
The resulting current can be expected to produce an asymmetry in the charge distribution of
particles
coming out, with more above the plane of collision than below it.\footnote{``Above" means in the direction of the magnetic field.} Such an asymmetry is indeed experimentally observed; however, there are other possible explanations for it. So it is not entirely clear if it can be attributed to the chiral magnetic effect. Nevertheless, let us see how this
 effect can be obtained form the Wess-Zumino term (\ref{fluid56}).

The original calculation of the chiral magnetic effect is via Feynman diagrams, but we can easily see it
from our action (\ref{fluid55}).
We calculate the electromagnetic current from the Wess-Zumino term, and then restrict to
two flavors, taking $U$ to be of the form
\beq
U = e^{i\theta}\, \left[ \begin{matrix}
V&0\\
0&1\\
\end{matrix}\right]
\label{fluid59}
\eeq
The current is given by
\beqar
J^\mu &=& J_3^{\mu} +
{e \over 48 \pi^2} \,\epsilon^{\mu\nu\alpha\beta}\, \Tr ( {\cal I}_\nu \, {\cal I}_\alpha \,{\cal I}_\beta)
+ i {e^2 \over 16 \pi^2} \epsilon^{\mu\nu\alpha\beta}\, \del_\nu A_\alpha\,
\Tr\left[ ( \Sigma_{3L} + \Sigma_{3R} ) \, I_\beta \right] ~+ J^\mu_\theta\nonumber\\
J^\mu_\theta &=& - {e^2 \over 4 \pi^2} \epsilon^{\mu\nu\alpha\beta}\, \del_\nu A_\alpha\,\del_\beta \theta\,
\left[ 2 + {1\over 4} \Tr \left(\Sigma_{3L} \,\Sigma_{3R} - 1\right)\right]
\label{fluid60}
\eeqar
where $J_3^\mu$ is the contribution from the usual non-anomalous terms
in (\ref{fluid55}) and
\beq
{\cal I}_\beta = g_L^{-1}\del_\beta g_L - g_R^{-1} \del_\beta g_R, \hskip .3in
\Sigma_{3L} = g_L^{-1} \sigma_3 g_L, \hskip .3in \Sigma_{3R} = g_R^{-1} \sigma_3 g_R
\label{fluid61}
\eeq
If we further restrict to $g_L = g_R$ (effectively setting $V =1$ at this stage)
we get 
\beqar
J^\mu_\theta &=& - {e^2 \over 2 \pi^2} \epsilon^{\mu\nu\alpha\beta} (\del_\nu A_\alpha) \, \del_\beta \theta\nonumber\\
J_i &=& - {e^2 \over 4 \, \pi^2} \, (\mu_L - \mu_R) \, B_i\, \label{fluid62}
\eeqar
In the second line, we wrote out $J_i$ and replaced ${\dot \theta}$ by 
$\half (\mu_L - \mu_R)$. This equation reproduces the chiral magnetic effect \cite{CME}.
The full set of equations are necessary to
describe full hydrodynamic transport of flavor charges.

There are many other anomaly related effects, such as a possible pion asymmetry \cite{NRR}
or chiral vorticity effects. But the present discussion suffices to illustrate the main
issues of principle.

\section{Comment on the metaplectic correction}

In subsection 7.1, we considered the quantization of the symplectic form
$i dz \wedge d\bz$, obtaining the standard coherent states. The quantum operator
corresponding to $\bz z$ was also identified  as $z \del_z + {\half}$, where the extra term $\half$
is due to the metaplectic correction.
We now consider a set of symplectic transformations which can elucidate the 
meaning and importance of the metaplectic structure.

The symplectic form $\Omega = i dz \wedge d\bz$ is invariant under
the infinitesimal transformations
\beq
z \rightarrow  z' = z + i \,A \, z + B \, \bz , \hskip .3in
\bz \rightarrow \bz' = \bz -i \, A\, \bz + B^* \, z
\label{com5}
\eeq
where $A$ is real. The finite version of these trasnformations form the
$Sp (1, \mathbb{R})$ group.
We can also introduce real variables $(p, q)$ by
\beq
z = {1\over \sqrt{2}} \, ( p + i q) , \hskip .3in
\bz = {1\over \sqrt{2}} \, ( p - i q) 
\label{com6}
\eeq
for which $\Omega = dp \wedge dq$. This would be convenient for choosing real polarizations
such as wave functions which only depend on $q$.
The transformations (\ref{com5}) do not preserve holomorphicity and these are what help to connect different polarizations. 
For example, consider for simplicity the case of $A =0$; then we have
\beq
\del_{\bz} \approx \del_{\bz'}  + {\bar B} \, \del_{z'}, \hskip .3in
\del_{\bz'} \approx \del_{\bz}  - {\bar B} \, \del_{z}
\label{com6a}
\eeq
We see that the holomorphic polarization in terms of $z, \, \bz$ is not the same
as the holomorphic polarization in terms of $z', \, \bz'$.
Thus the transformations (\ref{com5}) help implement infinitesimal
changes of polarization.
Classically, we have a closed Poisson bracket algebra for the generators of the transformations,
\beqar
&&\{ f_A\, , f_B \} = -i \, f_B, \hskip .3in 
\{ f_A\, , f_{\bar B} \} = i \, f_{\bar B}\nonumber\\
&&\{ f_{\bar B}\, , f_B \} = - 2i \, f_A
\label{com6b}
\eeqar
for $f_A = {\half} z \bz$, $f_B = -( i /2) z^2$, $ f_{\bar B} = (i/2) \bz^2$.

{\it In going to the quantum theory, since we need to have the facility of changing
polarizations, the unitary implementation
of (\ref{com5}) is important.}
The operators corresponding to $\bz$, $z$ are the annihilation and creation
operators $a$, $a^\dagger$, respectively, with $[ a, a^\dagger ] = 1$.
The quantum version of $f_B =  - (i/2) z^2$,
$f_{\bar B} = (i/2) \bz^2$ are unambiguously given by the prequantum operators
as 
\beq
{\hat f}_B = - {i\over 2}  a^{\dagger 2}, \hskip .3in 
{\hat f}_{\bar B} = {i \over 2}  a^2
\label{com6c}
\eeq
Their commutator is given by
\beqar
[ {\hat f}_{\bar B}, {\hat f}_{ B} ] &=&  \left( a^\dagger  a + {1\over 2}\right)
= i \, (- 2 i)  \left[ {\half} ( a^\dagger a + {\half} ) \right]\nonumber\\
&\equiv& - 2 i \, {\hat f}_A
\label{com7}
\eeqar
We see that the closure of the algebra and the quantum implementation of
(\ref{com6c}) requires us to identify ${\hat f}_A = a^\dagger a + {\half}$
as the quantum generator of the $A$-type transformations.
{\it The essence of the metaplectic correction is thus
the quantum realization of the $Sp (1, \mathbb{R})$.}

Notice that while the quantum operator corresponding to
$\bz z$ is identified as $a^\dagger  a + {\half}$, there is no statement about
whether one should use this operator for a Hamiltonian.
We bring up this point because, sometimes in the literature, one finds
the statement that the half-form quantization is needed as it
leads to
 the ``correct" quantization which should have
the zero-point energy if one applies this to the harmonic oscillator
(for which the classical Hamiltonian is $\bz z$). 
This statement certainly needs some clarification.
The classical Hamiltonian for the oscillator
is $\bz z+ C$ for any constant $C$, so the question of zero-point energy is
completely different.
To sharpen this point, consider the free relativistic scalar field
which can be considered as a collection of harmonic oscillators.
In fact, with the mode expansion
\beqar
\phi (x) &=& \sum_k Z_k \, u_k (x) + {\bar Z}_k \, u^*_k (x)\nonumber\\
{\dot \phi}(x)= \pi (x) &=& \sum_k (-i \omega_k) \left( Z_k \, u_k (x) - {\bar Z}_k \, u^*_k (x) \right)\nonumber\\
u_k (x) &=& {1\over \sqrt{2 \omega_k V} } \, e^{-ik\cdot x},
\hskip .3in \omega_k = \sqrt{k^2 + m^2}
\label{com2}
\eeqar
(for scalar fields in a cubical box of volume $V$ with periodic boundary conditions),
we find the symplectic structure and classical Hamiltonian
\beq
\Omega = i \prod_k d Z_k \wedge d {\bar Z}_k, \hskip .3in
H = \sum_k {\bar Z}_k Z_k + C
\label{com3}
\eeq
Classically, this is indeed a collection of harmonic oscillators.
In quantizing this, keeping any nonzero value for the zero-point 
energy is the wrong thing to do.
For this problem, we want to obtain a unitary realization of the
Poincar\'e group. One of the commutation rules for this group is
\beq
[ P_i, K_j ] = i\, \delta_{ij} \, H
\label{com4}
\eeq
where $P_i$ is the momentum operator and $K_j$ is the Lorentz boost generator. The Lorentz invariance of the vacuum requires $K_j \vert 0 \ra = 0$. As a result, we must have
$\la 0 \vert \,H \, \vert 0 \ra = 0$ (upon taking the expectation value of
(\ref{com4})), showing that the quantization we need should have no zero-point energy.
The generators of the symplectic transformations do have
the extra metaplectic correction, but the choice of the Hamiltonian
(and how it should represented as an operator) is determined by imposing
desirable symmetries.
More explicitly, the relevant algebraic relations for the symplectic transformations are
\beq
[ a_k a_l , a^\dagger_r a^\dagger_s ]
= \delta_{k r} a^\dagger_s a_l
+  \delta_{l r} a^\dagger_s a_k +  \delta_{k s} a^\dagger_r a_l +  \delta_{l s} a^\dagger_r a_k
+ \left( \delta_{ks} \delta_{lr} + \delta_{kr} \delta_{ls}\right)
\label{com8}
\eeq
One does realize this algebra unitarily on the Fock space of the theory.
(The finite transformations corresponding (\ref{com8}) are also what are
used to generate squeezed states in quantum optics.)
The Hamiltonian however is one of the generators of the Poincar\'e algebra,
given by
$H = \sum_k \omega_k \, a^\dagger_k a_k $ (with no term corresponding to the zero-point energy)
and $P_i = \sum_k k_i \, a^\dagger_k a_k$.


\end{document}